\documentclass{elsarticle}

\journal{Information and Computation}
\date{}

\usepackage{url}
\usepackage{amsmath}
\usepackage{amssymb}
\usepackage{verbatim}
\usepackage{hhline}
\usepackage{dcolumn}
\usepackage{ifthen}
\usepackage[thinlines]{easybmat}

\newboolean{LONGVERSION}
\setboolean{LONGVERSION}{true}

\newtheorem{theorem}{Theorem}[section]
\newtheorem{lemma}[theorem]{Lemma}
\newtheorem{corollary}[theorem]{Corollary}
\newdefinition{definition}[theorem]{Definition}
\newdefinition{example}[theorem]{Example}
\newproof{pf}{Proof}

\providecommand*{\Nset}{\mathbb{N}}            
\providecommand*{\Zset}{\mathbb{Z}}            
\providecommand*{\Qset}{\mathbb{Q}}            
\providecommand*{\Rset}{\mathbb{R}}            

\newcommand{\length}[1]{\lvert#1\rvert}

\newcommand*{\vect}[1]{\mathchoice{\mbox{\boldmath$\displaystyle#1$}}
{\mbox{\boldmath$\textstyle#1$}}
{\mbox{\boldmath$\scriptstyle#1$}}
{\mbox{\boldmath$\scriptscriptstyle#1$}}} 
\newcommand*{\mat}[1]{\vect{#1}}

\newcolumntype{.}{D{.}{.}{-1}}

\newcolumntype{,}{D{,}{,}{-1}}

\newcommand*{\transpose}{{\scriptscriptstyle\mathrm{T}}}

\newcommand{\kw}{\textbf}

\newcommand*{\assign}{\mathrel{\mathord{:}\mathord{=}}}

\renewcommand{\emptyset}{\mathord{\varnothing}}

\newcommand*{\union}{\cup}
\newcommand*{\inters}{\cap}

\newcommand*{\biginters}{\bigcap}
\newcommand*{\setdiff}{\setminus}

\newcommand*{\sseq}{\subseteq}

\newcommand*{\sslt}{\subset}

\newcommand{\sset}[2]{{\renewcommand{\arraystretch}{1.2}
                      \left\{\,#1 \,\left|\,
                               \begin{array}{@{}l@{}}#2\end{array}
                      \right.   \,\right\}}}

\newcommand{\defeq}{\mathrel{\mathord{:}\mathord{=}}}

\newcommand*{\reld}[3]{\mathord{#1}\subseteq#2\times#3}
\newcommand*{\fund}[3]{\mathord{#1}\colon#2\rightarrow#3}

\newcommand*{\Cplusplus}{{C\nolinebreak[4]\hspace{-.05em}\raisebox{.4ex}{\tiny\bf ++}}}

\newcommand{\st}{\mathrel{.}}
\newcommand{\itc}{\mathrel{:}}

\newcommand*{\pif}{\mathrel{\mathord{:}\mathord{-}}}

\newcommand*{\extend}[1]{\widehat{#1}}

\newcommand*{\cA}{\ensuremath{\mathcal{A}}}

\newcommand*{\cD}{\ensuremath{\mathcal{D}}}

\newcommand*{\cL}{\ensuremath{\mathcal{L}}}

\newcommand*{\cO}{\ensuremath{\mathcal{O}}}

\newcommand*{\cT}{\ensuremath{\mathcal{T}}}

\begin{document}

\begin{frontmatter}

\ifthenelse{\boolean{LONGVERSION}}{
\title%
{The Automatic Synthesis of Linear Ranking Functions:\\
The Complete Unabridged Version\tnoteref{th}}
}{
\title%
{A New Look at the Automatic Synthesis
 of Linear Ranking Functions\tnoteref{th}}
}

\tnotetext[th]{This work has been partly supported by PRIN project
``AIDA---Abstract Interpretation: Design and Applications''
and by the Faculty of Sciences of the University of Parma.}

\author[Parma,BUGSENG]{Roberto Bagnara}
\ead{bagnara@cs.unipr.it}
\author[Reunion]{Fred Mesnard}
\ead{Frederic.Mesnard@univ-reunion.fr}
\author[UNIPR]{Andrea Pescetti}
\ead{andrea.pescetti@unipr.it}
\author[Parma,BUGSENG]{Enea Zaffanella}
\ead{zaffanella@cs.unipr.it}

\address[Parma]{Dipartimento di Matematica, Universit\`a di Parma, Italy}
\address[BUGSENG]{BUGSENG srl, \textup{\url{http://bugseng.com}}}
\address[Reunion]{IREMIA, LIM, Universit\'e de la R\'eunion, France}
\address[UNIPR]{Universit\`a di Parma, Italy}

\begin{abstract}
The classical technique for
proving termination of a generic sequential computer program involves the
synthesis of a \emph{ranking function} for each loop of the program.
\emph{Linear} ranking functions are particularly interesting because
many terminating loops admit one and algorithms exist to automatically
synthesize it.  In this paper we present two such algorithms:
one based on work dated 1991 by Sohn and Van~Gelder; the other,
due to Podelski and Rybalchenko, dated 2004.
Remarkably, while the two algorithms will synthesize a linear ranking
function under exactly the same set of conditions, the former is
mostly unknown to the community of termination analysis and its general
applicability has never been put forward before the present paper.
In this paper we thoroughly justify both algorithms, we prove their
correctness, we compare their worst-case complexity and experimentally
evaluate their efficiency, and we present an open-source implementation
of them that will make it very easy to include termination-analysis
capabilities in automatic program verifiers.
\end{abstract}



\end{frontmatter}

\section{Introduction}

Termination analysis of computer programs (a term that here we interpret in
its broadest sense) consists in attempting to determine whether execution
of a given program will definitely terminate for a class of its possible
inputs.
The ability to anticipate the termination behavior of programs (or
fragments thereof) is essential to turn assertions of \emph{partial
correctness} (\emph{if} the program reaches a certain control point,
\emph{then} its state satisfies some requirements) into assertions
of \emph{total correctness} (the program \emph{will} reach that point
\emph{and} its state will satisfy those requirements).
It is worth observing that the property of termination of a program
fragment is not less important than, say,
properties concerning the absence of run-time errors.
For instance, critical reactive systems (such as fly-by-wire avionics
systems) must maintain a continuous interaction with the environment:
failure to terminate of some program components can stop the interaction
the same way as if an unexpected, unrecoverable run-time error occurred.

Developing termination proofs by hand is, as any other program verification
task, tedious, error-prone and, to keep it short, virtually impossible to
conduct reliably on programs longer than a few dozens of lines.
For this reason, automated termination analysis has been a hot research
topic for more than two decades.
Of course, due to well-known limitative results of computation theory,
any automatic termination analysis can only be expected to give the
correct answer (``the program does ---or does not--- terminate
on these inputs'') for some of the analyzed programs and inputs:
for the other programs and inputs the analysis will be inconclusive
(``don't know'').
It is worth noticing that there is no need to resort to the halting
problem to see how hard proving termination can be.
A classical example is the $3x+1$ problem,\footnote{Also known as
the Collatz problem, the Syracuse problem, Kakutani's problem,
Hasse's algorithm, and Ulam's problem: see, e.g.,
\cite{Lagarias85a}.}
whose termination for any $n$ has been a conjecture for more than 70 years:
\[
\begin{aligned}
  &\kw{while} \; n > 1 \; \kw{do} \\
  &\qquad \kw{if} \; (n \mathbin{\kw{mod}} 2) \neq 0 \; \kw{then} \; n \assign 3n+1 \\
  &\qquad \kw{else}\;  n \assign n \mathbin{\kw{div}} 2
\end{aligned}
\]

The classical technique for proving termination of a generic sequential computer program
consists in selecting, for each loop $w$ of the program:
\begin{enumerate}
\item
a set $S_w$ that is \emph{well-founded} with respect to a relation
$R_w \sseq S_w \times S_w$; namely, for each $U \sseq S_w$ such that
$U \neq \emptyset$, there exists $v \in U$ such that $(u, v) \notin R_w$
for each $u \in U$;
\item
a function $f_w$ from the set of program states that are relevant for $w$
(e.g., those concerning the head of the loop and that are reachable from
a designated set of initial states)
to the set $S_w$, such that the values of $f_w$ computed at any two subsequent
iterations of $w$ are in relation $R_w$.
\end{enumerate}
The function $f_w$ is called \emph{ranking function}, since it ranks program
states according to their ``proximity'' to the final states.
Let us focus on deterministic programs, and consider a loop $w$ and
a set of initial states $\Sigma_w^\mathrm{I}$ for $w$.
Assume further that the body of $w$ always terminates when
$w$ is initiated in a state $\sigma \in \Sigma_w^\mathrm{I}$ and that
$\Sigma_w^\mathrm{F}$ is a set of final states for $w$,
that is, $w$ immediately terminates when it is initiated
in a state $\sigma \in \Sigma_w^\mathrm{F}$.
If we fix any enumeration of
$\Sigma_w^\mathrm{I} = \{ \sigma_0^0, \sigma_1^0, \ldots \}$,
then the computations of $w$ we are interested in
can be represented by the (possibly infinite) sequence of (possibly infinite)
sequences
\begin{equation}
\label{eq:sequence-of-sequences}
\begin{matrix}
  \sigma_0^0 &\sigma_0^1 &\ldots \cr
  \vdots &\vdots&\ddots \cr
  \sigma_i^0 &\sigma_i^1 &\ldots \cr
  \vdots &\vdots&\ddots \cr
\end{matrix}
\end{equation}
Let $\Sigma_w$ be the set of all states that occur
in~\eqref{eq:sequence-of-sequences}.
Suppose that we succeed in finding a ranking function
$\fund{f_w}{\Sigma_w}{S_w}$,
where $S_w$ is well-founded with respect to $R_w$ and,
for each $m, n \in \Nset$, if $\sigma_m^n$ and $\sigma_m^{n+1}$ occur
in~\eqref{eq:sequence-of-sequences},
then $\bigl(f_w(\sigma_m^{n+1}), f_w(\sigma_m^n)\bigr) \in R_w$.
In this case we know that all the sequences
in~\eqref{eq:sequence-of-sequences}, and hence all the computations
they represent, are finite.

\begin{example}
\label{ex:rf-on-naturals}
Consider the following loop, where $x$ takes values in $\Zset$:
\[
\begin{aligned}
  &\kw{while} \; x \neq 0 \; \kw{do} \\
  &\qquad x \assign x-1
\end{aligned}
\]
Here the state at the loop head can be simply characterized by
an integer number: the value of $x$.
If we take $\Sigma^\mathrm{I} \defeq \Nset$ then the computation sequences
of interest are
\begin{equation*}
\begin{matrix}
  0      & \cr
  1      & 0      & \cr
  \vdots & \vdots & \ddots \cr
  n      & n-1    & \ldots & 0 \cr
  \vdots &\vdots  &        &   &   & \ddots \cr
\end{matrix}
\end{equation*}
We thus have $\Sigma = \Nset$ and $\Sigma^\mathrm{F} = \{ 0 \}$.
If we define $S \defeq \Nset$,
$f$ as the identity function over $\Nset$,
and $R \defeq \bigl\{\, (h, k) \bigm| h, k \in \Nset, h < k \,\bigr\}$,
then $S$ is well founded with respect to $R$ and
$f$ is a ranking function (with respect to $\Sigma$, $S$ and $R$).
\end{example}
\ifthenelse{\boolean{LONGVERSION}}{
Observe that, in the example above, taking $R$ as the predecessor
relation, i.e.,
$R \defeq \bigl\{\, (h, k) \bigm| h, k \in \Nset, h = k - 1 \,\bigr\}$,
would have worked too; or $f$ could have been defined as the
function mapping $h$ to $2h$, in which case $S$ could have been left
as before or defined as the set of even nonnegative integers\dots.
In general, if a ranking function exists, an infinite number of them do
exist.
}

\ifthenelse{\boolean{LONGVERSION}}{
The next example shows that freedom in the choice of the well-founded
ordering can be used to obtain simpler ranking functions.
\begin{example}
\label{ex:rf-on-more-complex-wfo}
Consider the following program, where variables take values in $\Nset$
and comments in braces describe the behavior of deterministic program
fragments that are guaranteed to terminate and whose details are unimportant:
\[
\begin{aligned}
  &\kw{var} \; a: \; \kw{array} [ 1 \mathrel{..} n ] \;
    \kw{of unsigned integer}; \\
  &\{\; \text{all elements of $a$ are written} \; \} \\
  & \kw{while} \; a[1] > 0 \; \kw{do} \\
  &\qquad \{\; \text{$i$ takes a value between $1$ and $n$
                     such that $a[i] \neq 0$} \; \} \\
  &\qquad a[i] \assign a[i]-1 \\
  &\qquad \{\; \text{positions $i+1$, $i+2$, \dots, $n$ of $a$
                     are possibly modified} \; \}
\end{aligned}
\]
Here we can take $\Sigma^\mathrm{I} = \Sigma = \Nset^n$
and $\Sigma^\mathrm{F} = \{ 0 \} \times \Nset^{n-1}$.
If we define $S \defeq \Nset^n$,
$f$ as the identity function over $\Nset^n$,
and $R \sslt \Nset^n \times \Nset^n$ as the lexicographic ordering
over $\Nset^n$, then $f$ is a ranking function with respect
to $\Sigma$, $S$ and $R$.
Finding a ranking function having $\Nset$ as codomain
would have been much more difficult and could not be done
without a complete knowledge of the program fragments we have
summarized with the comments between braces.
\end{example}
}

We have seen that, if there exists a ranking function, then all
computations summarized by~\eqref{eq:sequence-of-sequences} terminate.
What is interesting is that the argument works also the other way
around: if all the computations summarized
by~\eqref{eq:sequence-of-sequences} do terminate, then there exists a
ranking function (actually, there exists an infinite number of them).
In fact, suppose all the sequences in~\eqref{eq:sequence-of-sequences}
are finite.  Since the program is deterministic, any state occurs only
once in every sequence.  Moreover, if a state $\sigma$ occurs in more
than one sequence, then the suffixes of these sequences that immediately
follow $\sigma$ are all identical (since the future of any computation is
completely determined by its current state).
The function mapping each $\sigma$ in $\Sigma_w$ to the natural number
representing the length of such suffixes is thus well defined
and is a ranking function with respect to $\Sigma_w$ and $\Nset$
with the well-founded ordering given by the `$<$' relation.

It is worth observing that the above argument implies that
if any ranking function exists, then there exists a ranking function
over $(\Nset, \mathord{<})$.
This observation can be generalized to programs having bounded
nondeterminism~\cite{Dijkstra76}: therefore, ranking functions on
the naturals are sufficient, for instance, when modeling the input
of values for commonly available built-in data types.
\ifthenelse{\boolean{LONGVERSION}}{
However, as illustrated by Example~\ref{ex:rf-on-more-complex-wfo},
the use of more general well-founded orderings can simplify the
search for a ranking function.
}{
However,
the use of more general well-founded orderings can simplify the
search for a ranking function
(see, e.g., \cite[Example~1.2]{BagnaraMPZ12TR}).
}
Moreover, such a generalization is mandatory when dealing with
unbounded nondeterminism~\cite{Dijkstra76}
(see also~\cite[Section~10]{CousotC85}).

The termination of a set of computations and the existence of a
ranking function for such a set are thus completely equivalent.
On the one hand, this means that trying to prove that a ranking
function exists is, at least in principle, not less powerful than any
other method we may use to prove termination.
On the other hand, undecidability of the termination problem implies
that the existence of a ranking function is also undecidable.
An obvious way to prove the existence of a ranking function is
to \emph{synthesize} one from the program text and a description
of the initial states: because of undecidability, there exists
no algorithm that can do that in general.

The use of ranking functions as a tool to reason about termination
can be traced back to the seminal work of R.~W.~Floyd in \cite{Floyd67},
where they are introduced under the name of \emph{$W$-functions}.
Since then, several variations of the method have been proposed
so as to extend its applicability from the realm of classical sequential
programs to more general constructs (e.g., concurrency).
In particular, in \cite{CousotC85}, seven different `\emph{\`a la Floyd}'
induction principles for nondeterministic transition systems are formally
shown to be sound, semantically complete and equivalent.
For instance,
it is shown that it is sufficient to consider a single, global ranking
function, instead of a different ranking function for each program
control point, as originally proposed in \cite{Floyd67};
and that the decrease of such a global ranking function need not be
verified at all program control points, but it is enough to consider
a minimal set of \emph{loop cut-points};
moreover, when trying to prove properties
that only depend on the current state of the system
(e.g., termination of a deterministic program),
it is always possible to find a ranking function depending
on the current state only,
i.e., independent of the initial state of the system.
Note that these results have been implicitly exploited in the examples
above so as to simplify the presentation of the method.

In this paper we present, in very general terms so as to
encompass any programming paradigm, the approach to
termination analysis based on the explicit search of ranking
functions.  We then restrict attention to linear ranking functions
obtained from linear approximations of the program's semantics.
For this restriction, we present and fully justify two
methods to prove the existence of linear ranking functions:
one, based on work dated 1991 by Sohn and Van~Gelder, that is
almost unknown outside the field of logic programming even though,
as we demonstrate in the present paper, it is completely general;
the other, due to Podelski and Rybalchenko, dated 2004, was proved
correct by the authors but the reasons why it works were never
presented.  We then provide a proof of equivalence of the two
methods, thus providing an independent assessment of their
correctness and relative completeness.  We also compare their
theoretical complexity and practical efficiency on three related problems:
\begin{enumerate}
\item
proving that one linear ranking function exists;
\item
exhibiting one such function;
\item
computing the space of all linear ranking functions.
\end{enumerate}
The experimental evaluation is based on the implementation of
the two methods provided by the \emph{Parma Polyhedra Library}
\cite{BagnaraHZ08SCP}, a free software library of numerical
abstraction targeted at software/hardware analysis and verification.
These implementations are, to the best of our knowledge, the first ones
that are being made available, in source form, to the community.
For this reason, the implementations should be regarded as complementary
to the present paper in the common aim of making the automatic synthesis
of linear ranking functions known outside programming language barriers,
understandable and accessible.

The plan of the paper is as follows:
Section~\ref{sec:preliminaries} recalls preliminary notions and
introduces the notation used throughout the paper;
Section~\ref{sec:termination-analysis-of-individual-loops}
introduces the problem of automatic termination analysis of individual
loops and its solution technique based on the synthesis of ranking
functions;
Section~\ref{sec:sohn-and-van-gelder} presents a simple generalization
of the approach of \cite{SohnVG91} that is generally applicable to
termination analysis of any language;
Section~\ref{sec:podelski-and-rybalchenko} shows and fully justifies
the approach of \cite{PodelskiR04};
Section~\ref{sec:comparison} proves the two methods are equivalent
and compares them from the point of view of computational complexity;
Section~\ref{sec:implementation-and-experimental-evaluation}
presents the implementation of the two approaches offered by the
Parma Polyhedra Library and the corresponding experimental evaluation,
providing a comparison of their practical efficiency;
Section~\ref{sec:conclusions} concludes.

\ifthenelse{\boolean{LONGVERSION}}{
}{
Readers who are interested in a more detailed exposition of the themes
treated in this paper are referred to its technical
report version~\cite{BagnaraMPZ12TR}.
}

\section{Preliminaries}
\label{sec:preliminaries}

\ifthenelse{\boolean{LONGVERSION}}{
\subsection{Set Theory}

The set of all finite sequences of elements of $S$ is denoted by $S^\ast$.
The empty sequence is denoted by $\varepsilon$ and the length of a sequence
$w$ is denoted by $\length{w}$.

The set of non-negative integers, rationals and reals are denoted by
$\Nset$, $\Qset_+$ and $\Rset_+$, respectively.

\subsection{Linear Algebra}

For each $i \in \{1, \ldots, n\}$, $v_i$ denotes the $i$-th component
of the real (column) vector
$\vect{v} = \langle v_1, \ldots, v_n \rangle \in \Rset^n$.
A vector $\vect{v} \in \Rset^n$ can also be interpreted as a
matrix in $\Rset^{n \times 1}$ and manipulated accordingly with the
usual definitions for addition, multiplication (both by a scalar and
by another matrix), and transposition, which is denoted by
$\vect{v}^\transpose$, so that
\(
  \langle v_1, \ldots, v_n \rangle = (v_1, \ldots, v_n)^\transpose
\).
If $\vect{v} \in \Rset^n$ and $\vect{w} \in \Rset^m$, we will
write $\langle \vect{v}, \vect{w} \rangle$ to denote the column vector
in $\Rset^{n+m}$ obtained by ``concatenating'' $\vect{v}$ and $\vect{w}$,
so that
\(
  \langle \vect{v}, \vect{w} \rangle
    = \langle v_1, \ldots, v_n, w_1, \ldots, w_m \rangle
\).
The \emph{scalar product} of $\vect{v},\vect{w} \in \Rset^n$
is the real number
\(
  \vect{v}^\transpose \vect{w} = \sum_{i=1}^{n} v_i w_i
\).
The identity matrix in $\Rset^{n \times n}$ is denoted by $\mat{I}_n$.
We write $\vect{0}$ to denote a matrix in $\Rset^{n \times m}$
having all of its components equal to zero; the dimensions $n$ and $m$
will be clear from context. We sometimes treat scalars as vectors
in $\Rset^1$ or matrices in $\Rset^{1 \times 1}$.

For any relational operator
$\mathord{\bowtie} \in \{ <, \leq, =, \geq, > \}$, we write
$\vect{v} \bowtie \vect{w}$ to denote the conjunctive
proposition $\bigwedge_{i=1}^{n} (v_i \bowtie w_i)$. Moreover,
$\vect{v} \neq \vect{w}$ will denote the proposition
$\neg (\vect{v} = \vect{w})$.
We will sometimes use the convenient notation
$a \bowtie_1 b \bowtie_2 c$ to denote the conjunction
$a \bowtie_1 b \land b \bowtie_2 c$ and we will not distinguish
conjunctions of propositions from sets of propositions.
The same notation applies to vectors defined over other numeric
fields and, for the supported operations, to vectors defined
over numeric sets such as $\Nset$ and $\Qset_+$.

\subsection{First-Order Logic}

A triple $\Sigma = (S, F, R)$ is a \emph{signature}
if $S$ is a set of \emph{sort symbols},
$F \defeq (F_{w,s})_{w \in S^\ast, s \in S}$
is a family of sets of \emph{function symbols}
and $R \defeq (R_w)_{w \in S^\ast}$
is a family of sets of \emph{relation symbols} (or \emph{predicate symbols}).
If $F_{s_1 \cdots s_n, s} \ni f$ we use the standard notation for functions
and write $F \ni \fund{f}{s_1 \times \cdots \times s_n}{s}$.
Similarly, if $R_{s_1 \cdots s_n} \ni p$ we use the standard notation for
relations and write $R \ni p \subseteq s_1 \times \cdots \times s_n$.
A \emph{$\Sigma$-structure} $\cA = (S^\cA, F^\cA, R^\cA)$
consists of: a set $S^\cA$ containing one arbitrary set $s^\cA$ for each
sort symbol $s \in S$; a family $F^\cA$ of sets of functions such that,
for each $\fund{f}{s_1 \times \cdots \times s_n}{s}$, a function
$\fund{f^\cA}{s^\cA_1 \times \cdots \times s^\cA_n}{s^\cA}$ belongs to $F^\cA$;
a family $R^\cA$ of sets of relations defined similarly.

Let $X$ be a denumerable set of \emph{variable symbols}.
The set of \emph{$(\Sigma, X)$-terms} (or, briefly, \emph{terms})
is inductively defined as usual: elements of $X$ are terms and,
for each $f \in F_{w,s}$ with $\length{w} = k$, if $t_1$, \dots,~$t_k$
are terms, then $f(t_1, \ldots, t_k)$ is a term.
If $p \in R_w$ with $\length{w} = k$ and $t_1$, \dots,~$t_k$
are terms, then $p(t_1, \ldots, t_k)$ is an
\emph{atomic $(\Sigma, X)$-formula}.
\emph{$(\Sigma, X)$-formulas} are built as usual from atomic formulas and
logical connectives and quantifiers.
The first-order language $\cL(\Sigma, X)$ is the set of all
\emph{$(\Sigma, X)$-formulas}.
The notion of \emph{bound} and \emph{free variable occurrence} in a formula
are also defined in the standard way.
We will routinely confuse a tuple of variables with the set of its components.
So, if $\phi$ is a $(\Sigma, X)$-formula, we will write $\phi[\bar{x}]$
to denote $\phi$ itself, yet emphasizing that the set of free
variables in $\phi$ is included in $\bar{x}$.
Let $\bar{x}, \bar{y} \in X^\ast$ be of the same length
and let $\phi$ be a $(\Sigma, X)$-formula:
then $\phi[\bar{y}/\bar{x}]$ denotes the formula obtained by simultaneous
renaming of each free occurrence in $\phi$ of a variable in $\bar{x}$
with the corresponding variable in $\bar{y}$, possibly renaming bound
variable occurrences as needed to avoid variable capture.
Notice that $\phi[\bar{x}]$ implies
$\bigl(\phi[\bar{y}/\bar{x}]\bigr)[\bar{y}]$, for each
admissible $\bar{y} \in X^\ast$.

A formula with no free variable occurrences is termed \emph{closed}
or called a \emph{sentence}.
The \emph{universal closure} of a formula $\phi$ is denoted by
$\forall(\phi)$.
If $\phi$ is a closed $(\Sigma, X)$-formula and $\cA$ is a $\Sigma$-structure,
we write $\cA \models \phi$ if $\phi$ is satisfied when interpreting
each symbol in $\Sigma$ as the corresponding object in $\cA$.
A set $\cT$ of closed $(\Sigma, X)$-formulas is called a
\emph{$(\Sigma, X)$-theory}.  We write $\cA \models \cT$ if
$\cA \models \phi$ for each $\phi \in \cT$.
If $\phi$ is a closed $(\Sigma, X)$-formula and $\cT$ is
a $(\Sigma, X)$-theory, we write $\cT \models \phi$ if, for each
$\Sigma$-structure $\cA$, $\cA \models \cT$ implies $\cA \models \phi$.
In this case we say that $\phi$ is a \emph{logical consequence} of $\cT$.
}{
\paragraph*{Set Theory}

The set of all finite sequences of elements of $S$ is denoted by $S^\ast$.
The empty sequence is denoted by $\varepsilon$ and the length of a sequence
$w$ is denoted by $\length{w}$.
The set of non-negative integers, rationals and reals are denoted by
$\Nset$, $\Qset_+$ and $\Rset_+$, respectively.

\paragraph*{Linear Algebra}

For each $i \in \{1, \ldots, n\}$, $v_i$ denotes the $i$-th component
of the real (column) vector
$\vect{v} = \langle v_1, \ldots, v_n \rangle \in \Rset^n$.
A vector $\vect{v} \in \Rset^n$ can also be interpreted as a
matrix in $\Rset^{n \times 1}$ and manipulated accordingly with the
usual definitions for addition, multiplication (both by a scalar and
by another matrix), and transposition, which is denoted by
$\vect{v}^\transpose$, so that
\(
  \langle v_1, \ldots, v_n \rangle = (v_1, \ldots, v_n)^\transpose
\).
If $\vect{v} \in \Rset^n$ and $\vect{w} \in \Rset^m$, we will
write $\langle \vect{v}, \vect{w} \rangle$ to denote the column vector
in $\Rset^{n+m}$ obtained by ``concatenating'' $\vect{v}$ and $\vect{w}$,
so that
\(
  \langle \vect{v}, \vect{w} \rangle
    = \langle v_1, \ldots, v_n, w_1, \ldots, w_m \rangle
\).
The \emph{scalar product} of $\vect{v},\vect{w} \in \Rset^n$
is the real number
\(
  \vect{v}^\transpose \vect{w} = \sum_{i=1}^{n} v_i w_i
\).
The identity matrix in $\Rset^{n \times n}$ is denoted by $\mat{I}_n$.
We write $\vect{0}$ to denote a matrix in $\Rset^{n \times m}$
having all of its components equal to zero; the dimensions $n$ and $m$
will be clear from context. We sometimes treat scalars as vectors
in $\Rset^1$ or matrices in $\Rset^{1 \times 1}$.
For any relational operator
$\mathord{\bowtie} \in \{ <, \leq, =, \geq, > \}$, we write
$\vect{v} \bowtie \vect{w}$ to denote the conjunctive
proposition $\bigwedge_{i=1}^{n} (v_i \bowtie w_i)$. Moreover,
$\vect{v} \neq \vect{w}$ will denote the proposition
$\neg (\vect{v} = \vect{w})$.
We will sometimes use the convenient notation
$a \bowtie_1 b \bowtie_2 c$ to denote the conjunction
$a \bowtie_1 b \land b \bowtie_2 c$ and we will not distinguish
conjunctions of propositions from sets of propositions.
The same notation applies to vectors defined over other numeric
fields and, for the supported operations, to vectors defined
over numeric sets such as $\Nset$ and $\Qset_+$.

\paragraph*{First-Order Logic}

Let $\cL$ be a first-order language with variables in $X$.
We will routinely confuse a tuple of variables with the set of its components.
So, if $\phi$ is an $\cL$-formula, we will write $\phi[\bar{x}]$
to denote $\phi$ itself, yet emphasizing that the set of free
variables in $\phi$ is included in $\bar{x}$.
Let $\bar{x}, \bar{y} \in X^\ast$ be of the same length
and let $\phi$ be a $\cL$-formula:
then $\phi[\bar{y}/\bar{x}]$ denotes the formula obtained by simultaneous
renaming of each free occurrence in $\phi$ of a variable in $\bar{x}$
with the corresponding variable in $\bar{y}$, possibly renaming bound
variable occurrences as needed to avoid variable capture.
A formula with no free variable occurrences is termed \emph{closed}
or called a \emph{sentence}.
The \emph{universal closure} of a formula $\phi$ is denoted by
$\forall(\phi)$.
If $\phi$ is a closed $\cL$-formula and $\cA$ is an $\cL$-structure,
we write $\cA \models \phi$ if $\phi$ is satisfied in $\cA$.
A set $\cT$ of closed $\cL$-formulas is called an
\emph{$\cL$-theory}.  We write $\cA \models \cT$ if
$\cA \models \phi$ for each $\phi \in \cT$.
If $\phi$ is a closed $\cL$-formula and $\cT$ is
an $\cL$-theory, we write $\cT \models \phi$ if, for each
$\cL$-structure $\cA$, $\cA \models \cT$ implies $\cA \models \phi$.
In this case we say that $\phi$ is a \emph{logical consequence} of $\cT$.
}

\section{Termination Analysis of Individual Loops}
\label{sec:termination-analysis-of-individual-loops}

We will start by restricting our attention to individual loops of the
form
\begin{equation}
\label{eq:while-loop}
    \{\, I \,\} \; \kw{while}\;  B \; \kw{do} \; C
\end{equation}
where
\begin{itemize}
\item
$I$ is a loop invariant that a previous analysis phase has determined
to hold just before any evaluation of $B$;
\item
$B$ is a Boolean guard expressing the condition on the state upon
which iteration continues;
\item
$C$ is a command that, in the context set by \eqref{eq:while-loop},
is known to always terminate.
\end{itemize}
Notice that, for maximum generality, we do not impose any syntactic
restriction on $I$, $B$ and $C$ and will only observe their interaction
with the program state: $I$ and $B$ express conditions on the state,
and $C$ is seen as a state transformer, that is, a condition constraining
the program states that correspond to its initial and final states.
We assume that such conditions are expressed in a fragment of some
first-order language
\ifthenelse{\boolean{LONGVERSION}}{
$\cL = \cL(\Sigma, X)$
}{
$\cL$
}
that is closed under finite conjunction and implication (indeed
a limited form of implication is often enough).
We assume further that the meaning of the sentences in $\cL$ is given
by some theory $\cT$ for which we are given a sound inference procedure
denoted by `$\mathord{\vdash}$', that is, for each sentence $\phi \in \cL$,
if $\cT \vdash \phi$ then $\cT \models \phi$.
Finally, we fix
\ifthenelse{\boolean{LONGVERSION}}{
a $\Sigma$-structure
}{
an $\cL$-structure
}
$\cD$ such that $\cD \models \cT$,
which captures the domain over which computation and program reasoning
take place.
Let $\bar{x}$ be the tuple of variables containing (among possible others)
all the free variables of \eqref{eq:while-loop}.
The effect of $C$ within the loop can be captured by stipulating that
$\bar{x}$ characterizes the state \emph{before} execution of $C$,
introducing a tuple of new variables $\bar{x}'$ that characterizes the
state \emph{after} $C$'s execution, and by imposing restrictions on
the combined tuple $\bar{x}\bar{x}'$.
Our last assumption is that we are given formulas of $\cL$ that correctly
express the semantics of $I$, $B$, and $C$:
let us call these formulas $\phi_I$, $\phi_B$ and $\phi_C$, respectively.
With these definitions and assumptions, the semantics of
loop~\eqref{eq:while-loop} is correctly approximated as follows:
\begin{enumerate}
\item
whenever the loop guard $B$ is evaluated, $\phi_I[\bar{x}]$ holds;
\item
if $\phi_I[\bar{x}] \land \phi_B[\bar{x}]$ is inconsistent,
iteration of the loop terminates;
\item
just before execution of $C$,
$\phi_I[\bar{x}] \land \phi_B[\bar{x}]$ holds;
\item
just after execution of $C$,
$\phi_I[\bar{x}] \land \phi_B[\bar{x}] \land \phi_C[\bar{x}\bar{x}']$ holds.
\end{enumerate}

It is worth observing that the presence of the externally-generated
invariant $I$ is not restrictive: on the one hand, $\phi_I[\bar{x}]$
can simply be the ``true'' formula,
when nothing better is available; on the other hand, non trivial invariants
are usually a decisive factor for the precision of termination analysis.
As observed in \cite{ColonS02}, the requirement that $I$ must hold
before any evaluation of $B$ can be relaxed by allowing $I$ not to
hold finitely many times.\footnote{Such an invariant is called
\emph{tail invariant} in \cite{ColonS02}.}
The same kind of approximation can be applied to $\phi_I$, $\phi_B$
and $\phi_C$ by only requesting that they eventually hold.

We would like to stress that, at this stage, we have not lost
generality.  While the formalization of basic iteration units in terms
of while loops has an unmistakable imperative flavor, it is general
\ifthenelse{\boolean{LONGVERSION}}{
enough to capture iteration in other programming paradigms.
To start with, recall that a \emph{reduction system} is a pair
$(R, \mathord{\rightarrow})$,
where $R$ is a set and $\reld{\rightarrow}{R}{R}$.
A \emph{term-rewrite system} is a reduction system where $R$ is a set of
terms over some signature and `$\mathord{\rightarrow}$' is encoded by
a finite set of rules in such a way that, for each term $s$, the set
of terms $t$ such that $s \rightarrow t$ is finitely computable from
$s$ and from the system's rules.
Maximal reduction sequences of a term-rewrite system
can be expressed by the following algorithm, for each starting term $s$:
\[
\begin{aligned}
  &\mathrm{term} \assign s \\
  &\kw{while} \; \{\, t \mid \mathrm{term} \rightarrow t \,\} \neq \emptyset
    \; \kw{do} \\
  &\qquad \kw{choose} \; u \in \{\, t \mid \mathrm{term} \rightarrow t \,\}; \\
  &\qquad \mathrm{term} \assign u \\
\end{aligned}
\]
Here, the $\kw{choose}$ construct encodes the rewriting strategy of the system.
Let $R = \{\, t \mid s \rightarrow^\star t \,\}$ denote the set of terms
that can be obtained by any finite number of rewritings of the initial
term $s$. Then, the algorithm above can be transformed into the form
\eqref{eq:while-loop} by considering, as the invariant $I$, a property
expressing that variable `$\mathrm{term}$' can take values in any
over-approximation $S \supseteq R$ of all the possible rewritings.
Namely,
\[
\begin{aligned}
\label{eq:trs-as-while-loop}
  &\{\, \mathrm{term} \in S \,\} \\
  &\kw{while} \; \{\, t \mid \mathrm{term} \rightarrow t \,\} \neq \emptyset
    \; \kw{do} \\
  &\qquad \kw{choose} \; u \in \{\, t \mid \mathrm{term} \rightarrow t \,\}; \\
  &\qquad \mathrm{term} \assign u \\
\end{aligned}
\]
Termination of the rewritten while loop implies termination of the
original one; the reverse implication holds if $S = R$.

The semantics of logic programs, functional programs, concurrent programs
and so forth can be (and often are) formalized in terms of rewriting of
goals and various kinds of expressions: hence no generality is lost by
considering generic while loops of the form \eqref{eq:while-loop}.
}{
enough to capture iteration in other programming paradigms
\cite{BagnaraMPZ12TR}.
}

The approach to termination analysis based on ranking functions
requires that:
\begin{enumerate}
\item
a set $\cO$ and a binary relation $\reld{\prec}{\cO}{\cO}$
are selected so that $\cO$ is well-founded with respect to `$\mathord{\prec}$';
\item
a term $\delta[\bar{y}]$ of $\cL$ is found such that
\begin{equation}
\label{eq:general-rf-condition}
  \cT
    \vdash
      \forall
        \Bigl(
          \bigl(
            \phi_I[\bar{x}]
              \land \phi_B[\bar{x}]
              \land \phi_C[\bar{x}\bar{x}']
          \bigr)
            \rightarrow
              \omega\bigl(\delta[\bar{x}'/\bar{y}],
                          \delta[\bar{x}/\bar{y}]\bigr)
        \Bigr),
\end{equation}
where the interpretation of $\omega$ over $\cD$ corresponds to
`$\mathord{\prec}$';
the function associated to $\delta$ in $\cD$ is called
\emph{ranking function} for the loop~\eqref{eq:while-loop}.
\end{enumerate}
Termination of~\eqref{eq:while-loop} follows by the correctness of
$\phi_I$, $\phi_B$, $\phi_C$ and `$\mathord{\vdash}$', and by well-foundedness
of $\cO$ with respect to `$\mathord{\prec}$'.
To see this, suppose, towards a contradiction, that loop~\eqref{eq:while-loop}
does not terminate.  The mentioned soundness conditions would imply the
existence of an infinite sequence of elements of $\cO$
\begin{equation}
\label{eq:impossible-infinite-chain}
  o_0 \succ o_1 \succ o_2 \succ \cdots
\end{equation}
Let $U \sseq \cO$ be the (nonempty) set of elements in the sequence.
Since $\cO$ is well founded with respect to `$\mathord{\prec}$',
there exists $j \in \Nset$ such that, for each $i \in \Nset$,
$o_i \nprec o_j$.  But this is impossible, as,
for each $j \in \Nset$, $o_{j+1} \prec o_j$.
This means that the infinite chain~\eqref{eq:impossible-infinite-chain}
cannot exist and loop~\eqref{eq:while-loop} terminates.

\ifthenelse{\boolean{LONGVERSION}}{
\begin{example}
Let $\Sigma = (S, F, R)$ with $S = \{ \mathsf{i} \}$,
$F = F_{\varepsilon,\mathsf{i}} \union F_{\mathsf{i},\mathsf{i}}$,
$F_{\varepsilon,\mathsf{i}} = \{ \mathsf{0} \}$,
$F_{\mathsf{i},\mathsf{i}} = \{ \mathsf{s} \}$ and
\(
  R
    = R_{\mathsf{i}\cdot\mathsf{i}}
    = \{ =, < \}
\).
Let also $\cD = \bigl(\{ \Zset \}, \{ 0, s \}, \{ e, l \}\bigr)$
be a $\Sigma$-structure where
$s = \{\, (n, n+1) \mid n \in \Zset \,\}$,
$e = \{\, (n, n) \mid n \in \Zset \,\}$ and
$l = \{\, (n, m) \mid n, m \in \Zset, n < m \,\}$.
Let $X$ be a denumerable set of variable symbols and let $\cT$
be a suitable subtheory of arithmetic restricted to $\cL(\Sigma, X)$.
Consider now the loop
\[
\begin{aligned}
  &\{ x \geq 0 \} \\
  &\kw{while} \; x \neq 0 \; \kw{do} \\
  &\qquad x \assign x-1
\end{aligned}
\]
We have
$\phi_I = (x = \mathsf{0} \lor \mathsf{0} < x)$,
$\phi_B = \neg(x = \mathsf{0})$ and
$\phi_C = \bigl(\mathsf{s}(x') = x\bigr)$.
If we take $(\cO, \mathord{\prec}) = (\Nset, l \inters \Nset^2)$,
$\delta[y] = y$, and
\(
  \omega(\tau, \upsilon)
    =  \bigl(
         (\tau = \mathsf{0} \lor \mathsf{0} < \tau)
           \land
         \tau < \upsilon
       \bigr)
\),
we can substitute into~\eqref{eq:general-rf-condition} and obtain
\[
  \cT
    \vdash
      \forall x, x'
        \itc
          \bigl(
            (x = \mathsf{0} \lor \mathsf{0} < x)
              \land \neg(x = \mathsf{0})
              \land \mathsf{s}(x') = x
          \bigr)
            \rightarrow
              \bigl(
                (x' = \mathsf{0} \lor \mathsf{0} < x')
                  \land
                    x' < x
              \bigr)
\]
which simplifies to
\[
  \cT
    \vdash
      \forall x, x'
        \itc
          \bigl(
            \mathsf{0} < x \land \mathsf{s}(x') = x
          \bigr)
            \rightarrow
              \bigl(
                (x' = \mathsf{0} \lor \mathsf{0} < x')
                  \land
                    x' < x
              \bigr)
\]
which a reasonable inference engine can easily check to be true.
\end{example}
}

This general view of the ranking functions approach to termination
analysis allows us to compare the methods in the literature on a
common ground and focusing on what, besides mere presentation
artifacts, really distinguishes them from one another.
Real differences have to do with:
\begin{itemize}
\item
the choice of the well-founded ordering $(\cO, \prec)$;
\item
the class of functions in which the method ``searches'' for
the ranking functions;
\item
the choice of
\ifthenelse{\boolean{LONGVERSION}}{
the signature $\Sigma$,
}{
the first-order languae $\cL$,
}
the domain $\cD$ and theory $\cT$;
this has to accommodate the programming formalism at hand, the semantic
characterization upon which termination reasoning has to be based,
the axiomatization of $(\cO, \prec)$, and the representation of ranking
functions;
\item
the class of algorithms that the method uses to conduct such a search.
\end{itemize}
We now briefly review these aspects.

The most natural well-founded ordering is, of course,
$(\Nset, \mathord{<})$.
This is especially indicated when the termination arguments are based
on quantities that can be expressed by natural numbers.  This is the case,
for instance, of the work by Sohn and Van~Gelder for termination analysis
of logic programs \cite{SohnVG91,Sohn93th}.
Orderings based on $\Qset_+$ or $\Rset_+$ can be obtained by imposing
over them relations like those defined, for each $\epsilon > 0$, by
\(
  \mathord{<_\epsilon}
    \defeq
      \bigl\{\,
        (h, k) \in \mathbb{S}_+^2
      \bigm|
        h + \epsilon \leq k
      \,\bigr\}
\),
where $\epsilon \in \mathbb{S}_+$
and $\mathbb{S}_+ = \Qset_+$ or $\mathbb{S}_+ = \Rset_+$, respectively.
Of course, this is simply a matter of convenience:
a ranking function $f$ with codomain $(\Rset_+, \mathord{<_\epsilon})$
can always be converted into a ranking function $g$
with codomain $(\Nset, \mathord{<})$ by taking
$g(\bar{y}) = \lfloor f(\bar{y}) \epsilon^{-1} \rfloor$.
Similarly, any ranking function over $(\Rset_+, <_\epsilon)$
can be converted into a ranking function over $(\Rset_+, \mathord{<_1})$.
On tuples, the \emph{lexicographic} ordering is the most common choice
for a well-founded relation: given a finite number of well-founded
relations $\prec_i$ for $i=1, \ldots, n$ over a set $\mathbb{S}$,
the lexicographic ordering over
$\mathbb{S}^n$ is induced by saying that $\vect{s} \prec \vect{t}$
if and only if $s_i \prec_i t_i$ for an index $i$ and
$s_j = t_j$ for all indices $j < i$.
The termination analyzer of the Mercury programming language
\cite{Fischer02,SpeirsSS97} first attempts an analysis using
the $(\Nset, \mathord{<})$ ordering;  if that fails then it resorts
to lexicographic orderings.
Lexicographic orderings on Cartesian products of
$(\Rset_+, \mathord{<_\epsilon})$ are also used in \cite{BradleyMS05CAV}.

The synthesis of ranking functions is easily seen to be a search problem.
All techniques impose limits upon the universe of functions that is
the domain of the search. For instance, in the logic programming community,
the works in
\cite{Fischer02,SpeirsSS97,Verschaetse91,CodishT99}
use ranking functions of the form
$f(x_1, \ldots, x_n) = \sum_{i=1}^n \mu_i x_i$,
where, for $i=1$, \dots,~$n$,
$\mu_i \in \{0,1\}$ and the variable $x_i$ takes values in $\Nset$.
The method of Sohn and Van~Gelder \cite{SohnVG91,Sohn93th}
is restricted to linear functions of the form
$f(x_1, \ldots, x_n) = \sum_{i=1}^n \mu_i x_i$,
where, for $i=1$, \dots,~$n$,
$\mu_i \in \Nset$ and the variable $x_i$ takes values in $\Nset$.
Its generalization to $\Qset_+$ was proposed in~\cite{Mesnard96}
and further generalized by Mesnard and Serebrenik \cite{MesnardS05,MesnardS08}
to obtain affine functions of the form
$f(x_1, \ldots, x_n) = \mu_0 + \sum_{i=1}^n \mu_i x_i$,
where $\mu_i \in \Zset$ and $x_i$ take values in $\Qset$ or $\Rset$,
for $i=0$, \dots,~$n$.
Use of the method of Podelski and Rybalchenko~\cite{PodelskiR04} was
presented in~\cite{PodelskiR04TI} and is a component
of Terminator, a termination prover
of C systems code~\cite{CookPR05}.
Nguyen and De~Schreye \cite{NguyenDeS05} proposed,
in the context of logic
programming and following a thread of work in termination of
term rewrite systems that can be traced back to \cite{Lankford76}, to
use polynomial ranking functions.  These are of the basic form
$f(x_1, \ldots, x_n) = \mu_0 + \sum_{j=1}^m \mu_j \prod_{i=1}^n x_i^{k_{ij}}$
where $\mu_0 \in \Zset$ and,
for $i=1$, \dots,~$n$ and $j=1$, \dots,~$m$, $\mu_j \in \Zset$,
$k_{ij} \in \Nset$ and the variable $x_i$ takes values in $\Zset$
\cite{Zantema00}.  Several further restriction are usually imposed:
first a \emph{domain} $A \subseteq \Nset$ is selected;
then it is demanded that, for each $x_1, \ldots, x_n \in A$,
$f(x_1, \ldots, x_n) \in A$ and that $f$ is \emph{strictly monotone} over $A$
on all its arguments.  The set of all such polynomials is itself well-founded
with respect to `$\mathord{<_A}$': $f <_A g$ if and only if,
for each $x_1, \ldots, x_n \in A$, $f(x_1, \ldots, x_n) < g(x_1, \ldots, x_n)$.
The condition of strict monotonicity, namely,
for each $x_1, \ldots, x_n \in A$, each $i=1$, \dots,~$n$,
and each $y, z \in A$ with $y < z$,
\(
  f(x_1, \ldots, x_{i-1}, y, x_{i+1}, \ldots, x_n)
  <
  f(x_1, \ldots, x_{i-1}, z, x_{i+1}, \ldots, x_n)
\),
is ensured if, for each $j=1$, \dots,~$m$, we have $\mu_j \in \Nset$
and, for each $i=0$, \dots,~$n$, there exists $j$ such that
$\mu_j \neq 0$ and $k_{ij} \neq 0$.
Choosing $A \neq \Nset$ brings some advantages.
For example, if $A \sseq \{\, n \in \Nset \mid n \geq 2 \,\}$
then multiplication of polynomials is strictly
monotone on both its arguments (i.e., $f <_A f \cdot g$
and $g <_A f \cdot g$).
Additional restrictions are often imposed in order to make the search
of ranking functions tractable: both the maximum degree of polynomials
and their coefficients ---the $\mu_j$'s above--- can be severely limited
(an upper bound of~$2$ both on degrees and on coefficients is typical).
Quadratic ranking functions of the form
\(
  f(x_1, \ldots, x_n)
    = \langle x_1, \ldots, x_n, 1 \rangle^\transpose
        \mat{M} \langle x_1, \ldots, x_n, 1\rangle
\)
are considered in~\cite{Cousot05}, where the variables $x_i$ and
the unknown coefficients $\mu_{ij}$ of the $(n+1) \times (n+1)$
symmetric matrix $\mat{M}$ take values in $\Rset$.
\cite{BradleyMS05CAV} considers a search space of tuples of
(up to a fixed number of) linear functions.

The logic used in most papers about the synthesis of linear (or affine)
ranking functions (such as \cite{SohnVG91,PodelskiR04,ColonS01})
is restricted to finite conjunctions of linear equalities or inequalities
and simple implications (e.g., of a single inequality by a conjunction).
In \cite{BradleyMS05CAV} this logic is extended to include disjunction,
so as to capture precisely the effect of the loop body.

Concerning algorithms, the restriction to conjunctions of linear equalities
or inequalities allows the use of the simplex algorithm (or other algorithms
for linear programming) to prove the existence of linear ranking functions
in \cite{SohnVG91,PodelskiR04} or to synthesize one of them.
When a space of ranking functions is sought, these can be obtained by
projecting the systems of constraints onto a designated set of variables
using, for instance, Fourier-Motzkin elimination.
In these approaches, standard algorithms from linear programming work
directly on an abstraction of the loop to be analyzed and are able to
decide the existence of linear ranking functions for that abstraction.
The algorithms used in other approaches belong to the category of
``generate and test'' algorithms: the ``generate'' phase consists in
the selection, possibly guided by suitable heuristics, of candidate
functions, while the ``test'' phase amounts to prove that a candidate
is indeed a ranking function.  This is the case, for instance, of
\cite{BradleyMS05CAV}, where generation consists in the instantiation
of \emph{template functions} and testing employs an algorithm based
on a variant of Farkas' Lemma.
Non-linear constraints generated by the method described in~\cite{Cousot05}
are handled by first resorting to semidefinite
programming solvers and then validating the obtained results by using
some other tools, since these solvers are typically based on interior
point algorithms and hence may incur into unsafe rounding errors.
Note that, in principle, the very same observation would apply to
the case of linear constraints, if the corresponding linear programming
problem is solved using an interior point method or even a floating-point based
implementation of the simplex algorithm; however, there exist
implementations of the simplex algorithm based on exact arithmetic,
so that linear programming problems can be numerically solved
incurring no rounding errors at all and with a computational overhead
that is often acceptable.\footnote{In contrast,
an exact solver for non-linear constraints would probably require
a truly symbolic computation, incurring a much more significant
computational overhead.}

\ifthenelse{\boolean{LONGVERSION}}{
It should be noted that the fact that in this paper we only consider
simple while loops and linear ranking functions
is not as restrictive as it may seem.
Actually, one can trade the existence
of a potentially complex \emph{global} ranking function for
the whole program for the existence of elementary \emph{local}
ranking functions of some selected individual simple loops
appearing in a transformation of the whole program.
General formulations of this idea
are given in~\cite{PodelskiR04TI, DershowitzLSS01}
and provide a useful \emph{sufficient} condition for termination.
Now the question is: can one specify a particular class of programs
and a particular class of elementary local ranking functions
such that this sufficient condition
turns out to be a correct and complete decision procedure for termination
of this class of programs?
The \emph{Size-Change Principle}
proposed by \cite{LeeJB-A01} presents such a class of programs
where the local ranking functions can be safely restricted
to linear functions with 0/1 coefficients.
Moreover, in a generalization of this work presented in
\cite{CodishLS05, Ben-Amram10},
the authors prove that, under certain hypotheses,
linear functions are a large enough class of local ranking functions
for a sound and complete termination criterion.
Hence, for the class of programs they consider,
termination is a decidable property.
}

\section{The Approach of Sohn and Van Gelder, Generalized}
\label{sec:sohn-and-van-gelder}

As far as we know, the first approach to the automatic synthesis of
ranking functions is due to Kirack Sohn and Allen Van~Gelder
\cite{SohnVG91,Sohn93th}.
\ifthenelse{\boolean{LONGVERSION}}{
Possibly due to the fact that their original work concerned termination
of logic programs, Sohn and Van Gelder did not get the recognition we
believe they deserve.  In fact, as we will show, some key ideas of their
approach can be applied, with only rather simple modifications,
to the synthesis of ranking functions for any programming paradigm.
}{
Even though Sohn and Van Gelder's work concerned termination
of logic programs, we will show that the key ideas of their approach
can be applied, with only rather simple modifications, to the
synthesis of ranking functions for any programming paradigm, thus going
beyond what subsequent authors acknowledged.
}

\ifthenelse{\boolean{LONGVERSION}}{

In this section we present the essentials of the work of Sohn and Van~Gelder
in a modern setting: we will first see how the termination of logic programs
can be mapped onto the termination of \emph{binary} $\mathrm{CLP}(\Nset)$
programs;  then we will show how termination of these programs can be
mapped to linear programming;  we will then review the generalization
of Mesnard and Serebrenik to $\mathrm{CLP}(\Qset)$ and $\mathrm{CLP}(\Rset)$
programs and, finally, its generalization to the termination analysis
of generic loops.

\subsection{From Logic Programs to Binary $\mathrm{CLP}(\Nset)$ Programs}

Consider a signature $\Sigma_\mathsf{t} = \bigl(\{\mathsf{t}\}, F, R\bigr)$
and a denumerable set $X$ of variable symbols.
Let $T_\mathsf{t}$ be the set of all $(\Sigma_\mathsf{t}, X)$-terms.
A substitution $\theta$ is a total function $\fund{\theta}{X}{T_\mathsf{t}}$
that is the identity almost everywhere;
in other words, the set
$\bigl\{\, x \in X \bigm| \theta(x) \neq x \,\bigr\}$ is finite.
The \emph{application} of $\theta$ to $t \in T_\mathsf{t}$ gives the
term $\theta(t) \in T_\mathsf{t}$ obtained by simultaneously replacing
all occurrences of a variable $x$ in $t$ with $\theta(x)$.
Consider a system of term equations $E = \{ t_1 = u_1, \ldots, t_n = u_n \}$:
a substitution $\theta$ is a \emph{unifier} of $E$
if $\theta(t_i) = \theta(u_i)$ for $i = 1$, \dots,~$n$.
A substitution $\theta$ is a \emph{most general unifier} (mgu) of $E$
if it is a unifier for $E$ and, for any unifier $\eta$ of $E$, there exists
a substitution $\xi$ such that $\eta = \xi \circ \theta$.
Let $t$ and $u$ be terms: we say that $t$ and $u$ are \emph{variants}
if there exist substitutions $\theta$ and $\eta$ such that
$t = \theta(u)$ and $u = \eta(t)$.

A formula of the form $r(t_1, \ldots, t_n)$, where $r \in R$ and
$t_1, \ldots, t_n \in T_\mathsf{t}$ is called an \emph{atom}.
A \emph{goal} is a formula of the form
$B_1, \ldots, B_n$,
where $n \in \Nset$ and $B_1$, \dots,~$B_n$ are atoms.
The goal where $n = 0$, called the \emph{empty goal},
is denoted by $\square$.
A logic program is a finite set of \emph{clauses} of the form
$H \pif G$,
where $H$ is an atom, called the \emph{head} of the clause,
and $G$ is a goal, called its \emph{body}.
The notions of substitution, mgu and variant are generalized to atoms,
goals and clauses in the expected way.  For example, $\theta$ is an mgu
for atoms $r(t_1, \ldots, t_n)$ and $s(u_1, \ldots, u_m)$ if $r = s$,
$n = m$ and $\theta$ is an mgu for $\{ t_1 = u_1, \ldots, t_n = u_n \}$.

Left-to-right computation for logic programs can be defined in terms of
rewriting of goals.  Goal $B_1, \ldots, B_n$
can be rewritten to $C'_1, \ldots, C'_m, B'_2, \ldots, B'_n$
if there exists a variant of program clause $H \pif C_1, \ldots, C_m$
with no variables in common with $B_1$, \dots,~$B_n$,
the atoms $H$ and $B_1$ are unifiable with mgu $\theta$,
and $C'_i = \theta(C_i)$, for $i = 1$, \dots,~$m$,
$B'_j = \theta(B_j)$, for $j = 2$, \dots,~$n$.
Computation terminates if and when rewriting produces the empty goal.
Notice that the computation, due to the fact that there may be
several clauses that can be used at each rewriting step, is nondeterministic.

Let $\mathtt{nil}, \mathtt{cons} \in F$,
$\mathtt{perm}, \mathtt{select} \in R$ and
$v, w, x, y, z \in X$.
The following logic program defines relations over lists inductively
defined by the constant $\mathtt{nil}$, the empty list, and the binary
constructor $\mathtt{cons}$, which maps a term $t$ and a list $l$ to
the list whose first element is $t$ and the remainder is $l$:
\begin{equation}
\label{prog:perm}
\begin{aligned}
  \mathtt{list}(\mathtt{nil})
    &\pif \square; \\
 \mathtt{list}(\mathtt{cons}(x, y))
    &\pif \mathtt{list}(y); \\
  \mathtt{select}(x,\mathtt{cons}(x,y),y)
    &\pif \mathtt{list}(y) ; \\
  \mathtt{select}(x, \mathtt{cons}(y, z),\mathtt{cons}(y,w))
    &\pif \mathtt{select}(x, z,w); \\
  \mathtt{perm}(\mathtt{nil}, \mathtt{nil})
    &\pif \square; \\
  \mathtt{perm}(x, \mathtt{cons}(v, z))
    &\pif
          \mathtt{select}(v, x, y),
	  \mathtt{perm}(y, z).
\end{aligned}
\end{equation}
The program defines the unary relation $\mathtt{list}$ to be the
set of such lists. The ternary relation $\mathtt{select}$ contains
all $(x, y, z)$ such that $x$ appears in the list $y$, and $z$
is $y$ minus one occurrence of $x$.
The binary relation $\mathtt{perm}$ contains all the pairs of lists such
that one is a permutation of the other.

A computation of a logic program starting from some initial goal can:
terminate with success, when rewriting ends up with the empty goal;
terminate with failure, when rewriting generates a goal whose first atom
is not unifiable with the head of any (variant of) program clause;
loop forever, when the rewriting process continues indefinitely.
Because of nondeterminism, the same program and initial goal can
give rise to computations that succeed, fail or do not terminate.
A goal $G$ enjoys the \emph{universal termination} property with
respect to a program $P$ if all the computations starting from $G$ in
$P$ do terminate, either with success or failure.\footnote{The related concept
of \emph{existential termination} has a number of drawbacks and will
not be considered here.  See \cite{VasakP86} for more information.}

The idea behind this approach to termination analysis of logic programs is that
termination is often ensured by the fact that recursive
``invocations'' involve terms that are ``smaller''.
Rewriting of
\(
  \mathtt{list}(\mathtt{cons}(t_1, \mathtt{cons}(t_2, \mathtt{nil})))
\),
for example, results in
\(
  \mathtt{list}(\mathtt{cons}(t_2, \mathtt{nil}))
\)
and then
\(
  \mathtt{list}(\mathtt{nil})
\).
Various notions of ``smaller term'' can be captured by
\emph{linear symbolic norms} \cite{SohnVG91,LindenstraussS97}.
Consider the signature
\(
  \Sigma_\mathsf{e}
    =
      \bigl(
        \{\mathsf{e}\},
        \{ 0, 1, \mathord{+} \},
        P \union \{ \mathord{=}, \mathord{\leq} \}
      \bigr)
\).
The set $T_\mathsf{e}$ of $(\Sigma_\mathsf{e}, X)$-terms contains
affine expressions with natural coefficients.
A linear symbolic norm is a function of the form
$\fund{\Vert\cdot\Vert}{T_\mathsf{t}}{T_\mathsf{e}}$ such that
\[
  \Vert t \Vert
    \defeq
      \begin{cases}
        t,
          &\text{if $t \in X$}, \\
        c + \sum_{i=1}^n a_i \Vert t_i \Vert,
          &\text{if $t = f(t_1, \ldots, t_n)$},
      \end{cases}
\]
where $c$ and $a_1$, \dots,~$a_n$ are natural numbers that only depend
on $f$ and $n$.
The \emph{term-size} norm, for example, is characterized by $c = 0$ for each
$f \in F_{\varepsilon,\mathsf{t}}$ and by $c = 1$ and $a_i = 1$ for each
$f \in F_{w,\mathsf{t}} \subseteq F \setdiff F_{\varepsilon,\mathsf{t}}$
and $i = 1$, \dots,~$\length{w}$.\footnote{The variant used in \cite{SohnVG91},
called \emph{structural term size}, can be obtained by letting,
for each $f \in F_{w,\mathsf{t}}$,
$c = \length{w}$ and $a_i = 1$ for $i=1$, \dots,~$\length{w}$.}
The \emph{list-length} norm is, instead, characterized by $c = 0$ and
$a_i = 0$ for each
$f \neq \mathtt{cons} \in F_{\mathsf{t}\mathsf{t},\mathsf{t}}$,
and by $c = a_2 = 1$ and $a_1 = 0$ for the $\mathtt{cons}$ binary
constructor.

Once a linear symbolic norm has been chosen, a logic program can be converted
by replacing each term with its image under the norm.  For example, using
the list-length norm the above program becomes:
\begin{align}
\label{clause:list_1_1}
  \mathtt{list}(0)
    &\pif \square; \\
\label{clause:list_1_2}
  \mathtt{list}(1+y)
    &\pif \mathtt{list}(y); \\
\label{clause:select_3_1}
  \mathtt{select}(x, 1+y, y)
    &\pif \mathtt{list}(y); \\
\label{clause:select_3_2}
  \mathtt{select}(x, 1+z, 1+w)
    &\pif \mathtt{select}(x, z, w); \\
\label{clause:perm_2_1}
  \mathtt{perm}(0, 0)
    &\pif \square; \\
\label{clause:perm_2_2}
  \mathtt{perm}(x, 1+z)
    &\pif \mathtt{select}(v, x, y),
	  \mathtt{perm}(y, z).
\end{align}
The program obtained by means of this \emph{abstraction} process ---we
have replaced terms by an expression of their largeness--- is a
$\mathrm{CLP}(\Nset)$ program.
In the CLP (Constraint Logic Programming) framework \cite{JaffarM94},
the notion of unifiability is generalized by the one of solvability in
a given structure.  The application of most general unifiers is, in
addition, generalized by the collection of constraints into a
set of constraints called \emph{constraint store}.\footnote{We offer a
self-contained yet very simplified view of the CLP framework.  The
interested reader is referred to \cite{JaffarM94,JaffarMMS98}.}
In $\mathrm{CLP}(\Nset)$, the constraints are equalities between
affine expressions in $T_\mathsf{e}$ and computation proceeds by
rewriting a goal and augmenting a constraint store $\Gamma$, which is
initially empty, with new constraints.
Goal $B_1, B_2, \ldots, B_n$ can be rewritten to
$C_1, \ldots, C_m, B_2, \ldots, B_n$ if there exists a variant
$H \pif C_1, \ldots, C_m$ of some program clause with no variables
in common with $B_1$, \dots,~$B_n$
such that $H = p(t_1, \ldots, t_n)$, $B_1 = p(u_1, \ldots, u_n)$
and $\Gamma' \defeq \Gamma \union \{ t_1 = u_1, \ldots, t_n = u_n \}$
is satisfiable
over the $\Sigma_\mathsf{e}$-structure given by the naturals, the functions
given by the constants $0$ and $1$ and the binary sum operation,
and the identity relation over the naturals.  In this case $\Gamma'$
becomes the new constraint store.

The interesting thing about the abstract $\mathrm{CLP}(\Nset)$
program ---let us denote it by $\alpha(P)$--- is that the following holds:
if an abstract goal $\alpha(G)$ universally terminates with respect to
$\alpha(P)$, then the original goal $G$ universally terminates with
respect to the original program $P$, and this for each linear symbolic norm
that is used in the abstraction (see \cite[Section~6.1]{MesnardR03} for
a very general proof of this fact).  The converse does not hold because
of the precision loss abstraction involves.

We will now show, appealing to intuition, that the ability to approximate
the termination behavior of programs constituted by a single
\emph{binary} $\mathrm{CLP}(\Nset)$ clause,
that is, of the form
\begin{equation}
\label{eq:directly-recursive-clpn-clause}
  p(\bar{x}) \pif c[\bar{x}, \bar{x}'], p(\bar{x}'),
\end{equation}
where $p$ is a predicate symbol, gives a technique to approximate
the termination behavior of any $\mathrm{CLP}(\Nset)$ program.

The first step is to compute affine relations that correctly approximate
the \emph{success set} of the $\mathrm{CLP}(\Nset)$ program.
For our program, we can obtain (e.g., by standard abstract interpretation
techniques \cite{CousotC77,CousotC92lp})
\begin{align*}
  \text{$\mathtt{list}(x)$ succeeds}
    &\implies \mathrm{true}; \\
  \text{$\mathtt{select}(x, y, z)$ succeeds}
    &\implies z = y-1; \\
  \text{$\mathtt{perm}(x, y)$ succeeds}
    &\implies x = y.
\end{align*}
We now consider the clauses of the $\mathrm{CLP}(\Nset)$ program one by one.
Clause~\eqref{clause:list_1_1} does not pose any termination problem.
Clause~\eqref{clause:list_1_2} is already of the
form~\eqref{eq:directly-recursive-clpn-clause}: we can call the engine
described in the next section and obtain the ranking function
$f(x) = x$ for $\mathtt{list}(x)$, meaning that the argument
of $\mathtt{list}$ strictly decreases in the recursive call.
We thus note that
\begin{equation}
\label{cond:list-terminates}
\text{$\mathtt{list}(x)$ terminates if called with $x \in \Nset$.}
\end{equation}
Consider now clauses~\eqref{clause:select_3_1} and~\eqref{clause:select_3_2}:
for the former we simply have to note that we need to
satisfy~\eqref{cond:list-terminates} in order to guarantee termination;
for the latter, which is of the form~\eqref{eq:directly-recursive-clpn-clause},
we can obtain an infinite number of ranking functions for
$\mathtt{select}(x, y, z)$, among which are $f(x, y, z) = y$ (the second
argument decreases) and $f(x, y, z) = z$ (the third argument decreases).
Summing up, for the $\mathtt{select}$ predicate we have
\begin{equation}
\label{cond:select-terminates}
\text{$\mathtt{select}(x,y,z)$ terminates if called with $y \in \Nset$
      and/or $z \in \Nset$.}
\end{equation}
Now, clause~\eqref{clause:perm_2_1} does not pose any termination problem,
but clause~\eqref{clause:perm_2_2} is not of the
form~\eqref{eq:directly-recursive-clpn-clause}.
However we can use the computed model to ``unfold'' the invocation
to $\mathtt{select}$ and obtain
\begin{equation*}
\tag{\ref{clause:perm_2_2}'}
  \mathtt{perm}(x, 1+z)
    \pif y = x-1,
	  \mathtt{perm}(y, z),
\end{equation*}
which has the right shape and, as far as the termination behavior
of the entire program is concerned, is equivalent
to~\eqref{clause:perm_2_2} \cite{AptP93}.
From~(\ref{clause:perm_2_2}') we obtain, for
$\mathtt{perm}(x, y)$, the ranking functions $f(x, y) = x$
and $f(x, y) = y$.  We thus note:
\begin{equation}
\label{cond:perm-terminates}
\begin{aligned}
 \text{$\mathtt{perm}(x,y)$ terminates }
   &\text{if called with $x \in \Nset$ and/or $y \in \Nset$} \\
   &\text{and the call to $\mathtt{select}$
          in~\eqref{clause:perm_2_2} terminates.}
\end{aligned}
\end{equation}
Summarizing, we have that goals of the form
$\mathtt{perm}(k, y)$, where $k \in \Nset$,
satisfy~\eqref{cond:perm-terminates};
looking at clause~\eqref{clause:perm_2_2} it is clear that they
also satisfy~\eqref{cond:select-terminates}; in turn,
inspection of clause~\eqref{clause:select_3_1} reveals that
also~\eqref{cond:list-terminates} is satisfied.
As a result, we have proved that any invocation
in the original logic program~\eqref{prog:perm}
of $\mathtt{perm}(x, y)$ with $x$ bound to an
argument whose list-length norm is constant,
universally terminates.
It may be instructive to observe that this implementation of
$\mathtt{perm}$ is not symmetric: goals of the form
$\mathtt{perm}(x, k)$, where $k \in \Nset$, fail to
satisfy~\eqref{cond:perm-terminates} and, indeed,
it is easy to come up with goals $\mathtt{perm}(x, y)$ with $y$ bound to a
complete list that do not universally terminate in the original program.

The procedure outlined in the previous example can be extended (in
different ways) to any $\mathrm{CLP}(\Nset)$ programs.
As the precise details are beyond the scope of this paper, we only
illustrate the basic ideas and refer the interested reader to the literature.
The methodology is simpler for programs that are \emph{directly recursive},
i.e., such that all ``recursive calls'' to  $p$ only happen in clauses
for $p$.%
\footnote{%
For a $\mathrm{CLP}(\Nset)$ program $P$, let $\Pi_P$ be the set of predicate
symbols appearing in $P$.
On the set $\Pi_P$, we define the relation `$\to$' such that $p \to q$
if and only if $P$ contains a clause with $p$ as the predicate symbol
of its head and $q$ as the predicate symbol of at least one body atom.
Let `$\to^\star$' be  the reflexive and transitive closure of `$\to$'.
The relation defined by $p \simeq q$ if and only if $p \to^\star q$
and $q \to^\star p$ is an equivalence relation;
we denote by $[p]_\simeq$ the equivalence class including $p$.
A program $P$ is \emph{directly recursive} if and only if,
for each $p \in \Pi_P$, $[p]_\simeq = \{p\}$.
A program $P$ is \emph{mutually recursive} if it is not
directly recursive.}
Consider a directly recursive clause.  This has the general form
\[
  p(\bar{x})
    \pif
      c, \beta_0, p(\bar{x}_1), \beta_1, p(\bar{x}_2), \beta_2,
        \dots, p(\bar{x}_k), \beta_k,
\]
where the goals $\beta_0$, $\beta_1$, \dots,~$\beta_k$ do not contain atoms
involving $p$.
The computed model is used to ``unfold'' $\beta_0$ obtaining a sound
approximation, in the form of a conjunction of linear arithmetic constraints,
of the conditions upon which the first recursive call, $p(\bar{x}_1)$
takes place.  If we call $c_1$ the conjunction of $c$ with the constraint
arising from the unfolding of $\beta_0$, we obtain the binary,
directly recursive clause
\begin{align*}
  p(\bar{x}) &\pif c_1, p(\bar{x}_1). \\
\intertext{%
We can now use the model to unfold the goals $p(\bar{x}_1)$ and $\beta_1$
and obtain a constraint that, conjoined with $c_1$, gives us $c_2$, a sound
approximation of the ``call pattern'' for the second recursive call.
Repeating this process we will obtain the binary clauses
}
  p(\bar{x}) &\pif c_2, p(\bar{x}_2), \\
       &\;\vdots \\
  p(\bar{x}) &\pif c_k, p(\bar{x}_k).
\end{align*}
We repeat this process for each clause defining $p$ and end up with a set
of binary clauses, for which a set of ranking functions is computed,
using the technique to be presented in the next section.
The same procedure is applied to each predicate symbol in the program.
A final pass over the original $\mathrm{CLP}(\Nset)$ program is needed
to ensure that each body atom is called within a context that ensures
the termination of the corresponding computation.  This can be done
as follows:
\begin{enumerate}
\item
A standard global analysis is performed to obtain, for each predicate
that can be called in the original $\mathrm{CLP}(\Nset)$ program,
possibly approximated but correct information about which
arguments are known to be \emph{definite}, i.e., constrained to take
a unique value, in each call to that predicate (see, e.g., \cite{BakerS93}).
\item
For each recursive predicate that may be called, it is checked that,
for each possible combination of definite and not-known-to-be-definite
arguments, there is at least one ranking function that depends only
on the definite arguments.
\end{enumerate}
The overall methodology can be adapted to mutually recursive programs,
either by a direct extension of the above approach
(see, e.g., \cite{Sohn93th}) or by more advanced program transformations
(see, e.g., \cite{CodishGBGV03}).

\subsection{Ranking Functions for Binary, Directly Recursive
            $\mathrm{CLP}(\Nset)$ Programs}
\label{subsec:rf-for-binary-clp-n}

In order to show how ranking functions can be computed from directly
recursive binary $\mathrm{CLP}(\Nset)$ clauses,
we deal first with a single clause
\begin{equation*}
  p(\bar{x}) \pif c[\bar{x}, \bar{x}'], p(\bar{x}'),
\end{equation*}
}{
\subsection{Ranking Functions for Binary, Directly Recursive
            $\mathrm{CLP}(\Nset)$ Programs}
\label{subsec:rf-for-binary-clp-n}

We refer the interested reader to \cite{BagnaraMPZ12TR} for a more
complete reconstruction of the approach by Sohn and Van Gelder.
For the other readers, suffices it to say that termination analysis
of logic programs can be mapped onto termination analysis
of $\mathrm{CLP}(\Nset)$ programs, the termination of which implies
the termination of the original program.\footnote{See,
e.g., \cite{JaffarM94} for an introduction to CLP
---constraint logic programming--- languages, and \cite{Mesnard96}
for the mapping of logic programs to $\mathrm{CLP}(\Nset)$ programs.}
In turn, the termination analysis of general $\mathrm{CLP}(\Nset)$ programs
can be approximated by termination analysis of
\emph{directly recursive}, \emph{binary} $\mathrm{CLP}(\Nset)$ clauses.
In order to show how ranking functions can be computed from these,
we deal first with a single clause
\begin{equation}
\label{eq:directly-recursive-clpn-clause}
  p(\bar{x}) \pif c[\bar{x}, \bar{x}'], p(\bar{x}'),
\end{equation}
}
where $p$ is a predicate symbol,
$\bar{x}$ and $\bar{x}'$ are disjoint $n$-tuples of variables,
and $c[\bar{x}, \bar{x}']$ is a linear constraint involving
variables in $\bar{x} \union \bar{x}'$.\footnote{As usual, we abuse notation
by confusing a tuple with the set of its elements.}
The meaning of such a clause is that, if $p$ is called on some tuple
of integers $\bar{x}$, then there are two cases:
\begin{itemize}
\item
$c[\bar{x}, \bar{x}']$ is unsatisfiable (i.e., there does not exist
a tuple of integers $\bar{x}'$ that, together with $\bar{x}$,
satisfies it), in which case the computation will fail, and thus terminate;
\item
there exists $\bar{x}'$ such that $c[\bar{x}, \bar{x}']$ holds,
in which case the computation proceeds with the (recursive)
calls $p(\bar{x}')$, for each $\bar{x}'$ such that $c[\bar{x}, \bar{x}']$.
\end{itemize}
The question is now to see whether that recursive procedure is
\emph{terminating}, that is whether,
for each $\bar{x} \in \Nset^n$,
the call $p(\bar{x})$ will only give rise to chains of recursive calls
of finite length.
The approach of Sohn and Van Gelder allows to synthesize
a function $\fund{f_p}{\Nset^n}{\Nset}$ such that
\begin{equation}
\label{eq:svg-strict-decrease}
  \forall \bar{x}, \bar{x}' \in \Nset^n
    \itc c[\bar{x}, \bar{x}'] \implies f_p(\bar{x}) > f_p(\bar{x}').
\end{equation}
This means that the measure induced by $f_p$ strictly decreases
when passing from a call of $p$ to its recursive call.
Since the naturals are well founded with respect to `$<$',
this entails that $p$, as defined
in~\eqref{eq:directly-recursive-clpn-clause}, is terminating.

A very important contribution of Sohn and Van Gelder consists in
the algorithm they give to construct a class of
functions that satisfy~\eqref{eq:svg-strict-decrease}.
The class is constituted by linear functions of the form
\begin{equation}
\label{eq:svg-ranking-function-form}
  f_p(y_1, \ldots, y_n) = \sum_{i=1}^n \mu_i y_i,
\end{equation}
where $\mu_i \in \Nset$, for $i=1$, \dots,~$n$.
For this class of functions and by letting
$\bar{\mu} = (\mu_1, \ldots, \mu_n)$,
condition~\eqref{eq:svg-strict-decrease} can be rewritten as
\begin{equation}
\label{eq:svg-strict-decrease-rewritten}
  \exists \bar{\mu} \in \Nset^n
    \st
  \forall \bar{x}, \bar{x}' \in \Nset^n
    \itc c[\bar{x}, \bar{x}']
      \implies
        \sum_{i=1}^n \mu_i x_i - \sum_{i=1}^n \mu_i x'_i
          \geq 1.
\end{equation}
Given that $c[\bar{x}, \bar{x}']$ is a linear constraint,
for any choice of $\bar{\mu} \in \Nset^n$ we can easily express
\eqref{eq:svg-strict-decrease-rewritten} as an optimization problem
over the naturals.
In order to move from tuple notation to the more convenient vector
notation, assume without loss of generality that, for some $m \in \Nset$,
$\mat{A}_c \in \Zset^{m \times 2n}$
and $\vect{b}_c \in \Zset^{m}$ are such that
$\mat{A}_c \langle \vect{x}, \vect{x}' \rangle \geq \vect{b}_c$
is logically equivalent to $c[\bar{x}, \bar{x}']$ under the
obvious, respective interpretations.
Then, for any candidate choice of $\vect{\mu} \in \Nset^n$,
condition~\eqref{eq:svg-strict-decrease-rewritten} is equivalent
to imposing that the optimization problem
\begin{equation}
\label{eq:svg-optimization-problem}
\begin{aligned}
  \text{minimize }\quad
    & \theta = \langle \vect{\mu}, -\vect{\mu} \rangle^\transpose
                 \langle \vect{x}, \vect{x}' \rangle \\
  \text{subject to }\quad
    & \mat{A}_c \langle \vect{x}, \vect{x}' \rangle \geq \vect{b}_c \\
    & \vect{x}, \vect{x}' \in \Nset^n
\end{aligned}
\end{equation}
is either unsolvable or has an optimal solution whose \emph{cost}
$\hat{\theta}$ is such that $\hat{\theta} \geq 1$.
If this is the case, then $\vect{\mu}$ induces,
according to~\eqref{eq:svg-ranking-function-form},
a function $f_p$ satisfying~\eqref{eq:svg-strict-decrease}.
Notice that, for any fixed choice of $\vect{\mu} \in \Nset^n$,
$\theta$ is a linear expression and hence
\eqref{eq:svg-optimization-problem} is
an integer linear programming (ILP) problem.
This gives us an expensive way (since ILP is an $\mathrm{NP}$-complete problem
\cite{GareyJ90}) to test whether a certain $\vect{\mu} \in \Nset^n$
is a witness for termination of~\eqref{eq:directly-recursive-clpn-clause},
but gives us no indication about where to look for such
a tuple of naturals.

A first step forward consists in considering the relaxation
of~\eqref{eq:svg-optimization-problem} obtained by
replacing the integrality constraints $\vect{x}, \vect{x}' \in \Nset^n$
with $ \vect{x}, \vect{x}' \in \Qset_+^n$.
This amounts to trading precision for efficiency.
In fact, since any feasible solution
of~\eqref{eq:svg-optimization-problem} is also feasible for the
relaxed problem, if the optimum solution of the latter has a cost
greater than or equal to~$1$, then
either~\eqref{eq:svg-optimization-problem} is unfeasible or
$\hat{\theta} \geq 1$.  However, we may have $\hat{\theta} \geq 1$
even if the optimum of the relaxation is less than~$1$.\footnote{%
Let us consider the clause:
$p(x) \pif 2x \geq 2x'+1, p(x')$ with $\mu=1$.
The optimization over the integers leads to $\hat{\theta} = 1$,
whereas the optimization for the relaxation has
$\hat{\theta} = \frac12$.
}
On the other hand, the relaxed problem is a linear problem:
so by giving up completeness we have passed from an $\mathrm{NP}$-complete
problem to a problem in $\mathrm{P}$ for which we have, in addition,
quite efficient algorithms.%
\footnote{We denote by $\mathrm{P}$ the class of problems solvable
in weakly polynomial time. For a formal definition of $\mathrm{P}$
and the notion of $\mathrm{NP}$-completeness
we refer the reader to, e.g.,~\cite{Schrijver86}.}
Furthermore, we observe that although the parameters $\vect{\mu}$
are naturals in \eqref{eq:svg-strict-decrease-rewritten},
this condition can be relaxed as well:
if $\vect{\mu} \in \Qset_+^n$ gives a relaxed problem with optimum
greater than~$1$, then we can multiply this vector by a positive
natural so as to obtain a tuple of naturals
satisfying~\eqref{eq:svg-strict-decrease-rewritten}.
The relaxation can now be written using the standard linear programming
(LP) notation:
\begin{equation}
\label{eq:svg-relaxed-optimization-problem}
\begin{aligned}
  \text{minimize }\quad
    & \langle \vect{\mu}, -\vect{\mu} \rangle^\transpose
        \langle \vect{x}, \vect{x}' \rangle \\
  \text{subject to }\quad
    & \mat{A}_c \langle \vect{x}, \vect{x}' \rangle \geq \vect{b}_c \\
    & \langle \vect{x}, \vect{x}' \rangle \geq \vect{0}.
\end{aligned}
\end{equation}

We still do not know how to determine the vector of parameters $\vect{\mu}$
so that the optimum of~\eqref{eq:svg-relaxed-optimization-problem} is
at least~$1$, but here comes one of the brilliant ideas of Sohn and
Van Gelder: passing to the \emph{dual}.
It is a classical result of LP
theory that every LP problem can
be converted into an \emph{equivalent} dual problem.
The dual of~\eqref{eq:svg-relaxed-optimization-problem} is
\begin{equation}
\label{eq:svg-dual-relaxed-optimization-problem}
\begin{aligned}
  \text{maximize } \quad
    & \vect{b}_c^\transpose \vect{y} \\
  \text{subject to } \quad
    & \mat{A}_c^\transpose \vect{y}
        \leq
          \langle \vect{\mu}, -\vect{\mu} \rangle \\
    & \vect{y} \geq \vect{0},
\end{aligned}
\end{equation}
where $\vect{y}$ is an $m$-column vector of (dual) unknowns.
Duality theory ensures that if
both~\eqref{eq:svg-relaxed-optimization-problem}
and~\eqref{eq:svg-dual-relaxed-optimization-problem} have bounded
feasible solutions, then both of them have optimal solutions and these
solutions have the same cost.
More formally, for every choice of the parameters $\vect{\mu} \in \Qset_+^n$,
if $\langle \vect{\hat{x}}, \vect{\hat{x}}' \rangle \in \Qset^{2n}$
is an optimal solution
for~\eqref{eq:svg-relaxed-optimization-problem}
and $\vect{\hat{y}} \in \Qset^m$ is an optimal solution
for~\eqref{eq:svg-dual-relaxed-optimization-problem},
then
\(
  \langle \vect{\mu}, -\vect{\mu} \rangle^\transpose
    \langle \vect{\hat{x}}, \vect{\hat{x}}' \rangle
    =
      \vect{b}_c^\transpose \vect{\hat{y}}
\).
Moreover, if one of~\eqref{eq:svg-relaxed-optimization-problem}
and~\eqref{eq:svg-dual-relaxed-optimization-problem} is unfeasible,
then the other is either unbounded or unfeasible.
In contrast, if one of~\eqref{eq:svg-relaxed-optimization-problem}
and~\eqref{eq:svg-dual-relaxed-optimization-problem} is unbounded,
then the other is definitely unfeasible.

Thus, thanks to duality theory,
the LP problems~\eqref{eq:svg-relaxed-optimization-problem}
and~\eqref{eq:svg-dual-relaxed-optimization-problem} are equivalent
for our purposes and we can consider any one of them.
Suppose we analyze the dual
problem~\eqref{eq:svg-dual-relaxed-optimization-problem}:
\begin{itemize}
\item
If \eqref{eq:svg-dual-relaxed-optimization-problem} is unfeasible
then either \eqref{eq:svg-relaxed-optimization-problem} is unfeasible,
which implies trivial termination of~\eqref{eq:directly-recursive-clpn-clause},
or~\eqref{eq:svg-relaxed-optimization-problem} is unbounded,
in which case ---since we are working on relaxations--- nothing
can be concluded about whether $\vect{\mu}$ defines a ranking function
for \eqref{eq:directly-recursive-clpn-clause}.
\item
If~\eqref{eq:svg-dual-relaxed-optimization-problem} is feasible and unbounded
then~\eqref{eq:svg-relaxed-optimization-problem} is unfeasible
and~\eqref{eq:directly-recursive-clpn-clause}
trivially terminates.
\item
If~\eqref{eq:svg-dual-relaxed-optimization-problem} is feasible
and bounded, then we have proved termination
($\vect{\mu}$ induces a ranking function)
if the cost of the optimal solution is at least~$1$
(actually, any positive rational could be used instead of $1$).
The analysis is inconclusive otherwise.
\end{itemize}

The crucial point is that, in~\eqref{eq:svg-dual-relaxed-optimization-problem},
the parameters $\vect{\mu}$ occur linearly,
whereas in~\eqref{eq:svg-relaxed-optimization-problem}
they are multiplied by $\langle \vect{x}, \vect{x}' \rangle$.
So we can treat $\vect{\mu}$ as a vector of variables and
transform~\eqref{eq:svg-dual-relaxed-optimization-problem}
into the new LP problem in $m+n$ variables
\begin{equation}
\label{eq:svg-transformed-dual-relaxed-optimization-problem}
\begin{aligned}
  \text{maximize }\quad
    & \langle \vect{b}_c, \vect{0} \rangle^\transpose
        \langle \vect{y}, \vect{\mu} \rangle \\
  \text{subject to }\quad
    & \begin{pmatrix}
        \mat{A}_c^\transpose \;
          \begin{matrix}
            -\mat{I}_n \cr
            \mat{I}_n
          \end{matrix}
      \end{pmatrix}
        \langle \vect{y}, \vect{\mu} \rangle \leq \vect{0} \\
    & \langle \vect{y}, \vect{\mu} \rangle \geq \vect{0}.
\end{aligned}
\end{equation}
The requirement that, in order to guarantee termination
of~\eqref{eq:directly-recursive-clpn-clause},
the optimal solutions of~\eqref{eq:svg-relaxed-optimization-problem}
and~\eqref{eq:svg-dual-relaxed-optimization-problem} should not be
less than~$1$ can now be captured by incorporating
$\vect{b}_c^\transpose \vect{y} \geq 1$ into the constraints
of~\eqref{eq:svg-transformed-dual-relaxed-optimization-problem},
yielding
\begin{equation}
\label{eq:svg-final-transformed-dual-relaxed-optimization-problem}
\begin{aligned}
  \text{maximize }\quad
    & \langle \vect{b}_c, \vect{0} \rangle^\transpose
        \langle \vect{y}, \vect{\mu} \rangle \\
  \text{subject to }\quad
    & \begin{pmatrix}
        \mat{A}_c^\transpose
          &\begin{matrix}
             -\mat{I}_n \cr
             \mat{I}_n
           \end{matrix} \cr
        -\vect{b}_c^\transpose
          &\vect{0}
      \end{pmatrix}
        \langle \vect{y}, \vect{\mu} \rangle
          \leq
            \begin{pmatrix}
              \vect{0} \cr
              -1
            \end{pmatrix} \\
    & \langle \vect{y}, \vect{\mu} \rangle \geq \vect{0}.
\end{aligned}
\end{equation}
There are several possibilities:
\begin{enumerate}
\item If~\eqref{eq:svg-final-transformed-dual-relaxed-optimization-problem}
      is unfeasible, then:
  \begin{enumerate}
  \item if~\eqref{eq:svg-transformed-dual-relaxed-optimization-problem}
        is unfeasible, then, for each $\vect{\mu} \in \Qset_+^n$,
        \eqref{eq:svg-dual-relaxed-optimization-problem} is unfeasible
        and:
    \begin{enumerate}
    \item if~\eqref{eq:svg-relaxed-optimization-problem} is unfeasible,
          then~\eqref{eq:directly-recursive-clpn-clause} trivially terminates;
    \item otherwise~\eqref{eq:svg-relaxed-optimization-problem} is unbounded
          and we can conclude nothing about the termination
          of~\eqref{eq:directly-recursive-clpn-clause}.
    \end{enumerate}
  \item If~\eqref{eq:svg-transformed-dual-relaxed-optimization-problem}
        is feasible, then it is  bounded by a rational number $q < 1$.
        Thus, for each $\vect{\check{\mu}} \in \Qset_+^n$ extracted from
        a feasible solution
        $\langle \vect{\check{y}}, \vect{\check{\mu}} \rangle \in \Qset_+^{m+n}$
	of ~\eqref{eq:svg-transformed-dual-relaxed-optimization-problem},
        the corresponding LP
        problem~\eqref{eq:svg-dual-relaxed-optimization-problem} is also
        feasible, bounded, and its optimum $q' \in \Qset$ is such that
        $q' \leq q < 1$.  Moreover, we must have $q' \leq 0$.
        In fact, if $q' > 0$,
        problem~\eqref{eq:svg-relaxed-optimization-problem} instantiated
        over $\vect{\check{\mu}}' \defeq \vect{\check{\mu}}/q'$ would
        have an optimal solution of cost~$1$; the same would hold for the
        corresponding dual~\eqref{eq:svg-dual-relaxed-optimization-problem},
        but this would contradict the hypothesis
        that~\eqref{eq:svg-transformed-dual-relaxed-optimization-problem}
        is bounded by $q <1$.  Hence $q' \leq 0$.  Since by duality
        the optimum of problem~\eqref{eq:svg-relaxed-optimization-problem}
        is $q'$, the analysis is inconclusive.
  \end{enumerate}
\item If~\eqref{eq:svg-final-transformed-dual-relaxed-optimization-problem}
      is feasible, let
      $\langle \vect{\check{y}}, \vect{\check{\mu}} \rangle \in \Qset^{m+n}$
      be any of its feasible solutions. Choosing $\vect{\check{\mu}}$ for
      the values of the parameters,
      \eqref{eq:svg-dual-relaxed-optimization-problem} is feasible.
      There are two further possibilities:
  \begin{enumerate}
  \item either~\eqref{eq:svg-dual-relaxed-optimization-problem} is unbounded,
	so~\eqref{eq:directly-recursive-clpn-clause} trivially terminates;
  \item or it is bounded by a rational $q \geq 1$ and the same
	holds for its dual~\eqref{eq:svg-relaxed-optimization-problem}.
  \end{enumerate}
  In both cases, $\vect{\check{\mu}}$, possibly multiplied by a positive
  natural in order to get a tuple of naturals, defines,
  via~\eqref{eq:svg-ranking-function-form}, a ranking function
  for~\eqref{eq:directly-recursive-clpn-clause}.
\end{enumerate}
The above case analysis boils down to the following algorithm:
\begin{enumerate}
\item
Use the simplex algorithm to determine the feasibility
of~\eqref{eq:svg-final-transformed-dual-relaxed-optimization-problem},
ignoring the objective function.
If it is feasible, then any feasible solution induces a linear ranking
function for~\eqref{eq:directly-recursive-clpn-clause}; exit with success.
\item
If~\eqref{eq:svg-final-transformed-dual-relaxed-optimization-problem}
is unfeasible, then try to determine the feasibility
of~\eqref{eq:svg-optimization-problem} (e.g., by using the simplex
algorithm again to test whether the
relaxation~\eqref{eq:svg-relaxed-optimization-problem} is feasible).
If~\eqref{eq:svg-optimization-problem} is unfeasible
then~\eqref{eq:directly-recursive-clpn-clause} trivially terminates;
exit with success.
\item
Exit with failure (the analysis is inconclusive).
\end{enumerate}

An example should serve to better clarify the methodology we have employed.

\begin{example}
\label{ex:base-two-log}
In the $\mathrm{CLP}(\Nset)$ program
\begin{align*}
  p(x_1, x_2) &\pif x_1 \leq 1 \land x_2 = 0, \\
  p(x_1, x_2) &\pif x_1 \geq 2 \land 2x'_1 + 1 \geq x_1 \land 2x'_1 \leq x_1
                       \land x_2' + 1 = x_2, p(x'_1, x'_2),
\end{align*}
$p(x_1,x_2)$ is equivalent to
\[
  x_2
    =
      \begin{cases}
        \lfloor \log_2(x_1) \rfloor, & \text{if $x_1 \neq 0$;} \\
        0,                           & \text{otherwise}.
      \end{cases}
\]
The relaxed optimization problem in LP notation
\eqref{eq:svg-relaxed-optimization-problem} is%
\footnote{We will tacitly replace an equality in the form $\alpha =
\beta$ by the equivalent pair of inequalities $\alpha \geq \beta$ and
$-\alpha \geq -\beta$ whenever the substitution is necessary to fit
our framework.}
\begin{equation*}
\begin{aligned}
  \text{minimize }\quad
    & \langle \mu_1, \mu_2, -\mu_1, -\mu_2 \rangle^\transpose
        \langle x_1, x_2, x'_1, x'_2 \rangle \\
  \text{subject to }\quad
    & \begin{pmatrix}
         1 &  0 &  0 &  0 \\
        -1 &  0 &  2 &  0 \\
         1 &  0 & -2 &  0 \\
         0 &  1 &  0 & -1 \\
         0 & -1 &  0 &  1 \\
      \end{pmatrix}
      \begin{pmatrix}
	x_1 \\ x_2 \\ x'_1 \\ x'_2
      \end{pmatrix}
        \geq
      \begin{pmatrix}
        2 \\ -1 \\ 0 \\ 1 \\ -1
      \end{pmatrix} \\
    & \langle x_1, x_2, x'_1, x'_2 \rangle \geq \vect{0},
\end{aligned}
\end{equation*}
and the dual optimization problem
\eqref{eq:svg-dual-relaxed-optimization-problem} is
\begin{equation*}
\begin{aligned}
  \text{maximize }\quad
    & \langle 2, -1, 0, 1, -1 \rangle^\transpose
        \langle y_1, y_2, y_3, y_4, y_5 \rangle \\
  \text{subject to }\quad
    & \begin{pmatrix}
        1 &-1  &1  &0  &0\\
        0 &0   &0  &1 &-1\\
        0 &2  &-2  &0  &0\\
        0 &0   &0 &-1  &1\\
      \end{pmatrix}
        \begin{pmatrix} y_1 \\ y_2 \\ y_3 \\ y_4 \\ y_5 \end{pmatrix}
          \leq
            \begin{pmatrix} \mu_1 \\ \mu_2 \\ -\mu_1 \\ -\mu_2 \end{pmatrix} \\
    & \langle y_1, y_2, y_3, y_4, y_5 \rangle \geq \vect{0}.
\end{aligned}
\end{equation*}
Incorporation of the unknown coefficients of $\vect{\mu}$ among the
problem variables finally yields as the transformed
problem~\eqref{eq:svg-final-transformed-dual-relaxed-optimization-problem}:
\begin{equation}
\label{eq:base-two-log-lp-problem}
\begin{aligned}
  \text{maximize }\quad
    & \langle 2, -1, 0, 1, -1, 0, 0 \rangle^\transpose
        \langle y_1, y_2, y_3, y_4, y_5, \mu_1, \mu_2 \rangle \\
  \text{subject to }\quad
    & \begin{pmatrix}
        1  &-1 & 1 & 0 & 0 &-1 & 0 \\
        0  & 0 & 0 & 1 &-1 & 0 &-1 \\
        0  & 2 &-2 & 0 & 0 & 1 & 0 \\
        0  & 0 & 0 &-1 & 1 & 0 & 1 \\
       -2  & 1 & 0 &-1 & 1 & 0 & 0
     \end{pmatrix}
       \begin{pmatrix}
         y_1 \\ y_2 \\ y_3 \\ y_4 \\ y_5 \\ \mu_1 \\ \mu_2
       \end{pmatrix}
         \leq
           \begin{pmatrix} 0 \\ 0 \\ 0 \\ 0 \\ -1\end{pmatrix} \\
    & \langle y_1, y_2, y_3, y_4, y_5, \mu_1, \mu_2 \rangle
        \geq \vect{0}
\end{aligned}
\end{equation}
This problem is feasible so this $\mathrm{CLP}(\Nset)$
program terminates.
Projecting the constraints of \eqref{eq:base-two-log-lp-problem}
onto $\vect{\mu}$ we obtain, in addition, the knowledge
that every $\vect{\mu}$ with $\mu_1 + \mu_2 \geq 1$
gives a ranking function.
In other words, $\mu_1 x_1 + \mu_2 x_2$ is a ranking function
if the non-negative numbers $\mu_1$ and $\mu_2$
satisfy $\mu_1 + \mu_2 \geq 1$.
\end{example}

The following result illustrates the strength of the method:

\begin{theorem}
\label{th:svg-correctness-completeness}
Let $C$ be the binary $\mathrm{CLP}(\Qset_+)$ clause
$p(\bar{x}) \pif c[\bar{x}, \bar{x}'], p(\bar{x}')$,
where $p$ is an $n$-ary predicate and
$c[\bar{x}, \bar{x}']$ is a linear satisfiable constraint.
Let $\mathrm{plrf}(C)$ be the set of positive linear ranking functions
for $C$ and $\mathrm{svg}(C)$ be the set of solutions
of~\eqref{eq:svg-final-transformed-dual-relaxed-optimization-problem}
projected onto $\vect{\mu}$, that is,
\begin{align*}
  \mathrm{plrf}(C)
    &\defeq
      \biggl\{\,
        \vect{\mu} \in \Qset_+^n
      \biggm|
        \forall \bar{x}, \bar{x}' \in \Qset_+^n
          \itc
            c[\bar{x}, \bar{x}']
              \implies
                \sum_{i=1}^n \mu_i x_i - \sum_{i=1}^n \mu_i x'_i \geq 1
      \,\biggr\}, \\
  \mathrm{svg}(C)
    &\defeq
      \bigl\{\,
        \vect{\check{\mu}} \in \Qset_+^n
      \bigm|
        \text{%
          $\langle \vect{\check{y}}, \vect{\check{\mu}} \rangle$
          is a solution of
          \eqref{eq:svg-final-transformed-dual-relaxed-optimization-problem}%
         }
      \,\bigr\}.
\end{align*}
Then $\mathrm{plrf}(C) = \mathrm{svg}(C)$.
\end{theorem}

\begin{pf} As $c[\bar{x}, \bar{x}']$ is satisfiable,
problem~\eqref{eq:svg-relaxed-optimization-problem}
is feasible.  We prove each inclusion separately.
\paragraph*{$\mathrm{svg}(C) \subseteq \mathrm{plrf}(C)$}
Assume that~\eqref{eq:svg-final-transformed-dual-relaxed-optimization-problem}
is feasible and let
$\langle \vect{\check{y}}, \vect{\check{\mu}} \rangle$ be a
solution of~\eqref{eq:svg-final-transformed-dual-relaxed-optimization-problem}.
For this choice of $\vect{\check{\mu}}$,
the corresponding LP
problems~\eqref{eq:svg-relaxed-optimization-problem}
and~\eqref{eq:svg-dual-relaxed-optimization-problem}
are bounded by $q \geq 1$
(case 2b of the discussion above).
So $\vect{\check{\mu}} \in \mathrm{plrf}(C)$.
\paragraph*{$\mathrm{plrf}(C) \subseteq \mathrm{svg}(C)$}
Let us pick $\vect{\mu} \in \mathrm{plrf}(C)$.
For this choice, the corresponding LP
problem~\eqref{eq:svg-relaxed-optimization-problem} is bounded
by $r \geq 1$, so is its dual~\eqref{eq:svg-dual-relaxed-optimization-problem}.
Let $\vect{\hat{y}}$ be   an optimal solution for~\eqref{eq:svg-dual-relaxed-optimization-problem}.
Thus $\langle \vect{\hat{y}}, \vect{\mu} \rangle$ is a feasible solution
of~\eqref{eq:svg-transformed-dual-relaxed-optimization-problem}
and~\eqref{eq:svg-final-transformed-dual-relaxed-optimization-problem}.
Hence $\vect{\mu} \in \mathrm{svg}(C)$.
\end{pf}

As an immediate consequence, the question
\emph{``does a given binary recursive clause with linear constraint
admit a positive linear mapping?''}
can be solved in weakly polynomial time.

\begin{corollary}
\label{co:plrf-in-P}
Let $C$ be the binary $\mathrm{CLP}(\Qset_+)$ clause
$p(\bar{x}) \pif c[\bar{x}, \bar{x}'], p(\bar{x}')$,
where $c[\bar{x}, \bar{x}']$ is a linear satisfiable constraint.
The decision problem $\mathrm{plrf}(C) = \emptyset$ is in $\mathrm{P}$.
\end{corollary}

\begin{pf}
By Theorem~\ref{th:svg-correctness-completeness}
the problems $\mathrm{plrf}(C) = \emptyset$ and
$\mathrm{svg}(C) = \emptyset$ are equivalent.
So, if~\eqref{eq:svg-final-transformed-dual-relaxed-optimization-problem}
is feasible then the answer is \emph{no}: as $c[\bar{x}, \bar{x}']$ is
satisfiable, we are in case (2)(b).
Otherwise, again because of the satisfiability of $c[\bar{x}, \bar{x}']$,
either~\eqref{eq:svg-relaxed-optimization-problem} is
unbounded (case (1)(a)ii.\ or it is bounded by $q' < 0$ (case (1)(b)).
In both cases, the answer is \emph{yes}.
Finally, testing the satisfiability of a linear system,
as well as computing one of its solutions  ---and thus computing
one concrete linear ranking function---, is in $\mathrm{P}$
(see, e.g., \cite{Schrijver86}).
\end{pf}

For the case where we have more than one directly recursive
binary $\mathrm{CLP}(\Nset)$ clauses, $C_1$, \dots,~$C_n$,
the set of \emph{global} positive linear ranking functions,
i.e., that ensure termination whichever clause is selected
at each computation step, is given by $\biginters_{i=1}^n \mathrm{svg}(C_i)$.
This can be computed by taking the conjunction of the constraints
obtained, for each clause, from the projection of the constraints
of the corresponding linear
problem~\eqref{eq:svg-final-transformed-dual-relaxed-optimization-problem}
onto $\vect{\mu}$.

To summarize, the main contribution of Sohn and Van~Gelder lies
in their encoding of the ranking function search problem into
linear programming and their use of the duality theorem.
As we will see, this idea is amenable to a generalization that makes
it widely applicable to any programming paradigm, not just (constraint)
logic programming.

\subsection{The Generalization by Mesnard and Serebrenik}
\label{subsec:mesnard-and-serebrenik}

Fred Mesnard and Alexandre Serebrenik have generalized the method
of Sohn and Van Gelder from the analysis of logic programs to the
analysis of $\mathrm{CLP}(\Qset)$ and $\mathrm{CLP}(\Rset)$ programs
in \cite{MesnardS05,MesnardS08}.
In the following, for presentation purposes and without loss of generality,
we consider the case of rational-valued variables.
They use a class of affine ranking functions of the form
\begin{equation}
\label{eq:ms-ranking-function-form}
  f_p(y_1, \ldots, y_n) = \mu_0 + \sum_{i=1}^n \mu_i y_i,
\end{equation}
where $\mu_i \in \Qset$, for $i=0$, \dots,~$n$.
Allowing for rational-valued coefficients $\mu_i$ and variables $y_i$
(both the $\mu_i$'s and the $y_i$'s were naturals in~\cite{SohnVG91})
implies that~\eqref{eq:ms-ranking-function-form}
does not necessarily define a nonnegative function and that
Zeno sequences\footnote{Such as $1$, $\frac12$, $\frac14$, $\frac18$, \dots.}
are not automatically excluded.
Consequently, to avoid these two problems,
condition~\eqref{eq:svg-strict-decrease} is strengthened to\footnote{%
Our presentation is strictly more general than the formulation
in~\cite{MesnardS05,MesnardS08}, which imposes that
$f_p(\bar{x}) \geq 1 + f_p(\bar{x}') \land f_p(\bar{x}') \geq 0$.}
\begin{equation}
\label{eq:ms-strict-decrease}
  \forall \bar{x}, \bar{x}' \in \Qset^n
    \itc
      c[\bar{x}, \bar{x}']
        \implies
          \bigl(
            f_p(\bar{x}) \geq 1 + f_p(\bar{x}')
              \land
                f_p(\bar{x}) \geq 0
          \bigr).
\end{equation}
Note that the choice of the numbers $1$ and $0$ in the right hand side
of the above implication preserves generality: the
general form of the former condition, i.e.,
$f_p(\bar{x}) \geq \epsilon + f_p(\bar{x}')$
for a fixed and strictly positive~$\epsilon \in \Qset_+$,
can be transformed as shown in
Section~\ref{sec:termination-analysis-of-individual-loops}, and
the general form of the latter, i.e., $f_p(\bar{x}) \geq b$ for a
fixed~$b \in \Qset$, can be transformed into $f_p(\bar{x}) \geq 0$ by
a suitable choice of $\mu_0$.
Condition~\eqref{eq:ms-strict-decrease} can be rewritten as
\begin{equation}
\label{eq:ms-strict-decrease-matrix-form}
  \forall \bar{x}, \bar{x}' \in \Qset^n
    \itc
      c[\bar{x}, \bar{x}']
        \implies
          \biggl(
            \sum_{i=1}^n \mu_i x_i - \sum_{i=1}^n \mu_i x'_i \geq 1
              \land \mu_0 + \sum_{i=1}^n \mu_i x_i \geq 0
          \biggr).
\end{equation}

Using the same notation chosen for \eqref{eq:svg-optimization-problem},
the existence of a ranking function can now be equivalently expressed
as the existence of a solution of cost at least $1$ to the former and
a solution of cost at least $0$ to the latter of the following
optimization problems:
\begin{equation}
\label{eq:ms-optimization-problems}
\begin{aligned}
  \text{minimize }\quad
    & \langle \vect{\mu}, -\vect{\mu} \rangle^\transpose
        \langle \vect{x}, \vect{x}' \rangle \\
  \text{subject to }\quad
    & \mat{A}_c \langle \vect{x}, \vect{x}' \rangle
        \geq \vect{b}_c
\end{aligned}
\qquad\qquad
\begin{aligned}
  \text{minimize }\quad
    & \langle \vect{\tilde \mu}, \vect{0} \rangle^\transpose
        \langle \vect{\tilde x}, \vect{x}' \rangle \\
  \text{subject to }\quad
    & \mat{\tilde A}_c \langle \vect{\tilde x}, \vect{x}' \rangle
        \geq \vect{\tilde b}_c
\end{aligned}
\end{equation}
where the extended vectors
$\vect{\tilde \mu} \defeq \langle \mu_0, \vect{\mu} \rangle$ and
$\vect{\tilde x} \defeq \langle x_0, \vect{x} \rangle$
include the parameter $\mu_0$ and the new variable $x_0$,
respectively,
and the extended matrix and vector
\[
  \mat{\tilde A}_c
    \defeq
        \begin{pmatrix}
          1  & \vect{0} \cr
          -1 & \vect{0} \cr
          \vect{0} & \mat{A}_c
        \end{pmatrix}
\quad\text{and}\quad
  \vect{\tilde b}_c \defeq \langle 1, -1, \vect{b}_c \rangle
\]
encode the additional constraint $x_0 = 1$.

Reasoning as in Section \ref{subsec:rf-for-binary-clp-n}, the problems
\eqref{eq:ms-optimization-problems} can then be transformed, applying the
suitable form of the duality theorem, into the following dual problems
over new vectors of variables $\vect{y}$ and $\vect{z}$,
ranging over $\Qset^m$ and $\Qset^{m+2}$, respectively:
\begin{equation}
\label{eq:ms-dual-optimization-problems}
\begin{aligned}
  \text{maximize }\quad
    & \vect{b}_c^\transpose \vect{y} \\
  \text{subject to }\quad
    & \mat{A}_c^\transpose \vect{y}
        = \langle \vect{\mu}, -\vect{\mu} \rangle \\
    & \vect{y} \geq \vect{0}
\end{aligned}
\qquad\qquad
\begin{aligned}
  \text{maximize }\quad
    & \vect{\tilde{b}}_c^\transpose \vect{z} \\
  \text{subject to }\quad
    & \mat{\tilde A}_c^\transpose \vect{z}
        = \langle \vect{\tilde \mu}, \vect{0} \rangle \\
    & \vect{z} \geq \vect{0}
\end{aligned}
\end{equation}

Now the condition that the optimal solution is at least $1$ (resp., $0$)
can be added to the constraints, thus reducing the optimization problems
\eqref{eq:ms-optimization-problems} to testing the satisfiability
of the system:
\begin{equation*}
\left\{
  \begin{array}{l}
     \vect{b}_c^\transpose \vect{y} \geq 1\\
     \mat{A}_c^\transpose \vect{y} = \langle \vect{\mu}, -\vect{\mu} \rangle \\
     \vect{y} \geq \vect{0} \\
    \\
    \vect{\tilde{b}}_c^\transpose \vect{z} \geq 0\\
     \mat{\tilde A}_c^\transpose \vect{z} = \langle \vect{\tilde \mu}, \vect{0} \rangle \\
     \vect{z} \geq \vect{0} \\
  \end{array}
\right.
\end{equation*}
or equivalently, after incorporating the parameters $\vect{\mu}$
(resp., $\vect{\tilde{\mu}}$) into the variables, to the
generalization to $\Qset$ of problem
\eqref{eq:svg-final-transformed-dual-relaxed-optimization-problem}:
\begin{equation}
\label{eq:ms-final-transformed-dual-optimization-problems}
\left(
  \begin{BMAT}(c){c:c}{c:c:c:c}
    \mat{A}_c^\transpose
    & \begin{matrix}
        -\mat{I}_n \cr
        \mat{I}_n
      \end{matrix}
  \cr
    -\mat{A}_c^\transpose
    & \begin{matrix}
        \mat{I}_n \cr
        -\mat{I}_n
      \end{matrix}
  \cr
    -\mat{I}_m
    & \vphantom{\begin{matrix}
                 \mat{I}_n \cr
                 \mat{I}_n
               \end{matrix}}
      \vect{0}
  \cr
    -\vect{b}_c^\transpose
    &  \vect{0}
  \end{BMAT}
\right)
  \langle \vect{y}, \vect{\mu} \rangle
    \leq
\left(
  \begin{BMAT}(c){c}{c:c:c:c}
  \vphantom{\begin{matrix}
              \mat{I}_n \cr
              \mat{I}_n
            \end{matrix}}
    \vect{0} \cr
  \vphantom{\begin{matrix}
              \mat{I}_n \cr
              \mat{I}_n
            \end{matrix}}
    \vect{0} \cr
  \vphantom{\begin{matrix}
              \mat{I}_n \cr
              \mat{I}_n
            \end{matrix}}
    \vect{0} \cr
    -1
  \end{BMAT}
\right)
\bigwedge
\left(
  \begin{BMAT}(c){c:c}{c:c:c:c}
    \mat{\tilde{A}}_c^\transpose
    & \begin{matrix}
        -\mat{I}_{n+1} \cr
        \mat{0}
      \end{matrix}
  \cr
    -\mat{\tilde{A}}_c^\transpose
    & \begin{matrix}
        \mat{I}_{n+1} \cr
        \mat{0}
      \end{matrix}
  \cr
    \vphantom{\begin{matrix}
                \mat{I}_n \cr
                \mat{I}_n
              \end{matrix}}
    -\mat{I}_{m+2}
    & \vect{0}
  \cr
    -\vect{\tilde{b}}_c^\transpose
      & \vect{0}
  \end{BMAT}
\right)
  \langle \vect{z}, \vect{\tilde{\mu}} \rangle
    \leq \vect{0}.
\end{equation}

The following completeness result generalizes
Theorem~\ref{th:svg-correctness-completeness}:
\begin{theorem}
\label{th:ms-correctness-completeness}
Let $C$ be the binary $\mathrm{CLP}(\Qset)$ clause
$p(\bar{x}) \pif c[\bar{x}, \bar{x}'], p(\bar{x}')$,
where $p$ is an $n$-ary predicate and
$c[\bar{x}, \bar{x}']$ is a linear satisfiable constraint.
Let $\mathrm{lrf}(C)$ be the set of linear ranking functions for $C$
and $\mathrm{ms}(C)$ be the set of solutions
of~\eqref{eq:ms-final-transformed-dual-optimization-problems}
projected onto $\vect{\tilde{\mu}}$, that is,
\begin{align*}
  \mathrm{lrf}(C)
    &\defeq
      \sset{%
        \vect{\tilde{\mu}} \in \Qset^{n+1}
      }{%
        \forall \bar{x}, \bar{x}' \in \Qset^n
          \itc
            c[\bar{x}, \bar{x}']
              \implies \\\quad
                \sum_{i=1}^n \mu_i x_i - \sum_{i=1}^n \mu_i x'_i \geq 1
            \land
              \mu_0 + \sum_{i=1}^n \mu_i x_i \geq 0
      }, \\
  \mathrm{ms}(C)
    &\defeq
      \bigl\{\,
        \vect{\tilde{\mu}} \in \Qset^{n+1}
      \bigm|
        \text{%
          $\langle \vect{y}, \vect{\mu} \rangle$ and
          $\langle \vect{z}, \vect{\tilde{\mu}} \rangle$
          are solutions of the problems
          \eqref{eq:ms-final-transformed-dual-optimization-problems}%
        }
      \,\bigr\}.
\end{align*}
Then $\mathrm{lrf}(C) = \mathrm{ms}(C)$.
\end{theorem}

\begin{pf}
We use \emph{l} and \emph{r} as subscripts
of our references to the LP problems~\eqref{eq:ms-optimization-problems},
\eqref{eq:ms-dual-optimization-problems},
and~\eqref{eq:ms-final-transformed-dual-optimization-problems}
to denote the LP problems on the left and the LP problems on the right.
\paragraph*{$\mathrm{ms}(C) \subseteq \mathrm{lrf}(C)$}
Assume that~\eqref{eq:ms-final-transformed-dual-optimization-problems}
is feasible and let
$\langle \vect{\check{y}}, \vect{\check{\mu}} \rangle$ be a
solution of~\eqref{eq:ms-final-transformed-dual-optimization-problems}$_{l}$
and $\langle \vect{\check{z}}, \vect{\check{\mu}} \rangle$
be a solution of~\eqref{eq:ms-final-transformed-dual-optimization-problems}$_{r}$.
For this choice of $\vect{\check{\mu}}$,
the corresponding
LP problems~\eqref{eq:ms-dual-optimization-problems}$_{l}$
and~\eqref{eq:ms-optimization-problems}$_{l}$
are bounded by $1$ while
the corresponding LP
problems~\eqref{eq:ms-dual-optimization-problems}$_{r}$
and~\eqref{eq:ms-optimization-problems}$_{r}$
are bounded by $0$.
Hence we have
\[
   \forall \bar{x}, \bar{x}' \in \Qset^n
          \itc
            c[\bar{x}, \bar{x}']
              \implies
                \sum_{i=1}^n \mu_i x_i - \sum_{i=1}^n \mu_i x'_i \geq 1
\]
and
\[
  \forall \bar{x}, \bar{x}' \in \Qset^n
    \itc
      c[\bar{x}, \bar{x}']
        \implies
          \mu_0 + \sum_{i=1}^n \mu_i x_i \geq 0.
\]
Thus
\[
  \forall \bar{x}, \bar{x}' \in \Qset^n
    \itc
      c[\bar{x}, \bar{x}']
        \implies
          \sum_{i=1}^n \mu_i x_i - \sum_{i=1}^n \mu_i x'_i \geq 1
            \land
              \mu_0 + \sum_{i=1}^n \mu_i x_i \geq 0,
\]
so that $\vect{\check{\mu}} \in \mathrm{lrf}(C)$.
\paragraph*{$\mathrm{lrf}(C) \subseteq \mathrm{ms}(C)$}
Let us pick $\vect{\tilde{\mu}} \in \mathrm{lrf}(C)$.
For this choice, the corresponding LP problem~\eqref{eq:ms-optimization-problems}
are bounded
by $1$ and $0$, and so are their duals~\eqref{eq:ms-dual-optimization-problems}.
Let $\vect{\hat{y}}$ be an optimal solution
for~\eqref{eq:ms-dual-optimization-problems}$_{l}$.
Thus $\langle \vect{\hat{y}}, \vect{\mu} \rangle$ is a feasible solution
of~\eqref{eq:ms-final-transformed-dual-optimization-problems}$_{l}$.
Similarly, let $\vect{\hat{z}}$ be an optimal solution for~\eqref{eq:ms-dual-optimization-problems}$_{r}$.
Thus $\langle \vect{\hat{z}}, \vect{\tilde{\mu}} \rangle$ is a feasible solution
of~\eqref{eq:ms-final-transformed-dual-optimization-problems}$_{r}$.
Hence $\vect{\tilde{\mu}} \in \mathrm{ms}(C)$.
\end{pf}

Moreover, even for the case of the linear fragment of
$\mathrm{CLP}(\Qset)$ ---and $\mathrm{CLP}(\Rset)$---
checking for the existence of a linear ranking function
is a weakly polynomial problem.

\begin{corollary}
\label{co:lrf-in-P}
Let $C$ be the binary $\mathrm{CLP}(\Qset)$ clause
$p(\bar{x}) \pif c[\bar{x}, \bar{x}'], p(\bar{x}')$,
where $c[\bar{x}, \bar{x}']$ is a linear satisfiable constraint.
The decision problem $\mathrm{lrf}(C) = \emptyset$ is in~$\mathrm{P}$.
\end{corollary}

A space of ranking functions can be obtained (at a computational price
that is no longer polynomial) by projecting the constraints
of~\eqref{eq:ms-final-transformed-dual-optimization-problems}
onto $\vect{\tilde{\mu}}$.
Any $\vect{\tilde{\mu}}$ satisfying all the projected constraints
corresponds to one ranking function that, subject to
$c[\bar{x},\bar{x}']$, is bounded from below by $0$ and that decreases
by at least~$1$ at each iteration.
From these ``normalized'' ranking functions, the opposite of the
transformation outlined in
Section~\ref{sec:termination-analysis-of-individual-loops}
allows to recover all affine ranking functions: these are induced
by the set of parameters
\begin{equation}
\label{eq:ms-denormalization}
  \bigl\{\,
    \langle h, k\vect{\mu} \rangle
  \bigm|
    \langle \mu_0, \vect{\mu} \rangle \in \mathrm{lrf}(C),
    h \in \Qset,
    k \in \Qset_+\setdiff\{0\}
  \,\bigr\}.
\end{equation}

\subsection{Application to the Analysis of Imperative While Loops}
\label{sec:MS-applied-to-imperative}

The generalization of Mesnard and Serebrenik can be used, almost
unchanged, to analyze the termination behavior of imperative while
loops with integer- or rational-valued variables.
Consider a loop of the form~\eqref{eq:while-loop}, i.e.,
$\{\, I \,\} \; \kw{while}\;  B \; \kw{do} \; C$ where $I$ is known
to hold before any evaluation of $B$ and $C$ is known to always
terminate in that loop.
Termination analysis is conducted as follows:
\begin{enumerate}
\item
Variables are duplicated: if $\bar{x}$ are the $n$ variables of the
original loop, we introduce a new tuple of variables $\bar{x}'$.
\item
An analyzer based on convex polyhedra \cite{CousotH78}
is used to analyze the following program:
\begin{equation}
\label{prog:ms-loop}
\begin{aligned}
  &\{ I \} \\
  &x'_1 \assign x_1; \: \ldots; \; x'_n \assign x_n; \\
  &\kw{if} \; B[\bar{x}'/\bar{x}] \; \kw{then} \\
  &\qquad C[\bar{x}'/\bar{x}] \\
  &\qquad \bigstar
\end{aligned}
\end{equation}
Let the invariant obtained for the program point marked with `$\bigstar$'
be $c[\bar{x}, \bar{x}']$;  this is a finite conjunction of linear
constraints.
\item
The method of Mesnard and Serebrenik is now applied to the
$\mathrm{CLP}(\Qset)$
clause $p(\bar{x}) \pif c[\bar{x}, \bar{x}'], p(\bar{x}')$:
if termination can be established for that clause, then the while
loop we started with will terminate.
\end{enumerate}

Notice how the clause $p(\bar{x}) \pif c[\bar{x}, \bar{x}'], p(\bar{x}')$
approximates the termination behavior of the loop:
if we interpret the predicate $p$ applied to $\bar{x}$ as
``the loop guard is evaluated on values $\bar{x}$,''
then the clause can be read as ``if the loop guard is evaluated on values
$\bar{x}$, and $c[\bar{x}, \bar{x}']$ holds, then the loop guard will
be evaluated again on values $\bar{x}'$.''

We illustrate the overall methodology with an example.
\begin{example}
\label{ex:sm-example}
The following program, where $x_1$ and $y$ take values in $\Zset$,
computes and stores in $x_2$ the integer base-$2$ logarithm of $x_1$
if $x_1 > 0$, $0$ otherwise:
\[
\begin{aligned}
  &x_2 \assign 0; \\
  &\{ x_1 \geq 0 \land x_2 \geq 0 \} \\
  &\kw{while} \; x_1 \geq 2 \; \kw{do} \\
  &\qquad x_1 \assign x_1 \; \kw{div} \; 2; \\
  &\qquad x_2 \assign x_2+1
\end{aligned}
\]
where the loop invariant $\{ x_1 \geq 0 \land x_2 \geq 0 \}$ has been
obtained by static analysis.
After the duplication of variables, we submit to the analyzer the program
\[
\begin{aligned}
  &\{ x_1 \geq 0 \land x_2 \geq 0 \} \\
  &x'_1 \assign x_1; \; x'_2 \assign x_2; \\
  &\kw{if} \; x'_1 \geq 2 \; \kw{then} \\
  &\qquad x'_1 \assign x'_1 \; \kw{div} \; 2; \\
  &\qquad x'_2 \assign x'_2+1 \\
  &\qquad \bigstar
\end{aligned}
\]
and we obtain, for program point `$\bigstar$', the invariant
\[
  x_1 \geq 2 \land 2x'_1 + 1 \geq x_1 \land 2x'_1 \leq x_1
    \land x'_2 = x_2 + 1 \land x'_2 \geq 1.
\]
Applying the method of Mesnard and Serebrenik we obtain that, for each
$\mu_0, \mu_1, \mu_2 \in \Qset$ such that
$\mu_1 - \mu_2 \geq 1$,
$\mu_2 \geq 0$, and $\mu_0 + 2\mu_1 \geq 0$,
the function $f(x_1, x_2) \defeq \mu_0 + \mu_1 x_1 + \mu_2 x_2$
is a ranking function for the given while loop.
It is interesting to observe that the first constraint guarantees
strict decrease (at least $1$), the addition of the second constraint
guarantees boundedness from below, while the further addition of the
third constraint ensures nonnegativity, i.e., that $0$ is a lower bound.
\end{example}

\subsection{Application to Conditional Termination Analysis}
\label{sec:MS-applied-to-conditional-termination}

An important observation is that the method of Mesnard and Serebrenik
is immediately applicable in \emph{conditional termination analysis}.
This is the problem of (automatically) inferring the preconditions
under which code that does not universally terminate (i.e., there are
inputs for which it does loop forever) is guaranteed to terminate.
This problem has been recently studied in \cite{CookGL-ARS08},
where preconditions are inferred under which functions that are either
decreasing or bounded become proper ranking functions.
The two systems in~\eqref{eq:ms-final-transformed-dual-optimization-problems},
projected onto $\vect{\tilde{\mu}}$,
exactly define the space of non-negative candidate ranking functions
and the space of decreasing candidate ranking functions, respectively.
While this is subject for future research, we believe that the availability
of these two spaces allows to improve the techniques presented
in \cite{CookGL-ARS08}.

\section{The Approach of Podelski and Rybalchenko}
\label{sec:podelski-and-rybalchenko}

Andreas Podelski and Andrey Rybalchenko \cite{PodelskiR04} introduce
a method for finding linear ranking functions for a particular class
of unnested while loops that, with the help of a preliminary analysis phase,
is indeed completely general.

Consider a while loop of the form
\begin{equation}
\label{prog:pr-loop}
\begin{aligned}
  &\{ I \} \\
  &\kw{while} \; B \; \kw{do} \\
  &\qquad \blacktriangledown \\
  &\qquad C \\
  &\qquad \bigstar
\end{aligned}
\end{equation}
in which variables $x_1$, \dots,~$x_n$ occur.
Suppose we have determined (e.g., by a data-flow analysis based
on convex polyhedra) that the invariant
\begin{align}
\label{eq:pr-invariant-after-B}
  \sum_{i=1}^n g_{k,i} x_i &\leq b_k,
    &\text{for $k=1$, \dots,~$r$,} \\
\intertext{%
holds at the program point marked with `$\blacktriangledown$',
while the invariant
}
\label{eq:pr-invariant-after-C}
  \sum_{i=1}^n a_{k,i}' x_i' &\leq \sum_{i=1}^n a_{k,i} x_i + b_k,
    &\text{for $k=r+1$, \dots,~$r+s$,}
\end{align}
holds at the program point marked with `$\bigstar$',
where unprimed variables represent the values before the update
and primed variables represent the values after the update,
and all the coefficients and variables are assumed
to take values in $\Qset$.\footnote{In \cite{PodelskiR04} variables
are said to have integer domain, but this restriction seems unnecessary
and, in fact, it is not present in \cite{Rybalchenko04th}.}

The inequalities in~\eqref{eq:pr-invariant-after-B}
can be expressed in the form \eqref{eq:pr-invariant-after-C}
by just defining $a_{k,i}' \defeq 0$ and $a_{k,i} \defeq -g_{k,i}$
for $i=1$, \dots,~$n$ and $k=1$, \dots,~$r$.
The conjunction of~\eqref{eq:pr-invariant-after-B}
and~\eqref{eq:pr-invariant-after-C} can now be stated in matrix form as
\begin{equation}
\label{eq:pr-matrix-form}
  \begin{pmatrix} \mat{A} & \mat{A}' \end{pmatrix}
    \begin{pmatrix} \vect{x} \\ \vect{x}' \end{pmatrix}
      \leq \vect{b},
\end{equation}
where the matrix $\begin{pmatrix} \mat{A} & \mat{A}' \end{pmatrix}$
is obtained by juxtaposition of the two $(r+s)\times n$ matrices
$\mat{A} \defeq (-a_{k,i})$ and $\mat{A}' \defeq (a_{k,i}')$,
$\vect{b} \defeq \langle b_1, b_2, \ldots, b_{r+s} \rangle$ and,
as explained in Section~\ref{sec:preliminaries},
$\langle\vect{x}, \vect{x}'\rangle$ is obtained by juxtaposing the vectors
$\vect{x} \defeq \langle x_1, x_2, \ldots, x_n \rangle$ and
$\vect{x}' \defeq \langle x_1', x_2', \ldots, x_n' \rangle$.

Podelski and Rybalchenko have proved that~\eqref{prog:pr-loop}
is guaranteed to terminate on all possible inputs if there
exist two $(r+s)$-dimensional non-negative rational vectors
$\vect{\lambda}_1$ and $\vect{\lambda}_2$ such that:%
\footnote{For an informal justification of these equations,
see Section~\ref{sec:pr-justification}; a more detailed explanation
\ifthenelse{\boolean{LONGVERSION}}{
is provided in Section~\ref{sec:pr-from-lagrange}.
}{
is available in~\cite{BagnaraMPZ12TR}.
}
}

\begin{subequations}
\label{eq:pr-system}
\begin{align}
\label{eq:pr-system-a}
  \vect{\lambda}_1^\transpose \mat{A}'
    &= \vect{0}, \\
\label{eq:pr-system-b}
  (\vect{\lambda}_1^\transpose - \vect{\lambda}_2^\transpose) \mat{A}
    &= \vect{0}, \\
\label{eq:pr-system-c}
  \vect{\lambda}_2^\transpose (\mat{A}+\mat{A}')
    &= \vect{0}, \\
\label{eq:pr-system-d}
  \vect{\lambda}_2^\transpose \vect{b}
    &< 0.
\end{align}
\end{subequations}
Note that we have either zero or infinitely many solutions,
since if the pair of vectors $\vect{\lambda}_1$ and $\vect{\lambda}_2$
satisfies the constraints,
then the pair $k\vect{\lambda}_1$ and $k\vect{\lambda}_2$
satisfies them as well, for any $k \in \Qset_+\setdiff\{0\}$.
Podelski and Rybalchenko proved also the following completeness
result: if the iterations of \eqref{prog:pr-loop} are
\emph{completely characterized}
by conditions~\eqref{eq:pr-invariant-after-B}
and~\eqref{eq:pr-invariant-after-C}
---in which case they call it a ``simple linear loop''---
then $\vect{\lambda}_1, \vect{\lambda}_2 \in \Qset_+^{r+s}$
satisfying conditions \eqref{eq:pr-system-a}--\eqref{eq:pr-system-d}
exist \emph{if and only if} the program terminates for all inputs.

\subsection{Generation of Ranking Functions}

For each pair of vectors $\vect{\lambda}_1$ and $\vect{\lambda}_2$
satisfying the conditions~\eqref{eq:pr-system-a}--\eqref{eq:pr-system-d},
a linear ranking function for the considered program can be obtained as
\begin{align}
\label{eq:pr-ranking-function}
  f(\vect{x})
    &\defeq
      \vect{\lambda}_2^\transpose \mat{A}' \vect{x}. \\
\intertext{%
In \cite{PodelskiR04} a slightly more complex form is proposed, namely:
}
\label{eq:pr-ranking-function-with-cases}
  g(\vect{x})
    &\defeq
      \begin{cases}
        \vect{\lambda}_2^\transpose \mat{A}' \vect{x},
          &\text{if there exists $\vect{x}'$ such that
                 \(
                   \begin{pmatrix} \mat{A} & \mat{A}' \end{pmatrix}
                     \begin{pmatrix} \vect{x} \\ \vect{x}' \end{pmatrix}
                       \leq
                         \vect{b}
                 \)
           }, \\
        (\vect{\lambda}_2^\transpose - \vect{\lambda}_1^\transpose)\vect{b},
          &\text{otherwise,}
      \end{cases}
\end{align}
but the extra provisions are actually necessary only if one is interested
into an ``extended ranking function'' that is strictly decreasing
also on the very last iteration of the loop, that is, when the effect of
the command $C$ is such that $\vect{x}$ would violate the loop guard $B$
at the following iteration.
As this more complex definition does not seem to provide any
additional benefit, we disregard it and consider only
the linear ranking function~\eqref{eq:pr-ranking-function}.

\begin{example}
Consider again the program of Example~\ref{ex:sm-example}.
The invariants in the forms dictated by~\eqref{eq:pr-invariant-after-B}
and~\eqref{eq:pr-invariant-after-C} are given by the systems
$\{ -x_1 \leq -2, -x_2' \leq -1 \}$
and
\(
  \{
     2x_1'     \leq  x_1,     \,
    -2x_1' - 1 \leq -x_1,     \,
     -x_2'     \leq -x_2 - 1, \,
      x_2'     \leq  x_2 + 1
  \}
\),
respectively.
These can be expressed in the matrix form~\eqref{eq:pr-matrix-form}
by letting
\[
  \mat{A}
    \defeq
      \begin{pmatrix}
        \hfill -1 & \hfill  0 \\
        \hfill -1 & \hfill  0 \\
        \hfill  1 & \hfill  0 \\
        \hfill  0 & \hfill  1 \\
        \hfill  0 & \hfill -1 \\
        \hfill  0 & \hfill  0
      \end{pmatrix},
\quad
  \mat{A}'
    \defeq
      \begin{pmatrix}
        \hfill  0 & \hfill  0 \\
        \hfill  2 & \hfill  0 \\
        \hfill -2 & \hfill  0 \\
        \hfill  0 & \hfill -1 \\
        \hfill  0 & \hfill  1 \\
        \hfill  0 & \hfill -1
      \end{pmatrix},
\quad
  \vect{b}
    =
      \begin{pmatrix}
        \hfill -2 \\ \hfill 0 \\ \hfill 1 \\ \hfill -1 \\ \hfill 1 \\ \hfill -1
      \end{pmatrix}.
\]
Two non-negative rational vectors solving the system~\eqref{eq:pr-system}
are, for instance,
$\vect{\lambda}_1 = \langle 2, 0, 0, 0, 0, 0 \rangle^\transpose$ and
$\vect{\lambda}_2 = \langle 1, 1, 0, 0, 0, 0 \rangle^\transpose$.
\end{example}

\subsection{Justification of the Approach}
\label{sec:pr-justification}

A reader of \cite{PodelskiR04} wonders where the method of Podelski
and Rybalchenko comes from.  In fact, the paper does not give an
intuition about why conditions \eqref{eq:pr-system-a}--\eqref{eq:pr-system-d}
imply termination of~\eqref{prog:pr-loop}.
Those conditions can be mapped into a strengthening, tailored to the
linear case, of the well known Floyd termination verification conditions,%
\footnote{A.~Rybalchenko, personal communication, 2011.}
but such a higher level view needs to be extracted, with some effort,
from the details of the proof of \cite[Theorem~1]{PodelskiR04}.
The relative completeness of the approach is then proved
in \cite[Theorem~2]{PodelskiR04}
by exploiting the affine form of Farkas' Lemma, showing that such
a strengthening is unconsequential for the case of linear ranking
functions and simple linear loops.

\ifthenelse{\boolean{LONGVERSION}}{
}{
The intuitive reading hidden in the proof details is made explicit
in \cite[Section 6.2]{Cousot05}, where Patrick Cousot hints that
the method by Podelski and Rybalchenko can be derived from
the Floyd termination verification conditions by application of
Lagrangian relaxation. The interested reader can find more details
on this connection in the technical report version
of this paper~\cite{BagnaraMPZ12TR}.
}

\ifthenelse{\boolean{LONGVERSION}}{

\subsection{Interpretation in Terms of Lagrangian Relaxation}
\label{sec:pr-from-lagrange}

The intuitive reading hidden in the proof details mentioned above
is made explicit
in \cite[Section 6.2]{Cousot05}, where Patrick Cousot hints that
the method by Podelski and Rybalchenko can be derived from
the Floyd termination verification conditions by application of
Lagrangian relaxation.\footnote{Lagrangian relaxation is a standard device
to convert entailment into constraint solving: given a finite dimensional
vector space $\mathbb V$, a positive integer $n$ and functions
$\fund{f_k}{\mathbb{V}}{\Qset}$ for $k = 0$, \dots,~$n$,
the property that, for each $\vect{x} \in \mathbb V$,
$\bigwedge_{k=1}^n f_k(\vect{x}) \geq 0 \implies f_0(\vect{x}) \geq 0$
can be \emph{relaxed} to proving the existence of a vector
$\vect{a} \in \Qset_+^n$ such that, for all $\vect{x} \in \mathbb{V}$,
$f_0(\vect{x}) - \sum_{k=1}^n a_k f_k(\vect{x}) \geq 0$. If the
$f_k$ are affine functions, the latter condition is equivalent to the
former.}
We now show that this is indeed the case.

Assuming we are dealing with affine ranking functions and adding
the limitation that $r=1$ in~\eqref{eq:pr-invariant-after-B},
in~\cite{Cousot05} the existence of an affine ranking function
is formalized, following Floyd's method, by requiring the existence of
$\vect{\mu} \in \Qset^{n}$ and $\mu_0, \delta \in \Qset$ such that:
\begin{align*}
  \forall \vect{x} \in \Qset^n
    \itc
      \sigma_1(\vect{x}, \vect{x}') \geq 0
        &\implies
          \vect{\mu}^\transpose\vect{x} + \mu_0 \geq 0, \\
  \forall \vect{x}, \vect{x}' \in \Qset^n
    \itc
      \bigwedge_{k=1}^m
        \bigl( \sigma_k(\vect{x}, \vect{x}') \geq 0 \bigr)
          &\implies
            \vect{\mu}^\transpose(\vect{x} - \vect{x}') - \delta \geq 0, \\
  &\delta > 0,
\end{align*}
where the loop is described by the inequalities
$\sigma_k(\vect{x}, \vect{x}') \geq 0$,
with $\sigma_1$ being the inequality in~\eqref{eq:pr-invariant-after-B}.
These are provided with the usual intuitive reading:
the first implication states that the ranking function is
always nonnegative at the head of the loop body;
the second implication makes sure that the ranking function
is decreasing on each loop iteration;
the last constraint makes sure that such a decrease is strictly positive.

By applying Langrangian relaxation, the two implications can be simplified
away by introducing new, existentially quantified
variables $\alpha \in \Qset_+$ and $\vect{\beta} \in \Qset_+^m$, obtaining:
\begin{align*}
  \vect{\mu}^\transpose\vect{x}+\mu_0 - \alpha \sigma_1(\vect{x}, \vect{x}')
    &\geq 0, \\
  \vect{\mu}^\transpose(\vect{x}-\vect{x}')-\delta
      - \sum_{k=1}^m \beta_k \sigma_k(\vect{x}, \vect{x}')
    &\geq 0, \\
  \delta
    &> 0.
\end{align*}

The limitation that $r=1$ can actually be removed as long as $\alpha$ is
replaced by a vector $\vect{\alpha} \in \Qset^r_+$. The generalization yields
\begin{align*}
\vect{\mu}^\transpose\vect{x}+\mu_0 -
  \sum_{k=1}^r \alpha_k \sigma_k(\vect{x}, \vect{x}') &\geq 0 \\
\vect{\mu}^\transpose(\vect{x}-\vect{x}')-\delta
 - \sum_{k=1}^m \beta_k \sigma_k(\vect{x}, \vect{x}') &\geq 0 \\
\delta &> 0.
\end{align*}
If, and this is the case in the Podelski and Rybalchenko method, the
constraints $\sigma_k(\vect{x}, \vect{x}')$ for $k = 1$, \dots,~$m$ are
affine functions of $\langle \vect{x}, \vect{x}' \rangle$, the sums can be
interpreted as matrix products and the conditions rewritten as follows,
where $\mat{A}$, $\mat{A}'$ and $\vect{b}$ are the same as
in~\eqref{eq:pr-matrix-form}:
\begin{subequations}
\label{eq:lagrange-system}
\begin{align}
\label{eq:lagrange-system-a}
\Bigl(
  \langle\vect{\mu}, \vect{0}, \mu_0 \rangle^\transpose
  - \langle \vect{\alpha}, \vect{0} \rangle^\transpose
      \begin{pmatrix}
        -\mat{A} & -\mat{A}' & \vect{b}
      \end{pmatrix}
\Bigr)
  \langle \vect{x}, \vect{x}', 1 \rangle &\geq 0 \\
\label{eq:lagrange-system-b}
\Bigl(
  \langle\vect{\mu}, -\vect{\mu}, -\delta\rangle^\transpose
  - \vect{\beta}^\transpose
    \begin{pmatrix}
     -\mat{A} & -\mat{A}' & \vect{b}
    \end{pmatrix}
\Bigr)
  \langle\vect{x}, \vect{x}', 1\rangle &\geq 0 \\
\label{eq:lagrange-system-c}
\delta &> 0
\end{align}
\end{subequations}
Note that the inequalities~\eqref{eq:lagrange-system-a}--\eqref{eq:lagrange-system-c}
must hold for \emph{every} possible value of
$\vect{x}$ and $\vect{x}'$ in the whole space $\Qset^n$. Therefore, by
a suitable choice of $\vect{x}$ and $\vect{x}'$, each element of
the coefficient vectors in~\eqref{eq:lagrange-system-a}
and~\eqref{eq:lagrange-system-b} can be shown to be necessarily zero.
We define
$\vect{\lambda}_1 = \langle \vect{\alpha}, \vect{0} \rangle$
\footnote{We explicitly require that the extra coefficients added
to $\vect{\alpha}$ be zero for consistency with the derivation.
However, even though Podelski and Rybalchenko admit any
nonnegative rational numbers to appear in those positions of
$\vect{\lambda}_1$, there is no loss of generality: the synthesized
ranking functions~\eqref{eq:pr-ranking-function} do not depend on
these coefficients.}
and $\vect{\lambda}_2 = \vect{\beta}$, obtaining:
\begin{align*}
\vect{\mu} &= -\vect{\lambda}_1^\transpose \mat{A},
  &\vect{\mu} &= -\vect{\lambda}_2^\transpose \mat{A}, \\
\vect{0} &= -\vect{\lambda}_1^\transpose \mat{A}',
  &-\vect{\mu} &= -\vect{\lambda}_2^\transpose \mat{A'}, \\
\mu_0 &= \vect{\lambda}_1^\transpose \vect{b},
  &-\delta &= \vect{\lambda}_2^\transpose \vect{b},
    &\delta &> 0.
\end{align*}
These relations can finally be rearranged to yield:
\[
\begin{aligned}
  \vect{\lambda}_1^\transpose \mat{A}'
    &= \vect{0}, \\
  (\vect{\lambda}_1^\transpose - \vect{\lambda}_2^\transpose) \mat{A}
    &= \vect{0}, \\
  \vect{\lambda}_2^\transpose (\mat{A+A}')
    &= \vect{0}, \\
  \vect{\lambda}_2^\transpose \mat{b}
    &< 0,
\end{aligned}
\qquad
\begin{aligned}
\vect{\mu} &= \vect{\lambda}_2^\transpose \mat{A'}, \\
\mu_0 &= \vect{\lambda}_1^\transpose \vect{b}, \\
\delta &= -\vect{\lambda}_2^\transpose \vect{b},
\end{aligned}
\]
where the conditions \eqref{eq:pr-system-a}--\eqref{eq:pr-system-d}
appear on the left hand side and the conditions on the coefficients of the
synthesized ranking functions appear on the right hand side,
expressed in terms of $\vect{\lambda}_1$ and $\vect{\lambda}_2$.

}{
}

\subsection{An Alternative Implementation Approach}

As long as the distinction between
invariants~\eqref{eq:pr-invariant-after-B}
and~\eqref{eq:pr-invariant-after-C} is retained, the method of
Podelski and Rybalchenko can be implemented following an alternative
approach.
The linear invariants~\eqref{eq:pr-matrix-form} are more precisely described by
\begin{equation}
\label{eq:ipr-matrix-form}
\left(
  \begin{BMAT}(c){c:c}{c:c}
    \mat{A}_B
    & \mat{0}
  \cr
    \mat{A}_C
    & \mat{A}'_C
  \end{BMAT}
\right)
\begin{pmatrix}
 \vect{x} \\ \vect{x}'
\end{pmatrix}
\leq
\begin{pmatrix}
 \vect{b}_B \\ \vect{b}_C
\end{pmatrix}
\end{equation}
where
$\mat{A}_B \in \Qset^{r \times n}$,
$\mat{A}_C \in \Qset^{s \times n}$,
$\mat{A}'_C \in \Qset^{s \times n}$,
$\vect{b}_B \in \Qset^r$,
$\vect{b}_C \in \Qset^s$.
\ifthenelse{\boolean{LONGVERSION}}{
As shown in Section~\ref{sec:pr-from-lagrange},
}{
As shown in the technical report version of this paper
\cite[Section~5.3]{BagnaraMPZ12TR},
}
the existence of a linear ranking function
for the system~\eqref{eq:ipr-matrix-form} is equivalent to
the existence of three vectors
$\vect{v}_1 \in \Qset_+^r$, $\vect{v}_2 \in \Qset_+^r$,
$\vect{v}_3 \in \Qset_+^s$ such that
\begin{subequations}
\label{eq:ipr-system}
\begin{align}
\label{eq:ipr-system-a}
  (\vect{v}_1-\vect{v}_2)^\transpose \mat{A}_B - \vect{v}_3^\transpose \mat{A}_C
    &= \vect{0}, \\
\label{eq:ipr-system-b}
  \vect{v}_2^\transpose \mat{A}_B + \vect{v}_3^\transpose (\mat{A}_C+\mat{A}_C')
    &= \vect{0}, \\
\label{eq:ipr-system-c}
  \vect{v}_2 \vect{b}_B + \vect{v}_3 \vect{b}_C
    &< 0.
\end{align}
\end{subequations}
As already noted, the two vectors of the original Podelski and Rybalchenko
method can be reconstructed as
$\vect{\lambda}_1 = \langle \vect{v}_1, \vect{0} \rangle$ and
$\vect{\lambda}_2 = \langle \vect{v}_2, \vect{v}_3 \rangle$.

Note that the same approach is still valid when starting from the
single matrix form~\eqref{eq:pr-matrix-form} in full generality,
i.e., when we can't assume that the distinction between
invariants~\eqref{eq:pr-invariant-after-B} and~\eqref{eq:pr-invariant-after-C}
has been retained or that invariants are listed in the order we used to
build the matrix form~\eqref{eq:pr-matrix-form}: it is enough to apply a
straightforward permutation to~\eqref{eq:pr-matrix-form} to rearrange it in the
form~\eqref{eq:ipr-matrix-form}. In that case, due to the permutation
involved, we would solve a different linear programming problem; however,
we still obtain the same space of linear ranking functions we would have
obtained by applying the original method starting from the matrix
form~\eqref{eq:pr-matrix-form}, as we prove using the following lemma.

\begin{lemma}
Let $S$ be the space of linear ranking functions obtained by applying
the method of Podelski and Rybalchenko to
\(
\begin{pmatrix}
  \mat{A} & \mat{A}'
\end{pmatrix}
\langle \vect{x}, \vect{x}' \rangle
\leq
\vect{b}
\),
i.e.,
\[
 S
    \defeq
      \bigl\{\,
        \langle
          \vect{\lambda}_2^\transpose \mat{A}',
          \vect{\lambda}_1^\transpose\vect{b}
        \rangle
        \in \Qset^{n+1}
      \bigm|
        \text{%
          $\langle \vect{\lambda}_1, \vect{\lambda}_2 \rangle$
          is a solution of
          \eqref{eq:pr-system}%
         }
      \,\bigr\},
\]
and let $\mat{P} \in \Qset^{(r+s) \times (r+s)}$
be a permutation matrix.\footnote{We recall that a $k$-dimensional
\emph{permutation matrix} is a square matrix obtained by a
permutation of the rows or columns of the $k$-dimensional identity
matrix.} Then the application of the method of Podelski and
Rybalchenko to
\(
\mat{P}
\begin{pmatrix}
  \mat{A} & \mat{A}'
\end{pmatrix}
\langle \vect{x}, \vect{x}' \rangle
\leq
\mat{P}
\vect{b}
\)
yields the same space of linear ranking functions $S$.
\end{lemma}

\begin{pf}
  The system~\eqref{eq:pr-system} corresponding to
\(
\mat{P}
\begin{pmatrix}
  \mat{A} & \mat{A}'
\end{pmatrix}
\langle \vect{x}, \vect{x}' \rangle
\leq
\mat{P}
\vect{b}
\)
becomes
\begin{subequations}
\label{eq:pr-system-permuted}
\begin{align}
\label{eq:pr-system-permuted-a}
  \vect{\eta}_1^\transpose \mat{P} \mat{A}'
    &= \vect{0}, \\
\label{eq:pr-system-permuted-b}
  (\vect{\eta}_1^\transpose - \vect{\eta}_2^\transpose) \mat{P} \mat{A}
    &= \vect{0}, \\
\label{eq:pr-system-permuted-c}
  \vect{\eta}_2^\transpose \mat{P} (\mat{A}+\mat{A}')
    &= \vect{0}, \\
\label{eq:pr-system-permuted-d}
  \vect{\eta}_2^\transpose \mat{P} \vect{b}
    &< 0,
\end{align}
\end{subequations}
to be solved for the two $(r+s)$-dimensional non-negative rational
vectors $\vect{\eta}_1$ and $\vect{\eta}_2$.

Now, $\langle \vect{\lambda}_1, \vect{\lambda}_2 \rangle$
is a solution of~\eqref{eq:pr-system} if and only if
$\langle \vect{\lambda}_1, \vect{\lambda}_2 \rangle \mat{P}^{-1}$
is a solution of~\eqref{eq:pr-system-permuted}:
on one side, if $\langle \vect{\lambda}_1, \vect{\lambda}_2 \rangle$
is a solution of~\eqref{eq:pr-system} then
$\langle \vect{\eta}_1, \vect{\eta}_2 \rangle$ defined as
\(
\langle \vect{\eta}_1, \vect{\eta}_2 \rangle
  \defeq
\langle \vect{\lambda}_1, \vect{\lambda}_2 \rangle \mat{P}^{-1}
\)
is a solution of~\eqref{eq:pr-system-permuted}; on the other side,
if $\langle \vect{\eta}_1, \vect{\eta}_2 \rangle$ is a solution
of~\eqref{eq:pr-system-permuted} then
$\langle \vect{\lambda}_1, \vect{\lambda}_2 \rangle$ defined as
\(
\langle \vect{\lambda}_1, \vect{\lambda}_2 \rangle
  \defeq
\langle \vect{\eta}_1, \vect{\eta}_2 \rangle \mat{P}
\)
is a solution of~\eqref{eq:pr-system} and the desired property
can be verified by right-multiplying by $\mat{P}^{-1}$ both solutions.

The space of linear ranking functions for the permuted system is
\begin{align*}
  S_{\mat{P}}
    &=
      \bigl\{\,
        \langle
          \vect{\eta}_2^\transpose \mat{P} \mat{A}',
          \vect{\eta}_1^\transpose \mat{P} \vect{b}
        \rangle
        \in \Qset^{n+1}
      \bigm|
        \text{%
          $\langle \vect{\eta}_1, \vect{\eta}_2 \rangle$
          is a solution of
          \eqref{eq:pr-system-permuted}%
         }
      \,\bigr\} \\
    &=
      \bigl\{\,
        \langle
          \vect{\lambda}_2^\transpose \mat{P}^{-1} \mat{P} \mat{A}',
          \vect{\lambda}_1^\transpose \mat{P}^{-1} \mat{P} \vect{b}
        \rangle
        \in \Qset^{n+1}
      \bigm|
        \text{%
          $\langle \vect{\lambda}_1, \vect{\lambda}_2 \rangle$
          is a solution of
          \eqref{eq:pr-system}%
         }
      \,\bigr\} \\
    &= S,
\end{align*}
and thus it is unaltered with respect to the space of linear ranking
functions $S$ corresponding to the non-permuted system.
\end{pf}

Since the system~\eqref{eq:ipr-matrix-form} is obtained by applying
a suitable permutation to~\eqref{eq:pr-matrix-form}, a
straightforward application of this lemma proves that the space of
linear ranking functions obtained is the same in both cases.

Moreover, as
\(
  \vect{\lambda}_2^\transpose \mat{A}'
    =
  \vect{v}_3^\transpose \mat{A}_C'
\)
and $\vect{\lambda}_1 \vect{b} = \vect{v}_1 \vect{b}_B$, we can express
the space of linear ranking functions as
\[
 S
    \defeq
      \bigl\{\,
        \langle
          \vect{v}_3^\transpose \mat{A}_C',
          \vect{v}_1 \vect{b}_B
        \rangle
        \in \Qset^{n+1}
      \bigm|
        \text{%
          $\langle \vect{v}_1, \vect{v}_2, \vect{v}_3 \rangle$
          is a solution of
          \eqref{eq:ipr-system}%
         }
      \,\bigr\}.
\]

\section{Comparison of the Two Methods}
\label{sec:comparison}

In this section we compare the method by Mesnard and Serebrenik
with the method by Podelski and Rybalchenko: we first
prove that they have the same ``inferential power'', then
we compare their worst-case complexities, then we experimentally
evaluate them on a representative set of benchmarks.

\subsection{Equivalence of the Two Methods}

We will now show that the method proposed in \cite{PodelskiR04}
is equivalent to the one given in \cite{SerebrenikM05} on the class
of simple linear loops, i.e., that if one of the two methods
can prove termination of a given simple linear loop, then the other
one can do the same. This is an expected result since both methods
claim to be complete on the class of programs considered.

It is worth noting that a completeness result was already stated
in \cite[Theorem~5.1]{Mesnard96} for the case of \emph{single predicate}
$\mathrm{CLP}(\Qset_+)$ procedures, which can be seen to be a close
variant of the binary, directly recursive $\mathrm{CLP}(\Qset_+)$
programs considered in Theorem~\ref{th:svg-correctness-completeness}
and Corollary~\ref{co:plrf-in-P}.
Probably due to the programming paradigm mismatch, Podelski and Rybalchenko
\cite{PodelskiR04} fail to recognize the actual strength and generality
of the mentioned result, thereby claiming originality for their
completeness result.

\begin{theorem}
\label{th:pr-equals-ms}
Let $C$ be the binary $\mathrm{CLP}(\Qset)$ clause
$p(\bar{x}) \pif c[\bar{x}, \bar{x}'], p(\bar{x}')$,
where $p$ is an $n$-ary predicate and
$c[\bar{x}, \bar{x}']$ is a linear satisfiable constraint.
Let $\mathrm{pr}(C)$ and $\extend{\mathrm{ms}}(C)$ be the spaces of linear
ranking functions for $C$ obtained through the method of Podelski
and Rybalchenko and through the method of Mesnard and Serebrenik,
respectively, that is,
\begin{align*}
  \mathrm{pr}(C)
    &\defeq
      \bigl\{\,
        \langle
          \vect{\lambda}_2^\transpose \mat{A}',
          \vect{\lambda}_1^\transpose\vect{b}
        \rangle
        \in \Qset^{n+1}
      \bigm|
        \text{%
          $\langle \vect{\lambda}_1, \vect{\lambda}_2 \rangle$
          is a solution of
          \eqref{eq:pr-system}%
         }
      \,\bigr\}, \\
  \extend{\mathrm{ms}}(C)
    &\defeq\sset{%
        k\vect{\tilde{\mu}} \in \Qset^{n+1}
      }{%
        \text{%
          $\langle \vect{y}, \vect{\mu} \rangle$ and
          $\langle \vect{z}, \vect{\tilde{\mu}} \rangle$
          are solutions of
          \eqref{eq:ms-final-transformed-dual-optimization-problems},%
        } \\
        \text{%
          $k \in \Qset_+\setdiff\{0\}$%
        }%
      }.
\end{align*}
where $c[\bar{x}, \bar{x}']$ is equivalent to
\(
\begin{pmatrix}
  \mat{A} & \mat{A}'
\end{pmatrix}
\langle \vect{x}, \vect{x}' \rangle
\leq
\vect{b}
\)
or to $\mat{A}_c \langle \vect{x}, \vect{x}' \rangle \geq \vect{b}_c$,
respectively.
Then $\mathrm{pr}(C) = \extend{\mathrm{ms}}(C)$.
\end{theorem}

\begin{pf}
We will, as customary, prove the two inclusions
$\mathrm{pr}(C) \subseteq \extend{\mathrm{ms}}(C)$ and
$\mathrm{pr}(C) \supseteq \extend{\mathrm{ms}}(C)$.

\paragraph*{$\mathrm{pr}(C) \subseteq \extend{\mathrm{ms}}(C)$}
Suppose that there exist two non-negative rational vectors $\vect{\lambda}_1$
and $\vect{\lambda}_2$ satisfying~\eqref{eq:pr-system}, i.e.,
\(
\vect{\lambda}_1^\transpose \mat{A}'
  = (\vect{\lambda}_1^\transpose - \vect{\lambda}_2^\transpose) \mat{A}
  = \vect{\lambda}_2^\transpose(\mat{A}+\mat{A}') = \vect{0}
\)
and $\vect{\lambda}_2^\transpose \vect{b} < 0$.
We need to show that
\(
\langle
  \vect{\lambda}_2^\transpose \mat{A}', \vect{\lambda}_1^\transpose \vect{b}
\rangle
\in
\extend{\mathrm{ms}}(C)
\), which is equivalent to proving that there exists a positive coefficient
(that we can denote with $\frac{1}{k}$ without loss of generality)
$\frac{1}{k} \in \Qset_+\setdiff\{0\}$ such that
\(
\langle
  \frac{1}{k}\vect{\lambda}_2^\transpose \mat{A}',
  \frac{1}{k} \vect{\lambda}_1^\transpose \vect{b}
\rangle
\in
\mathrm{ms}(C),
\) or, by Theorem~\ref{th:ms-correctness-completeness}, that
\(
\langle
  \frac{1}{k}\vect{\lambda}_2^\transpose \mat{A}',
  \frac{1}{k} \vect{\lambda}_1^\transpose \vect{b}
\rangle
\in
\mathrm{lrf}(C),
\) which is in turn equivalent, by definition, to
\(
 \vect{\lambda}_2^\transpose \mat{A}' \vect{x}
 -\vect{\lambda}_2^\transpose \mat{A}' \vect{x}'
\geq k
\) and
\(
  \vect{\lambda}_1^\transpose \vect{b}
  + \vect{\lambda}_2^\transpose \mat{A}' \vect{x}
  \geq 0
\).
We have
\begin{align*}
  \begin{pmatrix}\mat{A} & \mat{A}'\end{pmatrix}
  \begin{pmatrix}\vect{x} \\ \vect{x}'\end{pmatrix}
    \leq \vect{b} \implies &
    \mat{A}\vect{x} + \mat{A}'\vect{x}' \leq \vect{b} & \\
    \implies &
      -\mat{A}\vect{x} \geq \mat{A}'\vect{x}' - \vect{b} & \\
    \implies &
      -\vect{\lambda}_2^\transpose \mat{A}\vect{x}
      \geq \vect{\lambda}_2^\transpose \mat{A}'\vect{x}'
      - \vect{\lambda}_2^\transpose \vect{b} &
      \text{[by $\vect{\lambda}_2 \geq \vect{0}$]} \\
    \implies &
      \vect{\lambda}_2^\transpose \mat{A}'\vect{x}
      \geq \vect{\lambda}_2^\transpose \mat{A}'\vect{x}'
      - \vect{\lambda}_2^\transpose \vect{b} &
      \text{[by~\eqref{eq:pr-system-c}]}
\end{align*}
and the former property is satisfied if we choose
$k = - \vect{\lambda}_2^\transpose \vect{b}$, which is nonnegative by
relation~\eqref{eq:pr-system-d}.
For the latter property, we have
\begin{align*}
  \mat{A}\vect{x} + \mat{A}'\vect{x}' \leq \vect{b}
  \implies &
    \vect{\lambda}_1^\transpose \mat{A}\vect{x} + \vect{\lambda}_1^\transpose
    \mat{A}'\vect{x}'
    \leq \vect{\lambda}_1^\transpose \vect{b} &
    \text{as } \vect{\lambda}_1^\transpose \text{ is non-negative}\\
  \implies &
    \vect{\lambda}_1^\transpose \mat{A}\vect{x}
      \leq \vect{\lambda}_1^\transpose \vect{b} &
    \text{[by~\eqref{eq:pr-system-a}]} \\
  \implies &
    \vect{\lambda}_2^\transpose \mat{A} \vect{x}
    \leq \vect{\lambda}_1^\transpose \vect{b} &
    \text{[by~\eqref{eq:pr-system-b}]} \\
  \implies &
    -\vect{\lambda}_2^\transpose \mat{A}'\vect{x}
    \leq \vect{\lambda}_1^\transpose \vect{b} &
    \text{[by~\eqref{eq:pr-system-c}]}
\end{align*}
and both properties are thus proved.

\paragraph*{$\mathrm{pr}(C) \supseteq \extend{\mathrm{ms}}(C)$}
In order to prove the inverse containment, we will need to recall
the affine form of Farkas' Lemma (see \cite{Schrijver86}).

\begin{lemma}[Affine form of Farkas' lemma]
Let $P$ be a nonempty polyhedron defined by the inequalities
$\mat{C}\vect{x}+\vect{d} \geq \vect{0}$.
Then an affine function $f(\vect{x})$ is non-negative
everywhere in $P$ if and only if it is a positive affine
combination of the columns of $\mat{C}\vect{x}+\vect{d}$:
$f(\vect{x}) = \lambda_0 + \vect{\lambda}^\transpose(\mat{C}\vect{x}+\vect{d})$
with $\lambda_0 \geq 0$, $\vect{\lambda} \geq \vect{0}$.
\end{lemma}

Let $\vect{\tilde{\mu}} \in \extend{\mathrm{ms}}(C)$. Then there exists
$h \in \Qset_+ \setdiff \{0\}$ such that $h \vect{\tilde{\mu}} \in \mathrm{lrf}(C)$
describes a linear ranking function $f$ for $C$.

The inequalities
\(
\begin{pmatrix}
  \mat{A} & \mat{A}'
\end{pmatrix}
\langle \vect{x}, \vect{x}' \rangle
\leq
\vect{b}
\)
define a polyhedron; according to the affine form of Farkas' lemma, a
function is non-negative on this polyhedron, i.e., throughout the loop,
if and only if it is a positive affine combination of the column
vectors
\(
\begin{pmatrix}
  \mat{A} & \mat{A}'
\end{pmatrix}
\langle \vect{x}, \vect{x}' \rangle
\leq
\vect{b}
\). In particular this
holds for the ranking function $f$ and its two properties:
$f(\vect{x}) \geq 0$ and
$f(\vect{x}) - f(\vect{x}') \geq 1$.

Hence there exist two non-negative rational vectors
$\vect{\lambda}_1$ and $\vect{\lambda}_2$ and two
non-negative numbers $\lambda_{0,1}$ and $\lambda_{0,2}$ such that
\begin{align*}
f(\vect{x}) &= \lambda_{0,1} + \vect{\lambda}_1^\transpose \left(
  - \begin{pmatrix}\mat{A}&\mat{A}'\end{pmatrix}
  \langle \vect{x}, \vect{x}' \rangle
  + \vect{b} \right)\\
\intertext{and}
f(\vect{x}) - f(\vect{x}') - 1 &= \lambda_{0,2} +
  \vect{\lambda}_2^\transpose \left(
  - \begin{pmatrix}\mat{A}&\mat{A}'\end{pmatrix}
  \langle \vect{x}, \vect{x}' \rangle
  + \vect{b} \right).
\end{align*}

Replacing $f(\vect{x})$ by $h\vect{\mu} \vect{x} + h\mu_0$, we
get two equalities ---one for the part containing variables and one for
the remaining part--- for each expression.
After simplification we obtain the following equalities:
\begin{subequations}
\begin{align}
\label{eq:ms-to-pr-a}
  - \vect{\lambda}_1^\transpose \begin{pmatrix}\mat{A}&\mat{A}'\end{pmatrix}
  \langle \vect{x}, \vect{x}' \rangle
  &=  h \vect{\mu} \vect{x} \\
\label{eq:ms-to-pr-b}
  - \vect{\lambda}_2^\transpose \begin{pmatrix}\mat{A}&\mat{A}'\end{pmatrix}
  \langle \vect{x}, \vect{x}' \rangle
  &=  h \vect{\mu} \vect{x} - h \vect{\mu} \vect{x}' \\
\label{eq:ms-to-pr-c}
  - \vect{\lambda}_2^\transpose \vect{b} &= 1 + \lambda_{0,2}
\end{align}
\end{subequations}

From~\eqref{eq:ms-to-pr-a} and~\eqref{eq:ms-to-pr-b} we obtain
$\vect{\lambda}_1^\transpose \mat{A}
  = - h \vect{\mu}^\transpose$,
$\vect{\lambda}_1^\transpose \mat{A}' = \vect{0}$,
$\vect{\lambda}_2^\transpose \mat{A}
  =  - h \vect{\mu}^\transpose$
and $\vect{\lambda}_2^\transpose \mat{A}'
  = h \vect{\mu}^\transpose$.
We can rewrite it as
$\vect{0} = \vect{\lambda}_1^\transpose \mat{A}'
  = (\vect{\lambda}_1^\transpose-\vect{\lambda}_2^\transpose)\mat{A}
  = \vect{\lambda}_2^\transpose(\mat{A}+\mat{A}')$.
From~\eqref{eq:ms-to-pr-c} we deduce
$\vect{\lambda}_2^\transpose \vect{b} < 0$.

The four conditions~\eqref{eq:pr-system} to prove termination
by \cite{PodelskiR04} are thus satisfied.
\end{pf}

The combination of Theorems~\ref{th:ms-correctness-completeness}
and~\ref{th:pr-equals-ms} gives:
\begin{theorem}
\label{th:pr-equals-ms-equals-lrf}
Let $C$ be the binary $\mathrm{CLP}(\Qset)$ clause
$p(\bar{x}) \pif c[\bar{x}, \bar{x}'], p(\bar{x}')$,
where $p$ is an $n$-ary predicate and
$c[\bar{x}, \bar{x}']$ is a linear satisfiable constraint.
Let $\extend{\mathrm{lrf}}(C)$ be the set of (positive multiples of)
linear ranking functions for $C$,
$\extend{\mathrm{ms}}(C)$ be the set of (positive multiples of) solutions
of the Mesnard and Serebrenik
system~\eqref{eq:ms-final-transformed-dual-optimization-problems}
projected onto $\mu$
and $\mathrm{pr}(C)$ be the set of the ranking function coefficients
obtained through the method of Podelski and Rybalchenko, that is,
\begin{align*}
  \extend{\mathrm{lrf}}(C)
    &\defeq
      \sset{%
        k\vect{\tilde{\mu}} \in \Qset^{n+1}
      }{%
        \forall \bar{x}, \bar{x}' \in \Qset^n
          \itc
            c[\bar{x}, \bar{x}']
              \implies \\
              \qquad
                \sum_{i=1}^n \mu_i x_i - \sum_{i=1}^n \mu_i x'_i \geq 1 \\
              \qquad
                {} \land \mu_0 + \sum_{i=1}^n \mu_i x_i \geq 0, \\
        k \in \Qset_+ \setdiff \{0\}%
      }, \\
  \extend{\mathrm{ms}}(C)
    &\defeq
      \sset{%
        k\vect{\tilde{\mu}} \in \Qset^{n+1}
      }{%
        \text{%
          $\langle \vect{y}, \vect{\mu} \rangle$ and
          $\langle \vect{z}, \vect{\tilde{\mu}} \rangle$
          are solutions of
          \eqref{eq:ms-final-transformed-dual-optimization-problems}},\\
          k \in \Qset_+ \setdiff \{0\}%
        }, \\
  \mathrm{pr}(C)
    &\defeq
      \bigl\{\,
        \langle
          \vect{\lambda}_2^\transpose \mat{A}',
          \vect{\lambda}_1^\transpose\vect{b}
        \rangle
        \in \Qset^{n+1}
      \bigm|
        \text{%
          $\langle \vect{\lambda}_1, \vect{\lambda}_2 \rangle$
          is a solution of
          \eqref{eq:pr-system}%
         }
      \,\bigr\},
\end{align*}
where $c[\bar{x}, \bar{x}']$ is equivalent to
\(
\begin{pmatrix}
  \mat{A} & \mat{A}'
\end{pmatrix}
\langle \vect{x}, \vect{x}' \rangle
\leq
\vect{b}
\)
or to $\mat{A}_c \langle \vect{x}, \vect{x}' \rangle \geq \vect{b}_c$,
respectively.
Then $\extend{\mathrm{lrf}}(C)$ = $\extend{\mathrm{ms}}(C)$ = $\mathrm{pr}(C)$.
\end{theorem}

\subsection{Worst-Case Complexity Using the Simplex Algorithm}

The computationally most expensive component in both methods is
the resolution of a linear optimization problem that can always be
expressed in the standard form
\[
\begin{aligned}
  \text{minimize }\quad   & \vect{c}^\transpose \vect{x} \\
  \text{subject to }\quad & \mat{A} \vect{x} = \vect{b} \\
                          & \vect{x} \geq \vect{0}
\end{aligned}
\]
by applying well known transformations: inequalities and
\emph{unconstrained} (i.e., not subject to lower or upper bounds)
variables can be replaced and the resulting equivalent problem in
standard form has one more variable for each inequality or unconstrained
variable appearing in the original problem.

The most common way to solve this linear optimization problems involves
using the simplex algorithm~\cite{PapadimitriouS82}, an iterative
algorithm that requires $\binom{e+u}{e}$ pivoting steps in the worst-case
scenario, where $e$ and $u$ denote the number of equalities in $\mat{A}$
and unknowns in $\vect{x}$ respectively.

For a simple linear loop of $m$ inequalities over $n$ variables,
Podelski and Rybalchenko require to solve a linear problem in standard
form having $3n$ equalities over $2m$ variables (the opposite of the
expression appearing in~\eqref{eq:pr-system-d} can be used as the
quantity to be minimized); this gives a worst-case complexity of
$\binom{3n+2m}{3n}$ pivoting steps, corresponding, by Stirling's formula,
to an exponential complexity of exponent $3n+2m$
\ifthenelse{\boolean{LONGVERSION}}{
approximately.\footnote{%
When $a + b \to \infty$, by Stirling's formula we
have $\binom{a + b} a \leq C 2^{a + b} (a + b)^{-1/2}$, where $C$ is an
absolute constant.  This inequality is sharp. Notice however that if
$a$, say, is known to be much smaller than $b$, a much stronger
inequality can be given, namely $\binom{a + b} a \leq (a + b)^a / a!$.}
}{
approximately.
}

If the alternative formalization of the Podelski and Rybalchenko method
is adopted for the same loop, then we will have the same $m$ constraints
as above for the `$\bigstar$' invariant, while the `$\blacktriangledown$'
invariant will be described by other $\ell$ constraints.
If redundant constraints are removed, we will have $\ell \leq m$.
Hence, the alternative approach will result in a linear programming
problem having $2n$ equalities over $m + 2\ell$ variables.
Hence, the worst-case number of pivoting steps will be an exponential
of exponent approximately $2n + m + 2\ell$.

For the same simple linear loop, Serebrenik and Mesnard require the
resolution of two linear problems, that can be rewritten to contain
$2n$ equalities over $m+n$ variables (with $n$ unconstrained variables)
and $2n+1$ equalities over $(m+2)+(n+1)$ variables (with $n+1$ unconstrained
variables), respectively. They can then be merged to generate a single
linear problem of $4n+1$ equalities over $m+(m+2)+(n+1)$ variables, $n+1$
of which unconstrained, and an extra inequality replacing one of the two
objective functions. In the end, we get a linear problem in standard form
with $4n+2$ equalities over $2m+2n+5$ variables. This means a worst-case
complexity of $\binom{6n + 2m + 7}{4n+2}$ pivoting steps and an exponential
complexity of exponent $6n + 2m$ approximately.

So the method proposed by Podelski and Rybalchenko has, in general,
a lower worst-case complexity than the one proposed by Mesnard
and Serebrenik, if the single linear problem approach is chosen.
The comparison of the two alternative implementation approaches for
the Podelski and Rybalchenko method depends on the relations between
quantities $n$, $m$ and $\ell$.
On the one hand, if $\ell$ is significantly smaller than $m$,
then the alternative approach could result in an efficiency improvement.
On the other hand, if the number of constraints is much higher than
the number of variables, then the original implementation approach
should be preferred.
Note that the need for two loop invariants instead of a single one
should not be seen as a big practical problem: in fact, most analysis
frameworks will provide the `$\blacktriangledown$' invariant
as the original input to the termination analysis tool,
which will then use it to compute the `$\bigstar$' invariant
(via the abstract execution of a single iteration of the loop);
that is, the computational cost for the `$\blacktriangledown$' invariant
is implicitly paid anyway.

It is well known, though, that the worst-case scenario for the
simplex algorithm is extremely uncommon in practice.
An average complexity analysis and, more recently, a smoothed complexity
analysis~\cite{SpielmanT04} have been carried out on the simplex
algorithm and showed why it usually takes polynomial time.
Besides the theoretical studies, several experimental evaluations
of implementations of the simplex algorithm reported that
the average number of pivoting steps seems to grow linearly with
the sum $e + u$ of the number of equalities and unknowns of the problem.
Therefore, for a more informative and meaningful comparison,
the next section presents an experimental evaluation of the methods
on a representative set of while loops.

\section{Implementation and Experimental Evaluation}
\label{sec:implementation-and-experimental-evaluation}

The \emph{Parma Polyhedra Library} (PPL) is a free software,
professional library for the handling of numeric approximations
targeted at static analysis and computer-aided verification of
hardware and software systems \cite{BagnaraHZ08SCP,BagnaraHZ09TCS}.
The PPL, which features several unique innovations
\cite{BagnaraHRZ05SCP,BagnaraHZ05FAC,BagnaraHZ06STTT,BagnaraHZ09FMSD,BagnaraHZ10CGTA},
is employed by numerous projects in this field, most notably by
GCC, the \emph{GNU Compiler Collection}, probably the most widely used
suite of compilers.\footnote{%
See \url{"http://bugseng.com/products/ppl} for more information.}

As an integral part of the overall project to which the present paper
belongs ---whose aim is to make the technology of the automatic
synthesis of linear ranking functions thoroughly explained and
generally available---, we have extended the PPL with all the methods
discussed in the present paper.  Previously, only a rather limited
demo version of \emph{RankFinder} was available, only in x86/Linux binary
format, implementing the method by Podelski and Rybalchenko.\footnote{%
See \url{http://www7.in.tum.de/~rybal/rankfinder/},
last checked on August 18th, 2011.}
In contrast, the PPL implementation is completely general and available,
both in source and binary formats, with high-level interfaces to
C, \Cplusplus{}, Java, OCaml and six different Prolog systems.

For each of the methods ---Mesnard and Serebrenik (MS) or Podelski
and Rybalchenko (PR)---, for each of the two possibilities
to encode the input ---either the single $\bigstar$ invariant
of~\eqref{prog:ms-loop} in Section~\ref{sec:MS-applied-to-imperative},
or the two $\blacktriangledown$ and $\bigstar$ invariants
of~\eqref{prog:pr-loop} in Section~\ref{sec:podelski-and-rybalchenko}---,
for each numerical abstractions supported by the PPL ---including
(not necessarily closed) convex polyhedra, bounded-difference shapes
and octagonal shapes---, the PPL provides three distinct functionalities
to investigate termination of the loop being analyzed:
\begin{enumerate}
\item
a Boolean termination test;
\item
a Boolean termination test that, in addition, returns the coefficients
of one (not further specified) affine ranking function;
\item
a function returning a convex polyhedron that encodes the space of
all affine ranking functions.
\end{enumerate}
In addition, using the MS method and for each input method, the
PPL provides
\begin{enumerate}
\item[4.]
a function returning two convex polyhedra that encode the space of
all decreasing functions (also known as \emph{quasi-ranking functions})
and all bounded functions, respectively,
for use in conditional termination analysis.
\end{enumerate}

We have evaluated the performance of the new algorithms implemented
in the PPL using the termination analyzer built into \emph{Julia},
a state-of-the-art analyzer for Java bytecode \cite{SpotoMP10}.
We have thus taken several Java programs in the Julia test suite
and, using Julia, we have extracted the constraint systems that
characterize the loops in the program that Julia cannot quickly
resolve with syntax-based heuristics.  This extraction phase
allowed us to measure the performance of the methods described
in the present paper, factoring out the time spent by Julia
in all the analyses (nullness, sharing, path-length,
unfolding, \dots) that allow to obtain such
constraint systems.

We first tested the performance (and correctness) of the new
PPL implementation with the implementation of the MS method,
based on CLP(Q), previously used by Julia and with the
implementation of PR, still based on CLP(Q), provided by
the demo version of RankFinder.
The reason we did this comparison is that, while we know that the
infinite precision implementation of the simplex algorithm available
in the PPL performs better than its direct competitors
\cite[Section~4, Table~3]{BagnaraHZ08SCP},\footnote{%
I.e., \emph{Cassowary}
(\url{http://www.cs.washington.edu/research/constraints/cassowary/}) and
\emph{Wallaroo} (\url{http://sourceforge.net/projects/wallaroo/}).
While GLPK, the \emph{GNU Linear Programming Toolkit}
(\url{http://www.gnu.org/software/glpk/}) includes
a solver that is termed ``exact,'' it still depends critically
on floating point computations; moreover, it has not yet been made available
in the public interface.}
we know there is much room for improvement: it could have been
the case that the constraint solver employed in modern CLP systems
made our implementation useless.  The result was quite satisfactory:
the PPL implementation is one to two orders of magnitude faster over
the considered benchmark suite.

The benchmark programs are:
\verb+CaffeineMark+, from Pendragon Software Corporation, measures
the speed of Java;
\verb+JLex+ is a lexical analyzer generator developed by Elliot Berk
and C.~Scott Ananian;
\verb+JavaCC+ is a parser generator from Sun Microsystems;
\verb+Java_CUP+ is a parser generator developed by Scott Hudson,
Frank Flannery and C. Scott Ananian;
\verb+Jess+ is a rule engine written by Ernest Friedman-Hill;
\verb+Kitten+ is a didactic compiler for a simple imperative
object-oriented language written by Fausto Spoto;\
\verb+NQueens+ is a solver of the n-queens problem which includes
a library  for binary decision diagrams;
\verb+Raytracer+ is a ray-tracing program;
\verb+Termination+ is a JAR file containing all the programs
of \cite[Figure~16]{SpotoMP10}.
In Table~\ref{tab:benchmarks} we report, for each benchmark,
the number of loops for which termination was investigated,
the interval, mean and standard deviations
---with two significant figures--- of the quantities
$n$ (number of variables) and $m$ (number of constraints)
that characterize those loops.
\begin{table}
\caption{Benchmarks used in the experimental evaluation}
\label{tab:benchmarks}
\centering
\begin{tabular}{lr,..,..}
benchmark &
  \multicolumn{1}{c}{loops} &
    \multicolumn{1}{c}{$n$} &
    \multicolumn{1}{c}{$\overline{n}$} &
    \multicolumn{1}{c}{$\sigma_n$} &
    \multicolumn{1}{c}{$m$} &
    \multicolumn{1}{c}{$\overline{m}$} &
    \multicolumn{1}{c}{$\sigma_m$} \\
\hhline{--------}
\verb+CaffeineMark+ &  151 & [1, 9]  &  6.0 & 1.3 & [2, 26] & 17. &  3.8 \\
\verb+JLex+         &  467 & [1, 14] &  7.2 & 2.5 & [2, 45] & 17. &  6.7 \\
\verb+JavaCC+       &  136 & [1, 14] &  8.6 & 4.1 & [1, 45] & 22. & 12.  \\
\verb+Java_CUP+     &   29 & [2, 14] &  8.3 & 4.3 & [5, 45] & 23. & 13.  \\
\verb+Jess+         &  151 & [1, 9]  &  6.0 & 1.3 & [2, 26] & 17. &  3.8 \\
\verb+Kitten+       & 1484 & [1, 15] & 11.  & 3.6 & [2, 45] & 29. & 10.  \\
\verb+NQueens+      &  359 & [1, 14] &  6.3 & 3.6 & [2, 45] & 17. & 10.  \\
\verb+Raytracer+    &    8 & [2, 9]  &  4.5 & 2.7 & [5, 26] & 11. &  7.8 \\
\verb+Termination+  &  121 & [1, 9]  &  4.2 & 3.5 & [2, 27] & 12. &  9.9 \\
\hhline{--------}
\end{tabular}
\end{table}

The results of the CPU-time comparison between the MS and PR methods are
reported in Table~\ref{tab:MS-vs-PR-analysis-time}.
\begin{table}
\caption{MS vs PR: CPU time in seconds}
\label{tab:MS-vs-PR-analysis-time}
\centering
\begin{tabular}{l......}
   \multicolumn{1}{c}{}
 & \multicolumn{2}{c}{term.~test}
 & \multicolumn{2}{c}{one r.~f.}
 & \multicolumn{2}{c}{all r.~f.} \\
\multicolumn{7}{c}{} \\[-4mm]
\hhline{-------}
benchmark &
  \multicolumn{1}{c}{MS} &
    \multicolumn{1}{c}{PR} &
      \multicolumn{1}{c}{MS} &
  \multicolumn{1}{c}{PR} &
    \multicolumn{1}{c}{MS} &
      \multicolumn{1}{c}{PR} \\
\hhline{-------}
\verb+CaffeineMark+ & 0.42 & 0.26 & 0.43 & 0.25 & 0.31 & 0.34 \\
\verb+JLex+ & 1.62 & 0.83 & 1.64 & 0.84 & 1.17 & 1.14 \\
\verb+JavaCC+ & 0.86 & 0.43 & 0.87 & 0.45 & 0.67 & 0.65 \\
\verb+Java_CUP+ & 0.35 & 0.14 & 0.35 & 0.14 & 0.29 & 0.22 \\
\verb+Jess+ & 0.42 & 0.26 & 0.43 & 0.26 & 0.29 & 0.34 \\
\verb+Kitten+ & 11.8 & 6.87 & 11.9 & 6.84 & 8.41 & 10.2 \\
\verb+NQueens+ & 1.43 & 0.76 & 1.44 & 0.74 & 0.99 & 1.03 \\
\verb+Raytracer+ & 0.04 & 0.03 & 0.04 & 0.03 & 0.03 & 0.03 \\
\verb+Termination+ & 0.25 & 0.15 & 0.25 & 0.15 & 0.18 & 0.21 \\
\hhline{-------}
\end{tabular}
\end{table}
Measurements took place on a GNU/Linux system equipped with an
Intel Core~2 Quad CPU Q9400 at 2.66~GHz
and 8~Gbytes of main memory;  a single core was used and the maximum
resident set size over the entire set of tests was slightly above 53~Mbytes.
From these we can conclude that the difference in performance
between the two methods is rather limited.  The PR method is
more efficient on the problem of semi-deciding termination,
with or without the computation of a witness ranking function,
while the MS method is superior on the problem of computing the
space of all affine ranking functions.

We also present, in Table~\ref{tab:precision-conditional-termination},
the precision results.  For each benchmark, along with
the total number of loops, we have the number of loops
for which termination is decided positively, either with the
MS or the PR method (column `term'); the remaining loops
are divided, using the MS method, between those that admit
a linear decreasing function (column `w/ d.f.')
and those who do not (column `w/o d.f.').
\begin{table}
\caption{Precision results and application to conditional termination}
\label{tab:precision-conditional-termination}
\centering
\begin{tabular}{lrrrr}
benchmark &
  \multicolumn{1}{c}{loops} &
    \multicolumn{1}{c}{term} &
    \multicolumn{1}{c}{w/ d.f.} &
    \multicolumn{1}{c}{w/o d.f.} \\
\hhline{-----}
\verb+CaffeineMark+  &  151  &  149 &  0  &  2 \\
\verb+JLex+  &  467  &  453 &  3  &  11 \\
\verb+JavaCC+  &  136  &  120 &  4  &  12 \\
\verb+Java_CUP+  &  29  &  27 &  0  &  2 \\
\verb+Jess+  &  151  &  149 &  0  &  2 \\
\verb+Kitten+  &  1484  &  1454 &  3  &  27 \\
\verb+NQueens+  &  359  &  271 &  4  &  84 \\
\verb+Raytracer+  &  8  &  6 &  0  &  2 \\
\verb+Termination+  &  121  &  119 &  0  &  2 \\
\hhline{-----}
\end{tabular}
\end{table}
It can be seen that the percentage of loops for which termination is
decided positively ranges from 75\% to 99\%, depending on the benchmark.
This means that we are conducting the experimental evaluation with a
termination analyzer, Julia, whose analysis algorithms ---though
certainly improvable--- very often provide enough information for
termination analysis.  This is crucial for the meaningfulness
of the experimental evaluation presented in this section.

\section{Conclusions}
\label{sec:conclusions}

Linear ranking functions play a crucial role in termination analysis,
as the termination of many programs can be decided by the existence
on one such function.
In this paper we have addressed the topic of the automatic synthesis
of linear ranking functions with the aim of clarifying its origins,
thoroughly explaining the underlying theory, and presenting new,
efficient implementations that are being made available to the general
public.

In particular, we have introduced, in general terms independent from
any programming paradigm, the problem of automatic termination
analysis of individual loops ---to which more general control flows
can be reconducted--- and its solution technique based on the synthesis
of ranking functions.

We have then presented and generalized a technique originally
due to Sohn and Van Gelder, that was virtually unknown outside
the logic programming field despite its general applicability
and its relative completeness (given a linear constraint system
approximating the behavior of a loop, if a linear ranking function
exists for that system, then the method will find it).
This method, due to its ability to characterize the spaces of all
the linear decreasing functions and all the linear bounded functions,
is also immediately applicable to \emph{conditional}
termination analysis; this theme is an excellent candidate for
future work.

We have also presented and, for the first time, fully justified,
a more recent technique by Podelski and Rybalchenko.
For this we also present an alternative formulation that can lead
to efficiency improvements.

We have compared the two methods, first proving their equivalence
---thus obtaining an independent confirmation on their correctness
and relative completeness--- and then studying their worst-case
complexity.

Finally, we have presented the implementation of all the techniques
described in the paper recently included in the Parma Polyhedra Library,
along with an experimental evaluation covering both the efficiency and
the precision of the analysis.

\section*{Acknowledgments}
We would like to express our gratitude to Amir Ben-Amram, David Merchat,
Andrey Rybalchenko, Alessandro Zaccagnini and the anonymous reviewers
for their useful comments and suggestions.
The connection between the MS method and conditional termination
analysis was indicated to us during a discussion with
Samir Genaim.


\begin{thebibliography}{10}
\expandafter\ifx\csname url\endcsname\relax
  \def\url#1{\texttt{#1}}\fi
\expandafter\ifx\csname href\endcsname\relax
  \def\href#1#2{#2} \def\path#1{#1}\fi

\bibitem{Lagarias85a}
J.~C. Lagarias, The $3x + 1$ problem and its generalizations, American
  Mathematical Monthly 92 (1985) 3--23.

\bibitem{Dijkstra76}
E.~W. Dijkstra, A Discipline of Programming, Prentice Hall, 1976.

\bibitem{CousotC85}
P.~Cousot, R.~Cousot, {`A la Floyd'} induction principles for proving
  inevitability properties of programs, in: M.~Nivat, J.~C. Reynolds (Eds.),
  Algebraic Methods in Semantics, Cambridge University Press, 1985, pp.
  277--312.

\bibitem{Floyd67}
R.~W. Floyd, Assigning meanings to programs, in: J.~T. Schwartz (Ed.),
  Mathematical Aspects of Computer Science, Vol.~19 of Proceedings of Symposia
  in Applied Mathematics, American Mathematical Society, Providence RI, New
  York City, USA, 1967, pp. 19--32.

\bibitem{BagnaraHZ08SCP}
R.~Bagnara, P.~M. Hill, E.~Zaffanella,
  \href{http://bugseng.com/products/ppl/documentation/BagnaraHZ08SCP.pdf}{The
  {Parma Polyhedra Library}: Toward a complete set of numerical abstractions
  for the analysis and verification of hardware and software systems}, Science
  of Computer Programming 72~(1--2) (2008) 3--21.

\bibitem{SohnVG91}
K.~Sohn, A.~{Van Gelder}, Termination detection in logic programs using
  argument sizes (extended abstract), in: Proceedings of the Tenth {ACM}
  {SIGACT-SIGMOD-SIGART} Symposium on Principles of Database Systems, ACM,
  Association for Computing Machinery, Denver, Colorado, United States, 1991,
  pp. 216--226.

\bibitem{PodelskiR04}
A.~Podelski, A.~Rybalchenko, A complete method for the synthesis of linear
  ranking functions, in: B.~Steffen, G.~Levi (Eds.), Verification, Model
  Checking and Abstract Interpretation: Proceedings of the 5th International
  Conference, VMCAI 2004, Vol. 2937 of Lecture Notes in Computer Science,
  Springer-Verlag, Berlin, Venice, Italy, 2004, pp. 239--251.

\bibitem{ColonS02}
M.~A. Col\'on, H.~B. Sipma, Practical methods for proving program termination,
  in: E.~Brinksma, K.~G. Larsen (Eds.), Computer Aided Verification:
  Proceedings of the 14th International Conference (CAV 2002), Vol. 2404 of
  Lecture Notes in Computer Science, Springer-Verlag, Berlin, Copenhagen,
  Denmark, 2002, pp. 442--454.

\bibitem{Sohn93th}
K.~Sohn, Automated termination analysis for logic programs, Ph.D. thesis,
  University of California Santa Cruz, Santa Cruz, CA, USA (1993).

\bibitem{Fischer02}
J.~Fischer, Termination analysis for {Mercury} using convex constraints,
  Honours report, Department of Computer Science and Software Engineering, The
  University of Melbourne, Melbourne, Australia (Aug. 2002).

\bibitem{SpeirsSS97}
C.~Speirs, Z.~Somogyi, H.~S{\o}ndergaard, Termination analysis for {Mercury},
  in: P.~{Van Hentenryck} (Ed.), Static Analysis: Proceedings of the 4th
  International Symposium, Vol. 1302 of Lecture Notes in Computer Science,
  Springer-Verlag, Berlin, Paris, France, 1997, pp. 157--171.

\bibitem{BradleyMS05CAV}
A.~R. Bradley, Z.~Manna, H.~B. Sipma, Linear ranking with reachability, in:
  Computer Aided Verification: Proceedings of the 15th International
  Conference, Vol. 3576 of Lecture Notes in Computer Science, Springer-Verlag,
  Berlin, Edinburgh, Scotland, UK, 2005, pp. 491--504.

\bibitem{Verschaetse91}
K.~Verschaetse, D.~{De Schreye}, Deriving termination proofs for logic
  programs, using abstract procedures, in: K.~Furukawa (Ed.), Proceedings of
  the 8th International Conference on Logic Programming, The MIT Press, Paris,
  France, 1991, pp. 301--315.

\bibitem{CodishT99}
M.~Codish, C.~Taboch, A semantic basis for the termination analysis of logic
  programs, Journal of Logic Programming 41~(1) (1999) 103--123.

\bibitem{Mesnard96}
F.~Mesnard, Inferring left-terminating classes of queries for constraint logic
  programs by means of approximations, in: M.~J. Maher (Ed.), Logic
  Programming: Proceedings of the Joint International Conference and Symposium
  on Logic Programming, MIT Press Series in Logic Programming, The MIT Press,
  Bonn, Germany, 1996, pp. 7--21.

\bibitem{MesnardS05}
F.~Mesnard, A.~Serebrenik, A polynomial-time decidable class of terminating
  binary constraint logic programs, Tech. Rep. 05-11, Universit\'e de la
  R\'eunion (2005).

\bibitem{MesnardS08}
F.~Mesnard, A.~Serebrenik, Recurrence with affine level mappings is {P}-time
  decidable for {CLP(R)}, Theory and Practice of Logic Programming 8~(1) (2008)
  111--119.

\bibitem{PodelskiR04TI}
A.~Podelski, A.~Rybalchenko, Transition invariants, in: Logic in Computer
  Science, Proceedings of the 19th IEEE Symposium, LICS 2004, IEEE Computer
  Society, Turku, Finland, 2004, pp. 32--41.

\bibitem{CookPR05}
B.~Cook, A.~Podelski, A.~Rybalchenko, Abstraction refinement for termination,
  in: C.~Hankin, I.~Siveroni (Eds.), Static Analysis: Proceedings of the 12th
  International Symposium, Vol. 3672 of Lecture Notes in Computer Science,
  Springer-Verlag, Berlin, London, UK, 2005, pp. 87--101.

\bibitem{NguyenDeS05}
M.~T. Nguyen, D.~{De Schreye}, Polynomial interpretations as a basis for
  termination analysis of logic programs, in: M.~Gabbrielli, G.~Gupta (Eds.),
  Proceedings of the 21st International Conference on Logic Programming, no.
  3668 in Lecture Notes in Computer Science, Springer-Verlag, Berlin, Sitges,
  Spain, 2005, pp. 311--325.

\bibitem{Lankford76}
D.~S. Lankford, A finite termination algorithm, Internal memo, Southwestern
  University, Georgetown, TX, USA (1976).

\bibitem{Zantema00}
H.~Zantema, Termination of term rewriting, Tech. Rep. UU-CS-2000-04, Department
  of Computer Science, Universiteit Utrecht, Utrecht, The Netherlands (2000).

\bibitem{Cousot05}
P.~Cousot, Proving program invariance and termination by parametric
  abstraction, lagrangian relaxation and semidefinite programming, in:
  R.~Cousot (Ed.), Verification, Model Checking and Abstract Interpretation:
  Proceedings of the 6th International Conference (VMCAI 2005), Vol. 3385 of
  Lecture Notes in Computer Science, Springer-Verlag, Berlin, Paris, France,
  2005, pp. 1--24.

\bibitem{ColonS01}
M.~A. Col\'on, H.~B. Sipma, Synthesis of linear ranking functions, in:
  T.~Margaria, W.~Yi (Eds.), Tools and Algorithms for Construction and Analysis
  of Systems, 7th International Conference, TACAS 2001, Vol. 2031 of Lecture
  Notes in Computer Science, Springer-Verlag, Berlin, Genova, Italy, 2001, pp.
  67--81.

\bibitem{DershowitzLSS01}
N.~Dershowitz, N.~Lindenstrauss, Y.~Sagiv, A.~Serebrenik, A general framework
  for automatic termination analysis of logic programs, Appl. Algebra Eng.
  Commun. Comput. 12~(1/2) (2001) 117--156.

\bibitem{LeeJB-A01}
C.~S. Lee, N.~D. Jones, A.~M. Ben-Amram, The size-change principle for program
  termination, in: C.~Norris, J.~J.~B. Fenwick (Eds.), Proceedings of the 28th
  ACM SIGPLAN-SIGACT Symposium on Principles of Programming Languages (POPL
  2001), Vol.~36 of ACM SIGPLAN Notices, ACM Press, London, UK, 2001, pp.
  81--92.

\bibitem{CodishLS05}
M.~Codish, V.~Lagoon, P.~Stuckey, Testing for termination with monotonicity
  constraints, in: M.~Gabbrielli, G.~Gupta (Eds.), Twenty First International
  Conference on Logic Programming, Vol. 3668 of Lecture Notes in Computer
  Science, Springer-Verlag, Berlin, Sitges, Spain, 2005, pp. 326--340.

\bibitem{Ben-Amram10}
A.~M. Ben-Amram, Size-change termination, monotonicity constraints and ranking
  functions, Logical Methods in Computer Science 6~(3:2) (2010) 1--32.

\bibitem{VasakP86}
T.~Vasak, J.~Potter, Characterisation of terminating logic programs, in:
  Proceedings of the 3rd IEEE Symposium on Logic Programming, IEEE Computer
  Society Press, Salt Lake City, Utah, USA, 1986, pp. 140--147.

\bibitem{LindenstraussS97}
N.~Lindenstrauss, Y.~Sagiv, Automatic termination analysis of logic programs,
  in: L.~Naish (Ed.), Logic Programming: Proceedings of the Fourteenth
  International Conference on Logic Programming, MIT Press Series in Logic
  Programming, The MIT Press, Leuven, Belgium, 1997, pp. 63--77.

\bibitem{JaffarM94}
J.~Jaffar, M.~J. Maher, Constraint logic programming: A survey, Journal of
  Logic Programming 19{\&}20 (1994) 503--582.

\bibitem{JaffarMMS98}
J.~Jaffar, M.~J. Maher, K.~Marriott, P.~J. Stuckey, The semantics of constraint
  logic programs, Journal of Logic Programming 37~(1-3) (1998) 1--46.

\bibitem{MesnardR03}
F.~Mesnard, S.~Ruggieri, On proving left termination of constraint logic
  programs, ACM Transactions on Computational Logic 4~(2) (2003) 207--259
  (paper) and 1--26 (electronic appendix).

\bibitem{CousotC77}
P.~Cousot, R.~Cousot, Abstract interpretation: A unified lattice model for
  static analysis of programs by construction or approximation of fixpoints,
  in: Proceedings of the Fourth Annual ACM Symposium on Principles of
  Programming Languages, ACM Press, Los Angeles, CA, USA, 1977, pp. 238--252.

\bibitem{CousotC92lp}
P.~Cousot, R.~Cousot, Abstract interpretation and applications to logic
  programs, Journal of Logic Programming 13~(2{\&}3) (1992) 103--179.

\bibitem{AptP93}
K.~R. Apt, D.~Pedreschi, Reasoning about termination of pure {Prolog} programs,
  Information and Computation 106~(1) (1993) 109--157.

\bibitem{BakerS93}
N.~Baker, H.~S{\o}ndergaard, Definiteness analysis for {CLP(${\cal R}$)}, in:
  G.~Gupta, G.~Mohay, R.~Topor (Eds.), Proceedings of the Sixteenth Australian
  Computer Science Conference, Vol.~15 of Australian Computer Science
  Communications, Brisbane, Australia, 1993, pp. 321--332.

\bibitem{CodishGBGV03}
M.~Codish, S.~Genaim, M.~Bruynooghe, J.~Gallagher, W.~Vanhoof, One loop at a
  time, in: A.~Rubio (Ed.), Proceedings of the 6th International Workshop on
  Termination (WST'03), Departamento de Sistemas Inform{\'a}ticos y
  Computaci{\'o}n, Universidad Polit{\'e}cnica de Valencia, Valencia, Spain,
  2003, pp. 1--4, published as Technical Report DSIC-II/15/03.

\bibitem{GareyJ90}
M.~R. Garey, D.~S. Johnson, Computers and Intractability; A Guide to the Theory
  of NP-Completeness, W. H. Freeman \& Co., New York, NY, 1990.

\bibitem{Schrijver86}
A.~Schrijver, Theory of linear and integer programming, Wiley, Chichester, New
  York, 1986.

\bibitem{CousotH78}
P.~Cousot, N.~Halbwachs,
  \href{http://www.di.ens.fr/~cousot/publications.www/CousotHalbwachs-POPL-78-%
ACM-p84--97-1978.pdf}{Automatic discovery of linear restraints among variables
  of a program}, in: Conference Record of the Fifth Annual ACM Symposium on
  Principles of Programming Languages, ACM Press, Tucson, Arizona, 1978, pp.
  84--96.

\bibitem{CookGL-ARS08}
B.~Cook, S.~Gulwani, T.~Lev-Ami, A.~Rybalchenko, M.~Sagiv, Proving conditional
  termination, in: A.~Gupta, S.~Malik (Eds.), Computer Aided Verification:
  Proceedings of the 20th International Conference (CAV 2008), Vol. 5123 of
  Lecture Notes in Computer Science, Springer-Verlag, Berlin, Princeton, NJ,
  USA, 2008, pp. 328--340.

\bibitem{Rybalchenko04th}
A.~Rybalchenko, Temporal verification with transition invariants, Ph.D. thesis,
  Universit{\"a}t des Saarlandes, Saarbr{\"u}cken, Germany (2004).

\bibitem{SerebrenikM05}
A.~Serebrenik, F.~Mesnard, On termination of binary {CLP} programs, in:
  S.~Etalle (Ed.), Logic Based Program Synthesis and Transformation: 14th
  International Symposium, Revised Selected Papers, no. 3573 in Lecture Notes
  in Computer Science, Springer-Verlag, Berlin, Verona, Italy, 2005, pp.
  231--244.

\bibitem{PapadimitriouS82}
C.~H. Papadimitriou, K.~Steiglitz, Combinatorial Optimization: Algorithms and
  Complexity, Prentice Hall, Englewood Cliffs, NJ, USA, 1982.

\bibitem{SpielmanT04}
D.~A. Spielman, S.-H. Teng, Smoothed analysis: Why the simplex algorithm
  usually takes polynomial time, Journal of the ACM 51 (2004) 385--463.

\bibitem{BagnaraHZ09TCS}
R.~Bagnara, P.~M. Hill, E.~Zaffanella,
  \href{http://bugseng.com/products/ppl/documentation/BagnaraHZ09TCS.pdf}{Appl%
ications of polyhedral computations to the analysis and verification of
  hardware and software systems}, Theoretical Computer Science 410~(46) (2009)
  4672--4691.

\bibitem{BagnaraHRZ05SCP}
R.~Bagnara, P.~M. Hill, E.~Ricci, E.~Zaffanella,
  \href{http://bugseng.com/products/ppl/documentation/BagnaraHRZ05SCP.pdf}{Pre%
cise widening operators for convex polyhedra}, Science of Computer Programming
  58~(1--2) (2005) 28--56.

\bibitem{BagnaraHZ05FAC}
R.~Bagnara, P.~M. Hill, E.~Zaffanella, Not necessarily closed convex polyhedra
  and the double description method, Formal Aspects of Computing 17~(2) (2005)
  222--257.

\bibitem{BagnaraHZ06STTT}
R.~Bagnara, P.~M. Hill, E.~Zaffanella,
  \href{http://bugseng.com/products/ppl/documentation/BagnaraHZ06STTT.pdf}{Wid%
ening operators for powerset domains}, Software Tools for Technology Transfer
  8~(4/5) (2006) 449--466.

\bibitem{BagnaraHZ09FMSD}
R.~Bagnara, P.~M. Hill, E.~Zaffanella,
  \href{http://bugseng.com/products/ppl/documentation/BagnaraHZ09FMSD.pdf}{Wea%
kly-relational shapes for numeric abstractions: Improved algorithms and proofs
  of correctness}, Formal Methods in System Design 35~(3) (2009) 279--323.

\bibitem{BagnaraHZ10CGTA}
R.~Bagnara, P.~M. Hill, E.~Zaffanella,
  \href{http://bugseng.com/products/ppl/documentation/BagnaraHZ10CGTA.pdf}{Exa%
ct join detection for convex polyhedra and other numerical abstractions},
  Computational Geometry: Theory and Applications 43~(5) (2010) 453--473.

\bibitem{SpotoMP10}
F.~Spoto, F.~Mesnard, {\'E}.~Payet, A termination analyzer for {Java} bytecode
  based on path-length, ACM Transactions on Programming Languages and Systems
  32~(3).

\end{thebibliography}

\begin{thebibliography}{10}
\expandafter\ifx\csname url\endcsname\relax
  \def\url#1{\texttt{#1}}\fi
\expandafter\ifx\csname href\endcsname\relax
  \def\href#1#2{#2} \def\path#1{#1}\fi

\bibitem{Lagarias85a}
J.~C. Lagarias, The $3x + 1$ problem and its generalizations, American
  Mathematical Monthly 92 (1985) 3--23.

\bibitem{Dijkstra76}
E.~W. Dijkstra, A Discipline of Programming, Prentice Hall, 1976.

\bibitem{BagnaraMPZ12TR}
R.~Bagnara, F.~Mesnard, A.~Pescetti, E.~Zaffanella, The automatic synthesis of
  linear ranking functions: The complete unabridged version,
  {\tt arXiv:cs.PL/1004.0944}, available from \url{http://arxiv.org/}. (2012).

\bibitem{CousotC85}
P.~Cousot, R.~Cousot, {`A la Floyd'} induction principles for proving
  inevitability properties of programs, in: M.~Nivat, J.~C. Reynolds (Eds.),
  Algebraic Methods in Semantics, Cambridge University Press, 1985, pp.
  277--312.

\bibitem{Floyd67}
R.~W. Floyd, Assigning meanings to programs, in: J.~T. Schwartz (Ed.),
  Mathematical Aspects of Computer Science, Vol.~19 of Proceedings of Symposia
  in Applied Mathematics, American Mathematical Society, Providence RI, New
  York City, USA, 1967, pp. 19--32.

\bibitem{BagnaraHZ08SCP}
R.~Bagnara, P.~M. Hill, E.~Zaffanella,
  \href{http://bugseng.com/products/ppl/documentation/BagnaraHZ08SCP.pdf}{The
  {Parma Polyhedra Library}: Toward a complete set of numerical abstractions
  for the analysis and verification of hardware and software systems}, Science
  of Computer Programming 72~(1--2) (2008) 3--21.

\bibitem{SohnVG91}
K.~Sohn, A.~{Van Gelder}, Termination detection in logic programs using
  argument sizes (extended abstract), in: Proceedings of the Tenth {ACM}
  {SIGACT-SIGMOD-SIGART} Symposium on Principles of Database Systems, ACM,
  Association for Computing Machinery, Denver, Colorado, United States, 1991,
  pp. 216--226.

\bibitem{PodelskiR04}
A.~Podelski, A.~Rybalchenko, A complete method for the synthesis of linear
  ranking functions, in: B.~Steffen, G.~Levi (Eds.), Verification, Model
  Checking and Abstract Interpretation: Proceedings of the 5th International
  Conference, VMCAI 2004, Vol. 2937 of Lecture Notes in Computer Science,
  Springer-Verlag, Berlin, Venice, Italy, 2004, pp. 239--251.

\bibitem{ColonS02}
M.~A. Col\'on, H.~B. Sipma, Practical methods for proving program termination,
  in: E.~Brinksma, K.~G. Larsen (Eds.), Computer Aided Verification:
  Proceedings of the 14th International Conference (CAV 2002), Vol. 2404 of
  Lecture Notes in Computer Science, Springer-Verlag, Berlin, Copenhagen,
  Denmark, 2002, pp. 442--454.

\bibitem{Sohn93th}
K.~Sohn, Automated termination analysis for logic programs, Ph.D. thesis,
  University of California Santa Cruz, Santa Cruz, CA, USA (1993).

\bibitem{Fischer02}
J.~Fischer, Termination analysis for {Mercury} using convex constraints,
  Honours report, Department of Computer Science and Software Engineering, The
  University of Melbourne, Melbourne, Australia (Aug. 2002).

\bibitem{SpeirsSS97}
C.~Speirs, Z.~Somogyi, H.~S{\o}ndergaard, Termination analysis for {Mercury},
  in: P.~{Van Hentenryck} (Ed.), Static Analysis: Proceedings of the 4th
  International Symposium, Vol. 1302 of Lecture Notes in Computer Science,
  Springer-Verlag, Berlin, Paris, France, 1997, pp. 157--171.

\bibitem{BradleyMS05CAV}
A.~R. Bradley, Z.~Manna, H.~B. Sipma, Linear ranking with reachability, in:
  Computer Aided Verification: Proceedings of the 15th International
  Conference, Vol. 3576 of Lecture Notes in Computer Science, Springer-Verlag,
  Berlin, Edinburgh, Scotland, UK, 2005, pp. 491--504.

\bibitem{Verschaetse91}
K.~Verschaetse, D.~{De Schreye}, Deriving termination proofs for logic
  programs, using abstract procedures, in: K.~Furukawa (Ed.), Proceedings of
  the 8th International Conference on Logic Programming, The MIT Press, Paris,
  France, 1991, pp. 301--315.

\bibitem{CodishT99}
M.~Codish, C.~Taboch, A semantic basis for the termination analysis of logic
  programs, Journal of Logic Programming 41~(1) (1999) 103--123.

\bibitem{Mesnard96}
F.~Mesnard, Inferring left-terminating classes of queries for constraint logic
  programs by means of approximations, in: M.~J. Maher (Ed.), Logic
  Programming: Proceedings of the Joint International Conference and Symposium
  on Logic Programming, MIT Press Series in Logic Programming, The MIT Press,
  Bonn, Germany, 1996, pp. 7--21.

\bibitem{MesnardS05}
F.~Mesnard, A.~Serebrenik, A polynomial-time decidable class of terminating
  binary constraint logic programs, Tech. Rep. 05-11, Universit\'e de la
  R\'eunion (2005).

\bibitem{MesnardS08}
F.~Mesnard, A.~Serebrenik, Recurrence with affine level mappings is {P}-time
  decidable for {CLP(R)}, Theory and Practice of Logic Programming 8~(1) (2008)
  111--119.

\bibitem{PodelskiR04TI}
A.~Podelski, A.~Rybalchenko, Transition invariants, in: Logic in Computer
  Science, Proceedings of the 19th IEEE Symposium, LICS 2004, IEEE Computer
  Society, Turku, Finland, 2004, pp. 32--41.

\bibitem{CookPR05}
B.~Cook, A.~Podelski, A.~Rybalchenko, Abstraction refinement for termination,
  in: C.~Hankin, I.~Siveroni (Eds.), Static Analysis: Proceedings of the 12th
  International Symposium, Vol. 3672 of Lecture Notes in Computer Science,
  Springer-Verlag, Berlin, London, UK, 2005, pp. 87--101.

\bibitem{NguyenDeS05}
M.~T. Nguyen, D.~{De Schreye}, Polynomial interpretations as a basis for
  termination analysis of logic programs, in: M.~Gabbrielli, G.~Gupta (Eds.),
  Proceedings of the 21st International Conference on Logic Programming, no.
  3668 in Lecture Notes in Computer Science, Springer-Verlag, Berlin, Sitges,
  Spain, 2005, pp. 311--325.

\bibitem{Lankford76}
D.~S. Lankford, A finite termination algorithm, Internal memo, Southwestern
  University, Georgetown, TX, USA (1976).

\bibitem{Zantema00}
H.~Zantema, Termination of term rewriting, Tech. Rep. UU-CS-2000-04, Department
  of Computer Science, Universiteit Utrecht, Utrecht, The Netherlands (2000).

\bibitem{Cousot05}
P.~Cousot, Proving program invariance and termination by parametric
  abstraction, lagrangian relaxation and semidefinite programming, in:
  R.~Cousot (Ed.), Verification, Model Checking and Abstract Interpretation:
  Proceedings of the 6th International Conference (VMCAI 2005), Vol. 3385 of
  Lecture Notes in Computer Science, Springer-Verlag, Berlin, Paris, France,
  2005, pp. 1--24.

\bibitem{ColonS01}
M.~A. Col\'on, H.~B. Sipma, Synthesis of linear ranking functions, in:
  T.~Margaria, W.~Yi (Eds.), Tools and Algorithms for Construction and Analysis
  of Systems, 7th International Conference, TACAS 2001, Vol. 2031 of Lecture
  Notes in Computer Science, Springer-Verlag, Berlin, Genova, Italy, 2001, pp.
  67--81.

\bibitem{JaffarM94}
J.~Jaffar, M.~J. Maher, Constraint logic programming: A survey, Journal of
  Logic Programming 19{\&}20 (1994) 503--582.

\bibitem{GareyJ90}
M.~R. Garey, D.~S. Johnson, Computers and Intractability; A Guide to the Theory
  of NP-Completeness, W. H. Freeman \& Co., New York, NY, 1990.

\bibitem{Schrijver86}
A.~Schrijver, Theory of linear and integer programming, Wiley, Chichester, New
  York, 1986.

\bibitem{CousotH78}
P.~Cousot, N.~Halbwachs,
  \href{http://www.di.ens.fr/~cousot/publications.www/CousotHalbwachs-POPL-78-%
ACM-p84--97-1978.pdf}{Automatic discovery of linear restraints among variables
  of a program}, in: Conference Record of the Fifth Annual ACM Symposium on
  Principles of Programming Languages, ACM Press, Tucson, Arizona, 1978, pp.
  84--96.

\bibitem{CookGL-ARS08}
B.~Cook, S.~Gulwani, T.~Lev-Ami, A.~Rybalchenko, M.~Sagiv, Proving conditional
  termination, in: A.~Gupta, S.~Malik (Eds.), Computer Aided Verification:
  Proceedings of the 20th International Conference (CAV 2008), Vol. 5123 of
  Lecture Notes in Computer Science, Springer-Verlag, Berlin, Princeton, NJ,
  USA, 2008, pp. 328--340.

\bibitem{Rybalchenko04th}
A.~Rybalchenko, Temporal verification with transition invariants, Ph.D. thesis,
  Universit{\"a}t des Saarlandes, Saarbr{\"u}cken, Germany (2004).

\bibitem{SerebrenikM05}
A.~Serebrenik, F.~Mesnard, On termination of binary {CLP} programs, in:
  S.~Etalle (Ed.), Logic Based Program Synthesis and Transformation: 14th
  International Symposium, Revised Selected Papers, no. 3573 in Lecture Notes
  in Computer Science, Springer-Verlag, Berlin, Verona, Italy, 2005, pp.
  231--244.

\bibitem{PapadimitriouS82}
C.~H. Papadimitriou, K.~Steiglitz, Combinatorial Optimization: Algorithms and
  Complexity, Prentice Hall, Englewood Cliffs, NJ, USA, 1982.

\bibitem{SpielmanT04}
D.~A. Spielman, S.-H. Teng, Smoothed analysis: Why the simplex algorithm
  usually takes polynomial time, Journal of the ACM 51 (2004) 385--463.

\bibitem{BagnaraHZ09TCS}
R.~Bagnara, P.~M. Hill, E.~Zaffanella,
  \href{http://bugseng.com/products/ppl/documentation/BagnaraHZ09TCS.pdf}{Appl%
ications of polyhedral computations to the analysis and verification of
  hardware and software systems}, Theoretical Computer Science 410~(46) (2009)
  4672--4691.

\bibitem{BagnaraHRZ05SCP}
R.~Bagnara, P.~M. Hill, E.~Ricci, E.~Zaffanella,
  \href{http://bugseng.com/products/ppl/documentation/BagnaraHRZ05SCP.pdf}{Pre%
cise widening operators for convex polyhedra}, Science of Computer Programming
  58~(1--2) (2005) 28--56.

\bibitem{BagnaraHZ05FAC}
R.~Bagnara, P.~M. Hill, E.~Zaffanella, Not necessarily closed convex polyhedra
  and the double description method, Formal Aspects of Computing 17~(2) (2005)
  222--257.

\bibitem{BagnaraHZ06STTT}
R.~Bagnara, P.~M. Hill, E.~Zaffanella,
  \href{http://bugseng.com/products/ppl/documentation/BagnaraHZ06STTT.pdf}{Wid%
ening operators for powerset domains}, Software Tools for Technology Transfer
  8~(4/5) (2006) 449--466.

\bibitem{BagnaraHZ09FMSD}
R.~Bagnara, P.~M. Hill, E.~Zaffanella,
  \href{http://bugseng.com/products/ppl/documentation/BagnaraHZ09FMSD.pdf}{Wea%
kly-relational shapes for numeric abstractions: Improved algorithms and proofs
  of correctness}, Formal Methods in System Design 35~(3) (2009) 279--323.

\bibitem{BagnaraHZ10CGTA}
R.~Bagnara, P.~M. Hill, E.~Zaffanella,
  \href{http://bugseng.com/products/ppl/documentation/BagnaraHZ10CGTA.pdf}{Exa%
ct join detection for convex polyhedra and other numerical abstractions},
  Computational Geometry: Theory and Applications 43~(5) (2010) 453--473.

\bibitem{SpotoMP10}
F.~Spoto, F.~Mesnard, {\'E}.~Payet, A termination analyzer for {Java} bytecode
  based on path-length, ACM Transactions on Programming Languages and Systems
  32~(3).

\end{thebibliography}

\ifthenelse{\boolean{LONGVERSION}}{
\hyphenation{ Ba-gna-ra Bie-li-ko-va Bruy-noo-ghe Common-Loops DeMich-iel
  Dober-kat Di-par-ti-men-to Er-vier Fa-la-schi Fell-eisen Gam-ma Gem-Stone
  Glan-ville Gold-in Goos-sens Graph-Trace Grim-shaw Her-men-e-gil-do Hoeks-ma
  Hor-o-witz Kam-i-ko Kenn-e-dy Kess-ler Lisp-edit Lu-ba-chev-sky
  Ma-te-ma-ti-ca Nich-o-las Obern-dorf Ohsen-doth Par-log Para-sight Pega-Sys
  Pren-tice Pu-ru-sho-tha-man Ra-guid-eau Rich-ard Roe-ver Ros-en-krantz
  Ru-dolph SIG-OA SIG-PLAN SIG-SOFT SMALL-TALK Schee-vel Schlotz-hauer
  Schwartz-bach Sieg-fried Small-talk Spring-er Stroh-meier Thing-Lab Zhong-xiu
  Zac-ca-gni-ni Zaf-fa-nel-la Zo-lo }\newcommand{\noopsort}[1]{}

}{
\hyphenation{ Ba-gna-ra Bie-li-ko-va Bruy-noo-ghe Common-Loops DeMich-iel
  Dober-kat Di-par-ti-men-to Er-vier Fa-la-schi Fell-eisen Gam-ma Gem-Stone
  Glan-ville Gold-in Goos-sens Graph-Trace Grim-shaw Her-men-e-gil-do Hoeks-ma
  Hor-o-witz Kam-i-ko Kenn-e-dy Kess-ler Lisp-edit Lu-ba-chev-sky
  Ma-te-ma-ti-ca Nich-o-las Obern-dorf Ohsen-doth Par-log Para-sight Pega-Sys
  Pren-tice Pu-ru-sho-tha-man Ra-guid-eau Rich-ard Roe-ver Ros-en-krantz
  Ru-dolph SIG-OA SIG-PLAN SIG-SOFT SMALL-TALK Schee-vel Schlotz-hauer
  Schwartz-bach Sieg-fried Small-talk Spring-er Stroh-meier Thing-Lab Zhong-xiu
  Zac-ca-gni-ni Zaf-fa-nel-la Zo-lo }\newcommand{\noopsort}[1]{}

}

\end{document}